\documentclass[a4paper,oneside,english]{elsart}
\usepackage[T1]{fontenc} 
\usepackage[latin1]{inputenc}
\usepackage{subfigure} 
\usepackage{float} 
\usepackage{graphicx}
\usepackage{setspace}
\usepackage{gensymb}
\usepackage{amsmath,xcolor,esint}
\usepackage[abs]{overpic}
\makeatletter
\newcommand*\dif{\mathop{}\!\mathrm{d}}

\usepackage{babel} \makeatother
\begin{document}

\begin{frontmatter}
\title{A five-equation model for the simulation of miscible and viscous compressible fluids}

\author{ Ben Thornber $^*$,  Michael Groom$^*$  David Youngs$^{\dag}$}

\address{$^*$ The University of Sydney, School of Aerospace, Mechanical and Mechatronic Engineering, Sydney}
\address{$^{\dag}$ University of Strathclyde, Department of Mechanical and Aerospace Engineering,Glasgow, G1 1XW, United Kingdom}

\newpage

\begin{abstract}

Typical multispecies compressible Navier-Stokes computations employ conservative equations for mass fraction transport. Upwind discretisations of these governing equations produce spurious pressure oscillations at diffuse contact surfaces between gases of differing ratio of specific heat capacities which degrade the convergence rate of the algorithm. Adding quasi-conservative equations for volume fraction can solve this error, however this approach has been derived only for immiscible fluids. Here, a five-equation quasi-conservative model is proposed that includes the effects of species diffusion, viscosity and thermal conductivity. The derivation of the model is presented, along with a numerical method to solve the governing equations at second order accuracy in space and time. Formal convergence studies demonstrate the expected order of accuracy is achieved for three benchmark problems, cross-validated against two standard mass fraction models. In these test cases, the new  model has between $2$ and $10$ times lower error for a given grid size. Simulations of a two-dimensional air-SF$_6$ Richtmyer-Meshkov instability demonstrate that the new model converges to the solution with four times fewer points in each direction when compared to the mass fraction model in an identical numerical framework. This represents an $\approx 40$ times lower computational cost for an equivalent error in two-dimensional computations. The proposed model is thus very suitable for Direct Numerical Simulation and Large Eddy Simulation of compressible mixing.
\end{abstract}
\end{frontmatter}

\newpage

\section{Introduction}

This paper is concerned with the simulation of miscible, multicomponent fluids with single momentum equation for the mass-weighted mean velocity of the mixture, including the effects of viscous, thermal and molecular diffusion which may be solved using shock-capturing numerical methods, also known as diffuse interface methods. By making standard simplifying assumptions for flows of this type (see e.g. \cite{Ramshaw2002}), the governing equations can be shown to be the compressible Navier-Stokes equations, complemented by an additional transport equation(s) that is used to determine the composition of the mixture. One possible choice is to add one or more additional equation(s) for the transport of (N-1) species mass fractions, resulting in a system of governing equations that is fully conservative, commonly referred to as the mass fraction or four-equation model. However, it has been well documented that when using the standard four-equation model to simulate flows where the ratio of specific heats varies with mixture composition, spurious pressure oscillations are generated when material interfaces are advected through the computational mesh \cite{Larrouturou1989}. Other types of conservative discretisations, such as the level-set model, also suffer from this error under the same conditions \cite{Abgrall2000}.

The primary motivation of this current research is to enable accurate, high-resolution viscous computations of compressible turbulent mixing problems. These types of problems are significant in applications such as inertial confinement fusion \cite{icfrmi}, supersonic combustion \cite{Yang1993} and astrophysics \cite{Arnett2000}. Of crucial importance is the understanding of fundamental flow instabilities that trigger transition to turbulence, such as the Richtmyer-Meshkov instability \cite{Richtmyer1960,Meshkov1969}. An example application is in the computation of the implosion of inertial confinement fusion capsules \cite{icfrmi,Clark2016}. Initially there are sharp interfaces at solid/solid or solid/gas interfaces. The very high temperatures and pressures generated by the implosion process convert all the materials to dense gaseous plasmas and at various stages of the implosion there may be both sharp and diffuse gas/gas interfaces. The need to model both sharp contact surfaces and diffuse boundaries is also a requirement in shock tube turbulent mixing experiments (see, e.g. \cite{Hahn2011}).

For these applications the numerical technique must be able to compute flows involving multiple gaseous species which are initially separate but become mixed as time proceeds due to various hydrodynamic instabilities. The diffuse boundary layer between components may initially be small compared to the mesh size i.e. there is a well-defined contact surface.  As the flow evolves, near-homogeneous mixing due to species diffusion may occur at a scale which is at sub-cell level at early times, then becomes of the order the mesh size at time evolves. When simulating these transitional, non-equilibrium flows numerically, stability and high order of accuracy are required in order to properly capture all of the necessary physics.  The purpose of the current paper is to derive an advection equation which may be added to the standard system of Navier-Stokes plus mass fraction equations to accurately model both of these limits, including species diffusion, without spurious numerically-generated pressure oscillations. 

For multiphase single velocity models, this problem has been thoroughly studied and has prompted a number of non-conservative or quasi-conservative approaches to be proposed. Karni showed that the pressure oscillations could be eliminated, either by using a primitive variable formulation (with corrections for leading order conservation errors) \cite{Karni1994}, or a conservative formulation augmented with a non-conservative pressure evolution equation \cite{Karni1996}. Jenny \textit{et al.} \cite{Jenny1997} proposed a correction for the energy equation, rendering the computation of the conserved variables a single-fluid computation and hence reducing (but not eliminating) pressure oscillations. The same paper also gave an expression for the relative error in pressure generated across a contact discontinuity, propagating with constant velocity $u$, that is initially aligned with the cell interface $x_{i-\frac{1}{2}}$ and has the CFL criterion $0 < \sigma=u\Delta t/\Delta x<1$,

\begin{equation}
\epsilon_p = \sigma(1-\sigma)\frac{(T_2-T_1)(\gamma_1-\gamma_2)}{\sigma(\gamma_2-1)T_2+(1-\sigma)(\gamma_1-1)T_1},
\label{eqn:Jenny}
\end{equation}

\noindent where $T_k$ are the temperatures for species $k$ on either side of the contact surface, and $\gamma_k$ the ratio of specific heats for each species. 

This applies to a conservative scheme that is first order accurate in time and space, includes a complete Riemann solver, with the ideal gas equation of state. Hence differences in temperature and specific heat ratio across the interface, as well as the convective velocity, all contribute to the size of the pressure error. Abgrall \cite{Abgrall1996} showed that such pressure errors could be eliminated by using an additional transport equation (in advection form) for a given function of the ratio of specific heats. This approach was dubbed quasi-conservative as it produces results with extremely small conservation errors. This class of algorithm was extended to a wider range of equations of state and multiple dimensions by Shyue \cite{Shyue1998} and Saurel and Abgrall \cite{Saurel1999}. A key idea behind these extensions was to have as many additional transport equations as there are parameters in the equation of state (e.g. $\gamma$ for an ideal gas), while not requiring any additional transport equations to simulate mixtures of more than two fluids. However this results in increasing complexity of the algorithm with an increasingly complex equation of state. Other approaches for eliminating pressure oscillations have also been proposed that involve non-conservative modifications to the solution procedure rather than the addition of transport equations (see e.g. \cite{Abgrall2000},\cite{Billet2003},\cite{Fedkiw1999}) however they will not be discussed in further detail here. 

Shyue \cite{Shyue1998} also gave a reformulation of the $\gamma$-based model of Abgrall in terms of the four-equation volume fraction ,which consists of (in 1D) the three conservative equations for density, momentum and energy plus a non-conservative advection equation for the volume fraction of one of the species. Allaire \textit{et al.} \cite{Allaire2002} and Massoni {\it et al.} \cite{Massoni2002} showed that by extending this to a five-equation model and providing suitable thermodynamic closure, any equation of state could be simulated whilst also maintaining pressure equilibrium across material interfaces.  Massoni {\it et al.} included heat conduction, and applied it to a range of multiphase problems with varying equations of state, maintaining pressure equilibrium. Murrone and Guillard presented a detailed study of a five equation model with excellent results at low and high Mach \cite{Murrone2005}. The numerical model, in which contacts can become smeared, is presented for inviscid, immiscible flows only.

More recent research on quasi-conservative formulations for compressible multicomponent flows has focused on extensions to high-order accuracy as well as the inclusion of viscous effects. By performing an asymptotic analysis on equations governing multi-phase flow, Perigaud and Saurel \cite{Perigaud2005} derived the same five-equation model as Allaire \textit{et al.} and Massoni \textit{et al.} but including the effects of viscosity. This model neglects the effects of species diffusion however. Johnsen and Colonius \cite{Johnsen2006} extended the models of \cite{Abgrall1996,Shyue1998} to high-order WENO finite-volume methods with the HLLC Riemann solver. An important result in this study was that in order to preserve pressure equilibrium across contact discontinuities, the reconstruction at cell interfaces must be performed using the primitive variables.  

It was shown by Johnsen and Ham \cite{Johnsen2012} that although the models of \cite{Abgrall1996,Shyue1998} conserve total mass, momentum and energy, they do not discretely conserve the mass of each species and also generate temperature errors at material interfaces. These temperature errors are irrelevant in inviscid flows, however they become important once physical diffusion effects are added. The authors proposed adding a transport equation for species mass fraction to the model of Abgrall \cite{Abgrall1996} (in conservative form, thus enforcing species mass conservation) and showed that temperature errors were prevented if the reconstruction and upwinding for both the continuity and mass fraction equations is consistent. A potential downside of this approach is that the location of the interface is not uniquely defined, although this difference is very small. Finally, Coralic and Colonius \cite{Coralic2014} applied the numerical framework of \cite{Johnsen2006} to the five-equation model of Allaire \textit{et al.}, with the inclusion of the effects of viscosity as in \cite{Perigaud2005}. Species diffusion was not included however. 

At very large scales it may be sufficient to assume that the dissipative effects of physical mechanisms are approximated by those of the numerics (referred to as ILES, see e.g. \cite{Grinstein2007,Drikakis2003}), however at smaller scales the effects of viscosity, thermal conduction and species diffusion become important \cite{Walchli2017} and require explicit modeling so as to gain confidence in the results. Previously, any computation of miscible compressible turbulent mixing with species diffusion has had to use the fully-conservative mass fraction model, even when the ratio of specific heats is not constant \cite{Tritschler2014pre,Ruiz2015,Thornber2011,Gicquel2012}. As outlined above, this approach is likely to suffer from errors in pressure and/or temperature generated at material interfaces. These errors are important even in relatively well mixed flows as species gradients are always present in the flow and thus spurious numerically generated pressure fluctuations may be large compared to physically generated pressure fluctuations.

Thus this paper derives a five-equation model incorporating viscosity and diffusion, where the additional equation is employed solely to ensure that spurious pressure oscillations are not generated within a mixed cell. Firstly, the new governing model is proposed which incorporates the effects of diffusivity, conduction and viscosity, and simplified to a form which is most amenable to numerical solution in Section \ref{derivation}. Section \ref{numdisc} presents the integral form of the governing equations, the closure for the volume averaged equation of state, and a second order in time and space discretisation of the governing equations. Section \ref{results} details three one dimensional test cases which (i) verify the observed order of accuracy against analytical incompressible solutions and (ii) illustrate clearly the superiority of the new set of governing equations against the classical mass fraction model implemented in an identical algorithmic framework and a second independent Lagrange-remap algorithm. Section \ref{2dresults} introduces a two-dimensional Direct Numerical Simulation of a shock-induced instability between two miscible gases, using a setup which is typical of shock-tube experiments exploring the physics of the Richtmyer-Meshkov instability and further demonstrates the advantages of the proposed model equations. The key conclusions of the paper are then summarised in Section \ref{concl}.

\section{A Five-Equation Model for Viscous, Miscible Gases \label{derivation}}

\subsection{Standard Four-Equation System}

The standard four-equation model describing advection and diffusion of multiple differing species may be written as follows:

\begin{eqnarray}
\frac{\rho Y_k}{\partial t}+\nabla\cdot (\rho Y_k {\bf u})&=&\nabla\cdot (\rho D_{12} \nabla{Y_k})\label{gov1}\\
\frac{\partial \rho {\bf u}}{\partial t}+\nabla\cdot (\rho {\bf u}.{\bf u})&=&-\nabla\cdot {\bf S}\label{gov2}\\
\frac{\partial \rho e}{\partial t}+\nabla\cdot (\rho e {\bf u})&=&-\nabla\cdot ({\bf S} \cdot {\bf u}+{\bf q}+{\bf q_d}),\label{gov3}
\end{eqnarray}

where,

\begin{equation}
{\bf S}=p{\bf I}-\lambda_b (\nabla\cdot {\bf u}){\bf I}-\mu\left[(\nabla {\bf u}+(\nabla {\bf u})^T\right], \,\,\,\,\lambda_b=-\frac{2}{3}\mu.
\end{equation}
 
Note that ${\bf q}=-\kappa \nabla T$ where $T=p\overline{m}/R\rho$ is the mixture temperature with mixture molecular weight $\overline{m}=\left(\sum_k Y_k/m_k\right)^{-1}$, $R$ the universal gas constant, $m_K$ the species molecular weights. The enthalpy diffusion term ${\bf q_d}=\sum_k \rho D_{12} \nabla{Y_k} h_{s,k}$, where species sensible enthalpy $h_{s,k}=c_{p,k}T$. The specific total energy $e=\epsilon+|{\bf u}|^2/2$, the internal energy $\rho \epsilon=\sum_k \rho Y_k \epsilon_k$, with $\epsilon_K=c_{v,k}T_K$ and each fluid satisfies the perfect gas equation of state

\begin{equation}
p_k=(\gamma_k -1)\rho_k \epsilon_k,\,\,\,\,\ \gamma_k=c_{pk}/c_{vk}.
\label{pgas}
\end{equation}

To close the above system without supplementary equations, it is assumed that the species are intimately mixed and are thus in pressure and temperature equilibrium. Given pressure and temperature equilibrium then $p=(\gamma-1)\rho \epsilon$, where $\gamma=c_p/c_v$ and $c_p=\sum_k c_{pk}Y_k$, $c_v=\sum_k c_{vk} Y_k$. 

As stated in the introduction, it has been demonstrated that this system of equations suffers from spurious pressure oscillations in the inviscid limit when used to model the transportation of species of differing thermodynamic properties in a finite volume Godunov-type approach \cite{Larrouturou1989,Abgrall1996,Abgrall2000}. This has a substantial practical impact on the accuracy of computations undertaken at high Reynolds numbers, for cases evolving from an initially sharp interface, or with strongly varying temperatures and fluid properties. It is essential that a governing model for diffusive problems be able to represent gradients which may be steep compared to the local grid size without suffering from these numerically-generated errors. Previous papers have added a extra equation for this purpose but only for the case when species diffusion is absent \cite{Massoni2002,Allaire2002}.

\subsection{The Five-Equation Model}

In order to model species diffusion in the presence of steep gradients in mixture composition, an additional equation is now derived which enables the preservation of pressure-constancy at contact surfaces in the sharp-interface limit, and is valid for problems with viscous, thermal and species diffusion. It should be emphasised that the basic physics is the same as for the standard four-equation model. However, the integral form of the equations used for numerical integration is different, as explained in Section \ref{numdisc}, and it is this which eliminates spurious numerical pressure oscillations.

During diffusion, as mixing occurs at a molecular level, both pressure and temperature equilibrium is expected during the mixing process. Avogadro's hypothesis, (\cite{chapman1970mathematical} 2.42), is assumed to hold: gases at the same pressure and temperature have the same molecular number density. As each molecule for a given species has the same mass, the number weighted mean velocity, the volume-weighted mean velocity and the mass-weighted mean velocity for species $k$ are all the same and denoted by ${\bf u_k}$. The equation for evolution of species $k$ particle number density $n_k$:

\begin{equation}
\frac{\partial n_k}{\partial t}+\nabla\cdot (n_k {\bf u_k})=0
\end{equation}

where ${\bf u_k}$ is now the mean velocity of each component. By summation, the equation for the evolution of total number density is:

\begin{equation}
\frac{\partial N}{\partial t}+\nabla\cdot (N\overline{{\bf u}})=0
\end{equation}

where $\overline{{\bf  u}}=\frac{\sum_k n_k {\bf u_k}}{N}$ is the number-weighted mean velocity of the mixture. Next, the individual number density equations are rewritten using the number fraction $X_k=\frac{n_k}{N}$.  Employing $\nabla \cdot (\phi {\bf F})=\nabla \cdot \phi {\bf F}+\phi\nabla \cdot {\bf F}$ where $\phi$ is a scalar, leads to:

\begin{eqnarray}
\frac{\partial N X_k}{\partial t}+\nabla\cdot (N X_k {\bf u_k})&=&0\\
N\frac{\partial X_k}{\partial t}-X_k\nabla\cdot (N\overline{{\bf u}})+\nabla\cdot (N X_k {\bf u_k})&=&0\\
N\frac{\partial X_k}{\partial t}+N\overline{{\bf u}}\cdot \nabla X_k-\nabla\cdot (X_k N\overline{{\bf u}})+ X_k {\bf u_k}\nabla N+N\nabla\cdot ( X_k {\bf u_k})&=&0\\
\frac{\partial X_k}{\partial t}+\overline{{\bf u}}\cdot \nabla X_k-\nabla\cdot (X_k \overline{{\bf u}})-X_k\overline{{\bf u}}\frac{\nabla N}{N}+&&\\\nonumber \nabla\cdot ( X_k {\bf u_k})+ X_k {\bf u_k}\frac{\nabla N}{N}&=&0
\end{eqnarray}

\begin{equation}
\frac{\partial X_k }{\partial t}+\overline{{\bf u}} \cdot \nabla X_k+\nabla\cdot [X_k ({\bf u_k- \overline{u}})]+X_k({\bf u_k-\overline{u}})\cdot \frac{\nabla N}{N}=0
\end{equation}

Thus there are three terms which modify $X_k$; advection with the number-weighted mean velocity, diffusive mixing, and pressure-temperature equilibriation. According to the last term, if a parcel of fluid $k$ moves into a region with different pressure and temperature the number fraction adjusts to the local value.

\subsection{Binary Mixtures}

For a binary mixture the terms in the number fraction equation can be specified in a simple form. If Fickian diffusion is assumed dominant \cite{chapman1970mathematical} then

\begin{equation}
{\bf u_1}-  {\bf u_2}=-\frac{D_{12}}{X_1 X_2} \nabla{X_1},
\label{fickbin}
\end{equation}

Substituting $\bf u_2 =\frac{\overline{{\bf u}}-X_1 {\bf u_1}}{X_2}$ into Eq. (\ref{fickbin}) implies $X_k({\bf u_k- \overline{u}})=-D_{12} \nabla{X_k}$ and the number fraction equation becomes:

\begin{equation}
\frac{\partial X_k }{\partial t}+\overline{{\bf  u}} \cdot \nabla X_k=\nabla\cdot (D_{12} \nabla{X_k})+D_{12} \nabla{X_k}\cdot \frac{\nabla N}{N}
\end{equation}


The number-weighted mean velocity of the fluid $\overline{{\bf u}}$ is related to the mass weighted velocity ${\bf u}$  as follows:

\begin{equation}
{\bf u}=\frac{\sum_k m_k X_k {\bf u_k}}{\sum_k m_k X_k}=\overline{{\bf u}}+\frac{\sum_k m_k X_k ({\bf u_k}-\overline{{\bf u}})}{\sum_k m_k X_k},
\end{equation}

where $m_k$ denotes the molecular mass for species $k$. Thus for two species,

\begin{equation}
\overline{{\bf u}}={\bf u}+\frac{m_1-m_2}{m_1 X_1+m_2 X_2}D_{12}\nabla X_1
\end{equation}

This form is useful as the further manipulations may result in numerical difficulties associated with dividing by number fractions or mass fractions which could be zero.

For binary diffusion, $D_{kj}$ are equal and gradient diffusion is dominant, then diffusive fluxes can be related to mass fraction gradients \cite{Williams1985}, and Eq. (\ref{fickbin}) can also be written as

\begin{equation}
{\bf u_1}-  {\bf u_2}=-\frac{D_{12}}{Y_1 Y_2} \nabla{Y_1}.
\label{fick2}
\end{equation}

This is employed as the diffusion term for the evolution of $\rho Y_k$. The full set of equations for compressible binary mixture advection diffusion problems is then:

\begin{eqnarray}
\frac{\partial \rho Y_k}{\partial t}+\nabla\cdot (\rho Y_k {\bf u})&=&\nabla\cdot (\rho D_{12} \nabla{Y_k}),\\
\frac{\partial \rho {\bf u}}{\partial t}+\nabla\cdot (\rho {\bf u}.{\bf u})&=&-\nabla\cdot {\bf S},\\
\frac{\partial \rho e}{\partial t}+\nabla\cdot (\rho e {\bf u})&=&-\nabla\cdot ({\bf S} \cdot {\bf u}+{\bf q}+{\bf q_d}),\\
\frac{\partial X_k }{\partial t}+{\bf u} \cdot \nabla X_k&=&\nabla \cdot (D_{12} \nabla{X_k})-\mathcal{M}D_{12} \nabla X_1 \cdot \nabla X_k+\nonumber\\ &&\hspace{5cm} D_{12} \nabla{X_k}\cdot\frac{\nabla N}{N},
\end{eqnarray}

where,

\begin{equation}
{\bf S}=p{\bf I}-\lambda_b (\nabla\cdot {\bf u}){\bf I}-\mu\left[(\nabla {\bf u}+(\nabla {\bf u})^T\right], \,\,\,\,\mathcal{M}=\frac{m_1-m_2}{m_1 X_1+m_2 X_2}.
\end{equation}

and $\lambda_b=-\frac{2}{3}\mu$. Finally ${\bf q}=-\kappa \nabla T$ where $T=p\overline{m}/R\rho$ is the mixture temperature in the current computations, ${\bf q_d}=\sum_k \rho D_{12} \nabla{Y_k} h_{sk}$ with $h_{s,k}=c_{p,k}T_k$, and number density $N=p/k_bT$ where $k_b$ is the Boltzmann constant. 

\subsection{Multiple Species}

The testcases presented in this paper will focus exclusively on binary mixtures, however the model can be extended to multiple species. In the case where diffusion is dominated by the $\nabla(X_k)$ terms \cite{chapman1970mathematical}, 

\begin{equation}
\nabla{X_k}=-\sum_j\frac{X_k X_j}{D_{kj}}({\bf u_k}-  {\bf u_j}),
\label{fickmultiple}
\end{equation}

The number weighted velocity is then given by $\overline{{\bf u}}-{\bf u}=\sum_{j,k} X_k Y_j (u_k-u_j)$. Here a simplified form of the number fraction equation is derived for the case where each species has an effective mixture diffusion coefficient $D_k$, such as the well known approximation of Hirschfelder and Curtiss\cite{Hirschfelder1969}. Giovangigli \cite{Giovangigli1991} demonstrated that the Hirschfelder and Curtiss approximated diffusion velocities can be defined as 

\begin{equation}
{\bf u_k}-{\bf u}=-\frac{D_k}{X_k}\nabla X_k+V_c,\hspace{0.2cm} D_k=\frac{1-Y_k}{\sum_{j \neq k} X_j/D_{kj}}
\label{uku}
\end{equation}

The velocity correction is employed to ensure that diffusive mass fluxes sum to zero, i.e. $\sum_k Y_k ({\bf u_k}-{\bf u})=0$. Thus

\begin{equation}
{\bf J}_k=\rho Y_k\left({\bf u_k}-  {\bf u}\right)=-\rho\left(D_k \frac{Y_k}{X_k}\nabla X_k -Y_k\sum_j D_j \frac{Y_j}{X_k}\nabla X_j\right)
\label{fickmultiple2}
\end{equation}

Now

\begin{equation}
Y_k=\frac{m_k X_k}{\overline {m}},\hspace{0.2cm} \overline {m}=\sum_k m_k X_k, \hspace{0.2cm} \frac{\nabla Y_k}{Y_k}=\frac{\nabla X_k}{X_k}-\frac{\nabla \overline {m}}{\overline {m}}
\end{equation}

From Eq. (\ref{uku}) it follows that the difference between the number weighted mean and the mass weighted mean is given by

\begin{equation}
{\overline {\bf u}}-{\bf u}=\sum_j X_j u_j-\sum_j Y_j u_j=\sum_j\left(\frac{Y_j}{X_j}-1\right)D_j \nabla X_j
\end{equation}

It also follows that 

\begin{equation}
{\bf u_k}-  {\overline{\bf u}}= -D_k \frac{\nabla X_k}{X_k} +\sum_j D_j \nabla X_j
\end{equation}

which leads to the following equation under the assumption of pointwise temperature and pressure equilibrium:

\begin{equation}
\frac{\partial X_k }{\partial t}+\overline{\bf u} \cdot \nabla X_k=-\nabla \cdot {\bf J^v_k}-{\bf J^v_k}\cdot\frac{\nabla N}{N}
\label{fickmultiple3}
\end{equation}

with

\begin{equation}
{\bf J^v_k}=X_k ({\bf u_k}-\overline {\bf u})=-D_k \nabla X_k+ X_k \sum_j D_j \nabla X_j,\hspace{0.2cm} \overline{\bf u}={\bf u}+\sum_j\left(\frac{Y_j}{X_j}-1\right)D_j \nabla X_j
\end{equation} 

Note that the number fraction diffusion fluxes sum to zero, as required. For numerical discretisation, it may be preferable to set $Y_j/X_j=m_j/{\overline m}$ which would be less likely to suffer from inaccuracy at very low number or mass fractions due to machine error.  Note that for fundamental studies of multispecies mixing using Direct Numerical Simulation it may be  useful to assume constant diffusivity , $D_{kj}=D$ \cite{Youngs2017}.  In that case Eq. (\ref{fickmultiple}) can be solved directly to give

\begin{eqnarray}
{\bf J_k}=\rho Y_k ({\bf u_k}-{\bf u})&=&-\rho D \nabla Y_k\\
{\bf J^v_k}=X_k ({\bf u_k}-  {\overline{\bf u}})&=&-D \nabla X_k
\end{eqnarray}

\section{Integral Form and Spatially Averaged Equation of State \label{numdisc}}

\subsection{Integral Form}

The new set of governing equations for $n$ fluid species can be written in integral form as:

\begin{equation*}
\begin{aligned}
& \frac{\partial}{\partial t}\iiint_{V}\rho Y_k \dif V+\oiint_{A}(\rho Y_k\mathbf{u})\cdot\mathbf{n}\dif A=\oiint_{A}(\rho D_{12}\nabla Y_k)\cdot\mathbf{n}\dif A \\
& \frac{\partial}{\partial t}\iiint_{V}\rho\mathbf{u} \dif V+\oiint_{A}(\rho\mathbf{u}\cdot\mathbf{u})\cdot\mathbf{n}\dif A=-\oiint_{A}\mathbf{S}\cdot\mathbf{n}\dif A \\
& \frac{\partial}{\partial t}\iiint_{V}\rho e \dif V+\oiint_{A}(\rho e\mathbf{u})\cdot\mathbf{n}\dif A=-\oiint_{A}(\mathbf{S}\cdot\mathbf{u}+\mathbf{q}+\mathbf{q_d})\cdot\mathbf{n}\dif A \\
& \frac{\partial}{\partial t}\iiint_{V}X_k \dif V+\iiint_{V}(\mathbf{u}+\mathcal{M}D_{12}\nabla X_1-D_{12}\frac{\nabla N}{N})\cdot\nabla X_k\dif V=\oiint_{A}(D_{12}\nabla X_k)\cdot\mathbf{n}\dif A \\
\end{aligned}
\end{equation*} 
\noindent

Note that the final equation need only be solved for $n-1$ components. Writing $\mathcal{U}=\mathbf{u}+\mathcal{M}D_{12}\nabla X_1-D_{12}\frac{\nabla N}{N}$ and using the identity $\nabla\cdot(\psi\mathbf{V})=\mathbf{V}\cdot\nabla\psi+\psi\nabla\cdot\mathbf{V}$, the number fraction equation may be written as a conservative equation minus a correction:

\begin{equation*}
\frac{\partial}{\partial t}\iiint_{V} X_k \dif V+\oiint_{A}(\mathcal{U}X_k)\cdot\mathbf{n}\dif A-\iiint_{V}X_k(\nabla\cdot\mathcal{U})\dif V=\oiint_{A}(D_{12}\nabla X_k)\cdot\mathbf{n}\dif A
\end{equation*} 

For the applications considered here, interfaces may be initially sharp (sub-cell), then become diffuse as time proceeds. Thus it is important to clarify the definition of the volume averaged $\overline{X_k}$-variable. For a computational cell

\begin{equation}
\overline{X_k} =\frac{\iiint_{V} X_k \dif V}{V^{cell}}
\end{equation}

\noindent where $\overline{X_K}$ is thus the volume averaged number fraction for species $k$. This differs from the number fraction based on the number of molecules of species $k$ in the computational cell:-

\begin{equation}
\widetilde{X_k}=\frac{\iiint_{V}N X_k \dif V}{\iiint_{V}N \dif V}
\end{equation}

\noindent where $N$ is the total number density. If there is a sharp interface within a cell, $\overline{X_k}$ is equal to the cell averaged volume fraction $z_k$ commonly employed in multiphase systems \cite{Allaire2002,Massoni2002,Murrone2005}. If there is homogeneous mixing within a cell, $\overline{X_k}=\tilde{X_k}$. Moreover, if pressure and temperature are uniform within a cell, then so is $N$ (according to Avogadro's hypothesis: equal volumes of gas at the same temperature and pressure contain the same number of molecules) and in this more general case $\overline{X_k}=\tilde{X_k}$. However, in typical cases $\overline{X_k}$ and $\tilde{X_k}$ will not be equivalent, as volume averaged species temperatures can vary.  It should be emphasised that the purpose of using the $X_k$-equations is to give improved numerical accuracy; the basic physics is the same as that used in the mass-fraction based model.

\subsection{Equation of State for a Finite Volume}

\begin{figure}\centering
  \begin{overpic}[width=0.55\textwidth]{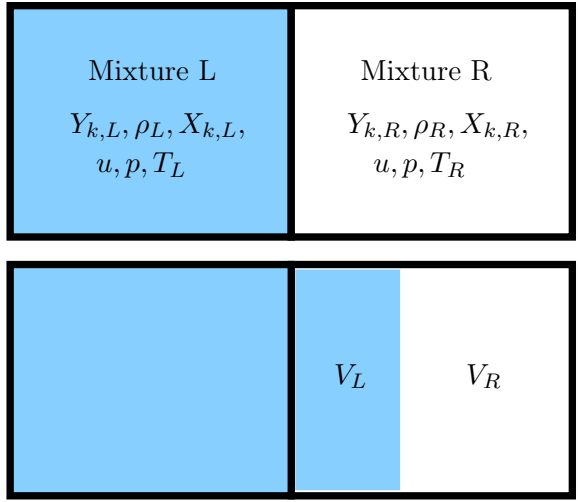}
    \small
    \put(32,160){Mixture L}
    \put(25,140){$Y_{k,L},\rho_L,X_{k,L},$}
    \put(35,125){$u,p,T_L$}
    \put(135,160){Mixture R}
    \put(130,140){$Y_{k,R},\rho_R,X_{k,R},$}
    \put(140,125){$u,p,T_R$}
    \put(125,45){$V_L$}
    \put(175,45){$V_R$}
\end{overpic}

\caption{Advection of an isolated contact surface separating two miscible gas mixtures at velocity $u>0$ and constant pressure $p$. Initial condition at time $t$ (top) and final condition at time $t+\Delta t$ (bottom). \label{contactsurface}}
\end{figure}

The governing equations must be closed by an equation of state for the mixture within a finite volume. For cases where diffusion is negligible, or the interface is sharp relative to the mesh size, then the algorithm must robustly advect this interface without spurious numerically generated pressure waves. Here it is important to consider the `volume-averaged' equation of state, rather than a `mixture' equation of state, since the latter implies a mixture in the sense of Avogadro's hypothesis, whereas the former does not require equal species pressures and temperatures since gases may occupy separate regions of space. 

In this paper, the system is closed by assuming that each species has the same volume-averaged pressure, however, volume-averaged species temperatures are different. At first glance this appears contrary to the application of Avogadro's hypothesis during the derivation of the system of equations, but is a consequence of solving the integral form of the governing equations in a finite-volume framework. The fact that a flow where all species are in local pressure and temperature equilibrium can exhibit different volume-averaged species temperatures is entirely physical as will now be outlined.

The classical approach is to consider the advection of an isolated contact surface at very high Reynolds number where the effects of diffusion, viscosity and conduction may be neglected. The Riemann problem is shown schematically in Fig. \ref{contactsurface}.  In this case, the sharp interface is perfectly aligned to an interface between two finite volumes. The exact solution of this simple advection problem is a constant pressure contact surface moving at velocity $u$, and represents the simplest case of inviscid advection. This sub-section will demonstrate the form of the volume-averaged equation of state which does not produce spurious numerically generated pressure oscillations in this limit. 

The contact surface separates a left gas mixture $L$ and right gas mixture $R$ with properties:

\begin{equation}
(\rho,u,p,Y_{k},X_{k},T)= 
\begin{cases}
    (\rho_L,u,p,Y_{k,L},X_{k,L},T_L),& \text{for left cell }\\
     (\rho_R,u,p,Y_{k,R},X_{k,R},T_R),              & \text{for right cell},
\end{cases}
\end{equation}

\noindent where the velocity $u>0$, the  mixture pressures are equal but have different temperatures and mixture compositions. All species are in local pressure and temperature equilibrium. 

At time $t+\Delta t$ the contact surface has swept out a volume $V_L=u\Delta t$ of the downstream cell. This volume now contains mixture $L$ at a temperature $T_L$, and the remaining volume of the cell $V_R=V-V_L$ contains mixture $R$ at a temperature T$_R$. The fraction of the cell volume which is occupied by mixture $L$ and mixture $R$ is defined as $v_L=V_L/V$ and $v_R=V_R/V$ respectively, thus $v_L+v_R=1$. Volume-averaged number fractions $\overline{X_k}$ can be defined in the downstream cell at $t+\Delta t$ as:

\begin{equation}
\overline{X_k}=v_L X_{k,L}+v_R X_{k,R},
\label{xkn+1}
\end{equation}

\noindent and internal energy is given by,

\begin{equation}
\overline{\rho \epsilon}=\sum_k \rho Y_k \epsilon_k=v_L \sum_k \rho_{L} Y_{k,L}\epsilon_{k,L}+v_R\sum_k \rho_{R} Y_{k,R}\epsilon_{k,R}.
\end{equation}

Defining a species partial density 

\begin{equation}
\rho_{k,(L,R)}=\frac{p}{T_{(L,R)} R_k}, 
\end{equation}

and note that as each individual mixture is in local pressure and temperature equilibrium then

\begin{equation}
\rho_{k,(L,R)}=\frac{\rho_{(L,R)} Y_{(L,R)}}{X_{(L,R)}}.
\end{equation}

\noindent Given $\epsilon_{k,(L,R)}=c_{v,k}T_{(L,R)}$ then internal energy may be written as:

\begin{equation}
\overline{\rho \epsilon}=p \sum_k\frac{ v_L X_{k,L}+v_R X_{k,R}}{\gamma_k-1}=p\sum_k \frac{\overline{X_k}}{\gamma_k-1}.
\end{equation}

The analytical solution to this problem is a constant pressure contact surface. To satisfy this condition, it is clear that the appropriate volume-averaged effective ratio of specific heats is: 

\begin{equation}
\frac{1}{\gamma-1}=\sum_k\frac{\overline{X_k}}{\gamma_k-1}
\end{equation}

With this condition, the downstream internal energy $\rho \epsilon=p/(\gamma-1)$, and pressure in the mixed cell is maintained at the original pressure $p$ following the advection step as required by the analytical solution. This simple analysis can be extended to more complex equations of state following analagous derivations for immiscible fluids \cite{Allaire2002,Murrone2005,Massoni2002}, and illustrates the key useful property of the governing model which will hold when the equations are discretised with a contact-resolving Riemann solver. Subsequent time steps may be treated in a similar fashion by considering the advection of sub-components of mixtures each at their respective pressure and temperature equilibria. 

Note that the above derivation is equivalent to specifying that the appropriate volume-averaged temperature for each species is $\overline{T_K}=p \overline{X_k}/R_k \overline{\rho Y_k}$. In the limit of an infinitely small cell then $T_k=T$ as expected from the assumptions underlying the governing equations. Assuming constant specific heats for each species such that $T_k=\epsilon_k/c_{vk}$, the volume-averaged species temperatures $\overline{T_k}$  are mass-weighted averages of the initial temperatures $T_{L,R}$. Enforcing temperature equilibrium would change the number fractions, $\overline{X_k}$, and this will give a different mean $\gamma$ for the cell and give rise to a change in pressure if $\gamma_1$  and $\gamma_2$  are different. Hence homogenizing a mixed cell will, in general, give a non-physical change in pressure at the contact (unless $\gamma_1=\gamma_2$). 

The isobaric closure proposed above is designed to correct this problem, which manifests itself clearly in the standard mass fraction model, where even if the species mixtures are initially in local pressure and temperature equilibrium (as in the current problem), the analytical cell averaged species temperatures can be expected to differ following an advection step. The above closure is only possible with the addition of the equation for transport of $X_k$.

\subsection{Numerical Methods \label{nummeth}}

The governing equations are implemented within the University of Sydney code Flamenco \cite{Garcia2014}. This code has an existing inviscid implementation of the volume fraction model of Allaire {\it et al.} which has been well documented in previous publications \cite{Allaire2002,Probyn2014}. Given that the inviscid part of the governing model proposed here is in the same mathematical form as the model of Allaire {\it et al.}, the discretisation may be undertaken in an analogous fashion. 

The existing scheme is based on a Godunov-type method of lines approach in a structured multiblock framework, which has been modified to provide reduced dissipation in low Mach number regions of the flow \cite{Thornber2007b,Thornber2007c,Thornber2007d}. This modification is low cost but restores the accuracy of the compressible algorithm demonstrated for $M>10^{-4}$. Most computations with this algorithm utilise a nominally fifth order reconstruction for the inviscid component \cite{Kim2005}, and second order central differences for the viscous and diffusive terms. Temporal integration is achieved via a second order accurate in time TVD Runge-Kutta method \cite{Spiteri2002}. Note that in more than one dimension the formal order of accuracy would be second, given the one dimensional reconstruction stencils. This approach is applied in a straightforward manner to all of the conservative equations, however a different approach must be chosen for the number fraction model.

The Jacobian for the two-species system written in primitive variables $\left[\rho Y_1, \rho Y_2, u, P, X_1\right]^T$ is as follows:

\begin{equation}
A(W)=\left[
\begin{matrix}
u & 0 & \rho Y_1 & 0 & 0\\
0 & u & \rho Y_2 & 0 & 0\\
0 & 0 & u & 1/\rho & 0\\
0 & 0 & \rho c^2 & u & 0\\
0 & 0 & 0 & 0 & \mathcal{U}\\
\end{matrix}
\right]
\end{equation}

\noindent where for an ideal gas the speed of sound $c^2=\sum_k \rho Y_k h_k/ \rho \xi$,  $\xi=1/(\gamma-1)=\sum_k X_k/(\gamma_k-1)$. The system has a full set of eigenvectors, with the corresponding eigenvalues:

\begin{equation}
\lambda_1=\lambda_2=u,\lambda_3=u-c,\lambda_4=u+c, \lambda_5=\mathcal{U}.
\end{equation}

The form of the Jacobian given above is not conventional as it depends on derivatives of the solution variables. These terms are included here on the left hand side of the system of equations for numerical reasons (discussed in detail in Section \ref{nummeth}), namely that they modify the upwind direction of $X_k$. A justification for this is given by applying Cattaneo's relaxation approach \cite{ToroSIAM2014,MontecinosJCP2014}, whereby a new variable $\psi$ is introduced in place of $\frac{\partial X_1}{\partial x}$ along with a transport equation that relaxes $\psi$ towards $\frac{\partial X_1}{\partial x}$ by means of a stiff source term, thus enabling the Jacobian to be written in terms of solely the solution variables. Further exploration of using this type of relaxation approach to allow for a more efficient discretisation will be the subject of future work.

Here, the $X_k$ equation is discretised following Abgrall \cite{Abgrall1996}, where the cell interface
flux $F^{i+1/2}$ and $F^{i-1/2}$ in physical space for cell $i$ is computed as

\begin{eqnarray}
F^{i+1/2}&=&(X_k \mathcal{U})^{RS,i+1/2}-X^i_k \mathcal{U}^{RS,i+1/2}\\
F^{i-1/2}&=&(X_k \mathcal{U})^{RS,i-1/2}-X^i_k \mathcal{U}^{RS,i-1/2}.
\end{eqnarray}

\noindent where $(.)^{RS}$ indicates a term arising from the solution of the Riemann problem at the cell interface, which in the current algorithm employs the HLLC approximate solver.

The key problematic terms are the diffusion and equilibriation terms which have now been moved to the left hand side. These modify the upwind direction of the advection of number fraction dependent on the gradient of the number density $N$ and the gradient of the number fraction itself. Thus, these gradients are computed using second order accurate central differences, centred on the cell interface. The required values of $\mathcal{M}$ and $N$ are computed as an average of the two interface reconstructed values. As an example, in one dimension with constant grid spacing $\Delta x$, $\mathcal{U}$ is given by:

\begin{equation}
\mathcal{U}^{RS,i+1/2}=u^{RS,i+1/2}+D_{12} \frac{\mathcal{M}^L+\mathcal{M}^{R}}{2} \frac{X_k^{i+1}-X_k^{i}}{\Delta x}-D_{12} \frac{2(N^{i+1}-N^{i})}{(N^{R}+N^{L})\Delta x},
\end{equation} 

\noindent where $u^{RS}$ is gained from the solution of the classical Reimann problem. The upwind number fraction ($X_k^{RS}$) required to compute the number fraction fluxes is determined via a modified HLLC approach where the signal speeds incorporate the additional diffusion velocities, i.e. the contact surface in the Riemann problem for the number fraction is assumed to advect at a velocity $\mathcal{U}$, with the gradients given by the aforementioned second order central difference approximation. It is worth noting that recent advances could permit an arbitrarily higher order discretisation, which will be the subject of future work \cite{dumbser2011simple}.

Next, the diffusion terms must be discretised in a manner which is consistent with the underlying physics. As diffusive fluxes are computed at the cell interfaces, Avogadro's hypothesis must be applied. This is a consequence of the hypothesis of local pressure and temperature equilibrium invoked in the derivation of the governing equations, i.e. the species diffusion terms require pressure and temperature equilibrium at the cell interface. If pressure and temperature equilibrium is not enforced during the computation of the diffusive fluxes, then the resultant species densities and temperatures will be in error. Within the current algorithm, this is addressed by rewriting the mass fraction gradient in the diffusive flux as:

\begin{equation}
\nabla\cdot (\rho D_{12} \nabla{Y_k})=\nabla\cdot (\rho D_{12} \nabla{\frac{W_k X_k}{W}}),
\end{equation}

where $W_k$ is the species $k$ molecular weight and $W$ is the mixture molecular weight.

All cases use a fifth-order accurate limited scheme to interpolate from cell averaged quantities to cell interface values in the computation of the inviscid terms \cite{Kim2005}, however it must be noted that the current discretisation is formally second order accurate due to the discretisation of the quasi-conservative number fraction equation, viscous and diffusive terms. 

For explicit time integration, as employed here, the stable time step is chosen based on the minimum in the whole domain of the following: 

\begin{equation}
\Delta t=CFL\times \textrm{min} \left(\frac{\Delta x}{\textrm{max}_k (\lambda_k)}, \frac{\rho \Delta x}{2\mu}, \frac{\rho \Delta x^2}{2D}, \frac{\rho c_v \Delta x^2}{2\kappa}\right).
\end{equation}

\section{One Dimensional Results \label{results}}

This section presents three validation cases where analytical incompressible solutions are available which can be used to verify the observed order of accuracy of the key components of the algorithm. The results using the new number fraction approach are compared with the classical mass fraction approach implemented within Flamenco (the discretisation approach can be found in \cite{Thornber2011}) using the same reconstruction and time stepping scheme, and results from an entirely independent one-dimensional Lagrange-remap algorithm which used the method of Turmoil3D \cite{Youngs1982,Youngs1994} and employs mass fractions to track individual species.

Turmoil3D employs mass fraction equations to track individual species, utilising a Lagrange-remap scheme  \cite{Grinstein2007}. This is fundamentally completely different from the Godunov-type approach employed in Flamenco, thus providing an independent benchmark solution for non-analytical fields such as the pressure field in problems at finite Mach number. Turmoil3D does not use the current number fraction based model. 

Although the model equations and algorithm are compressible, this section employs exact incompressible analytical solutions to verify and validate each of the key terms in the model equations. In many applications species diffusion occurs at a very small scale, over which the pressure is near uniform. For example, in the highly compressible two dimensional Richtmyer-Meshkov problem detailed in Section \ref{2dresults}, a diffuse boundary forms after shock passage, and there is little pressure variation across the this boundary layer. Hence 1D test cases based on diffusion at a boundary between two gases at uniform pressure provide highly relevant test cases of the new model. If diffusive velocities are small compared the sound speed, the mixing process is near incompressible and results can be checked against analytic solutions for the incompressible limit. In applications, the diffuse boundary layer will be advected through the mesh and this is a more challenging situation for the numerical method. Hence, results are also given for a case when a uniform velocity is added to the flow. Note that the numerical results are expected to agree with the analytical results only up to the point where the difference between the numerical and analytical results is not dominated by compressibility effects, which are absent from the analytical solution. This occurs at the finest grid resolutions where the error due to the discretisation (truncation error) becomes lower than the difference due to compressibility and thus the convergence rate measured relative to the analytical solution stalls. It occurs at the same error magnitude regardless of whether the mass or number fraction models are employed.

The first three cases are one-dimensional, where Case 3 is the same as Case 1 but with a net mean velocity added. The one-dimensional problem of diffusion at a plane boundary between two gases is a useful test case for numerical implementations. For mixing between two gases at the same pressure to be an incompressible process, there should be no change in pressure when the two gases mix. This will be true for two different gases at the same temperature or for two identical gases at different temperatures. In Case 1 and Case 3 heat conduction is unimportant. In Case 2 heat conduction is a dominant process. Case 1 and 2 are used to verify the algorithms for near-pure diffusion cases, whereas Case 3 tackles a more realistic case with both advection and diffusion. In these three one-dimensional cases viscosity is assumed to be zero, and equal diffusivities are used ($D_{kj}=D$ and $\kappa=\rho c_p D$).

Note that the variable reconstruction employed within Flamenco includes limiting at maxima and minima for all test cases presented in this Section.

\subsection{Case 1: Diffusion of an Isothermal Contact Surface between two Different Species}

This test case has been employed in Kokkinakis {\it et al.} \cite{kokkinakis2015}, however here the initial conditions are described in full, along with a rigorous initialisation process which enables a formal convergence study, up to the limit of assumed incompressibility. 

The computation domain is $0 \le x \le 1$, and grid sizes from $32 \rightarrow 2048$ cells have been employed. Reflective boundary conditions are used at the left and right hand boundaries. The two fluids have the properties $\rho_1=20$, $\rho_2=1$, $\gamma_1=2$, $\gamma_2=1.4$. The specific heats satisfy the requirement that $(\gamma_1-1)\rho_1 c_{v1}=(\gamma_2-1)\rho_2 c_{v2}$, which gives temperature equilibrium if the two fluids have the same pressure.

If the initial uniform pressure is sufficiently high, the mixing process is quasi- incompressible and the analytic solution for the number fraction distribution is given by 

\begin{equation}
X_1=\frac{1}{2} \left[1-\textrm{erf} \left(\mathcal{Z}\right)\right], \mathcal{Z}=\frac{x-x_0}{\sqrt{4Dt+h_0^2}},
\end{equation}

\noindent where $x_0=0.5$, $h_0=0.02$ and the diffusion coefficient $D=0.01$. 

In the initialisation of the problem, compressibility effects are minimised by assuming that the volume weighted mean velocity is initially zero. Hence the initial mass weighted mean velocity is given by $u=-\frac{D}{\rho}\frac{\partial \rho}{\partial x}$.

Noting that a finite volume algorithm requires the cell averaged quantities for the initialisation,

\begin{equation}
\overline q=\frac{1}{x_{i+1/2}-x_{i-1/2}}\int^{x_{i+1/2}}_{x_{i-1/2}} q(x) dx
\end{equation}

\noindent the following exact initialisations are given for completeness:

\begin{equation}
\overline \rho=\frac{1}{x_{i+1/2}-x_{i-1/2}}\left[\frac{\rho_1+\rho_2}{2}x +\frac{\rho_1-\rho_2}{2}\mathcal{A}\right]_{x_{i-1/2}}^{x_{i+1/2}}
\end{equation}

\begin{equation}
\overline {\rho Y_1}=\frac{1}{x_{i+1/2}-x_{i-1/2}}\left[\frac{\rho_1}{2}x -\frac{\rho_1}{2}\mathcal{A}\right]_{x_{i-1/2}}^{x_{i+1/2}}
\end{equation}

\begin{equation}
\overline {X_1}=\frac{1}{x_{i+1/2}-x_{i-1/2}}\left[\frac{1}{2}x -\frac{1}{2}\mathcal{A}\right]_{x_{i-1/2}}^{x_{i+1/2}}
\end{equation}

\begin{equation}
\overline {\rho u}=\frac{1}{x_{i+1/2}-x_{i-1/2}}\left[\frac{-D(\rho_2-\rho_1)}{2} \textrm{erf}\left(\mathcal{Z}\right)\right]_{x_{i-1/2}}^{x_{i+1/2}}
\end{equation}

\begin{equation}
\mathcal{A}=\{x-x_0\}\textrm{erf}\left(\mathcal{Z}\right)+\frac{\sqrt{4Dt+h_0^2}}{\sqrt{\pi}}\textrm{e}^{-\mathcal{Z}^2}
\end{equation}

For the standard test case, the initial uniform pressure is $p_0=10000$. The peak initial Mach number, based on the mass-weighted mean velocity is then $M=0.019$. In order to show the effect of compressibility, calculations have also been performed with $p_0=10$. The initial Mach number is then $M=0.6$. For $p_0=10000$, the flow is near-incompressible with pressure and temperature remaining approximately constant, thus the incompressible analytical solution may be employed as a reference solution to demonstrate the scheme's order of accuracy. Note that for Case 1, the $(\overline{u}-u)\cdot \nabla X_1$ term in the number fraction equation is essential, but the $\nabla N/N$ term vanishes in the incompressible limit (Case 2 will validate this term).

\subsubsection{Quasi-Incompressible Diffusion}

\begin{table*}[]
\centering
\caption{Case 1 convergence rates for the number fraction model.}
\label{tab:Case1VFConvergence}
\begin{tabular}{lllllll}
\hline
N & $L^1$  & $L^2$  & $L^\infty$  & $\mathcal{O} (L^1)$  & $\mathcal{O} (L^2)$  & $\mathcal{O} (L^\infty)$  \\
\hline
32 & 6.8188e-04 & 9.7707e-04 &  2.5498e-03 & - & - & - \\
64 & 1.7441e-04 & 2.4761e-04 & 5.8434e-04  & 1.9670 & 1.9804 & 2.1255 \\
128 & 4.6246e-05 & 6.6306e-05 & 1.6157e-04   & 1.9151 & 1.9009  & 1.8547 \\
256 & 1.2132e-05 & 1.7464e-05 & 4.1914e-05 & 1.9305 & 1.9248 & 1.9466  \\
512 & 3.3962e-06 & 4.9435e-06 & 1.0798e-05 & 1.8368 & 1.8207 & 1.9567 \\
1024 & 1.3170e-06 & 2.1030e-06 & 5.6508e-06 & 1.3666 & 1.2331 & 0.9342 \\
2048 & 9.4227e-07 & 1.6173e-06 & 4.3847e-06 & 0.4831 & 0.3789 & 0.3660 \\
4096 & 9.2548e-07 & 1.5375e-06 & 4.0689e-06 & 0.0259 & 0.0730 & 0.1078 \\
\hline
\end{tabular}
\end{table*}

\begin{table*}[]
\centering
\caption{Case 1 convergence rates for the mass fraction model.}
\label{tab:Case1MFConvergence}
\begin{tabular}{lllllll}
\hline
N & $L^1$  & $L^2$  & $L^\infty$  & $\mathcal{O} (L^1)$  & $\mathcal{O} (L^2)$  & $\mathcal{O} (L^\infty)$  \\
\hline
32 & 2.6562e-03 & 4.0717e-03 &  1.0109e-02 & - & - & - \\
64 & 6.6652e-04 & 1.0133e-03 & 2.4719e-03  & 1.9947 & 2.0066 & 2.0319 \\
128 & 1.6521e-04 & 2.5061e-04 & 6.0838e-04   & 2.0124 & 2.0156  & 2.0226 \\
256 & 4.1252e-05 & 6.2547e-05 & 1.5160e-04 & 2.0017 & 2.0024 & 2.0047  \\
512 & 9.8440e-06 & 1.4912e-05 & 3.6097e-05 & 2.0672 & 2.0685 & 2.0704 \\
1024 & 2.0744e-06 & 3.1381e-06 & 7.4040e-06 & 2.2466 & 2.2485 & 2.2855 \\
2048 & 5.3978e-07 & 1.0996e-06 & 3.4383e-06 & 1.9422 & 1.5129 & 1.1066 \\
4096 & 7.9846e-07 & 1.3563e-06 & 3.7717e-06 & -0.5648 & -0.3027 & -0.1335 \\
\hline
\end{tabular}
\end{table*}

\begin{figure}
\begin{centering}
\includegraphics[width=\textwidth]{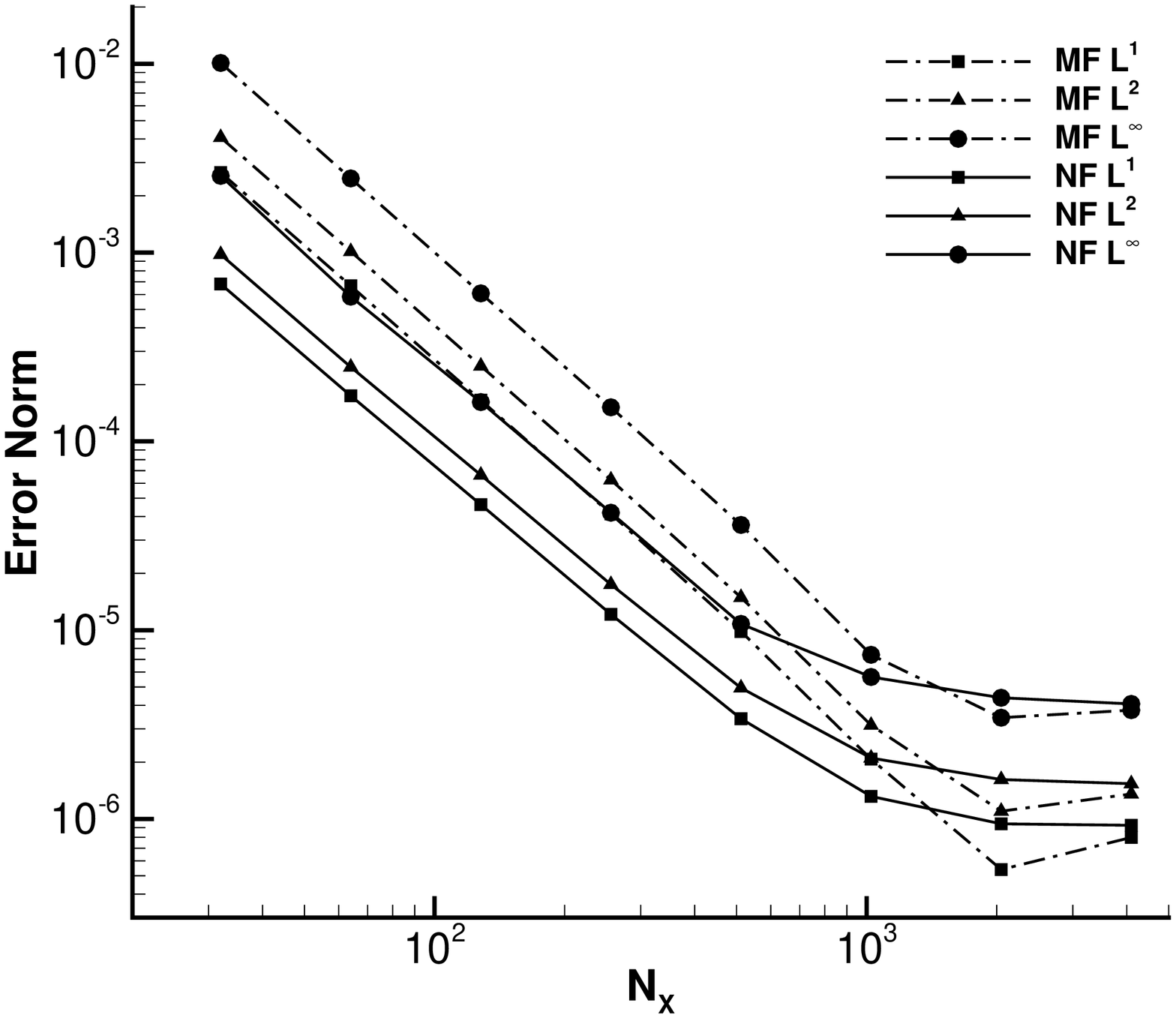}
\caption{Plot of Case 1 convergence rates, showing comparison between the mass fraction and number fraction approaches. \label{fig:Case1Convergence}}
\end{centering}
\end{figure}

This section focusses on the results for the quasi-incompressible case where $p_0=10000$. Tables \ref{tab:Case1VFConvergence} and \ref{tab:Case1MFConvergence} document the $L^1$, $L^2$ and $L^\infty$ errors in the simulated number fraction profile at $t=0.5$ for the number fraction and mass fraction formulations compared to the incompressible analytical solution, along with the observed convergence rates. Figure \ref{fig:Case1Convergence} plots the error norms as a function of number of points $N_x$ for the mass fraction and number fraction equations. 

There are several important points to make. Firstly, the results  produced here  demonstrate that a formal convergence at approximately second order accuracy is achieved. Convergence to the analytical solution stalls at error norms of $\approx 10^{-5} \rightarrow 10^{-6}$ as here the effects of compressibility are no longer negligible, thus the incompressible analytical solution is not the exact solution to this (marginally) compressible problem. 

Secondly, for a given grid resolution, the number fraction model has a substantially lower actual error, on the order of one third of the errors for the mass fraction formulation. Thus for a constant error, the number fraction equations may be run on a mesh of nearly $1/2$ the number of points at the coarser resolutions.

\begin{figure}
\begin{centering}
\includegraphics[width=0.48\textwidth]{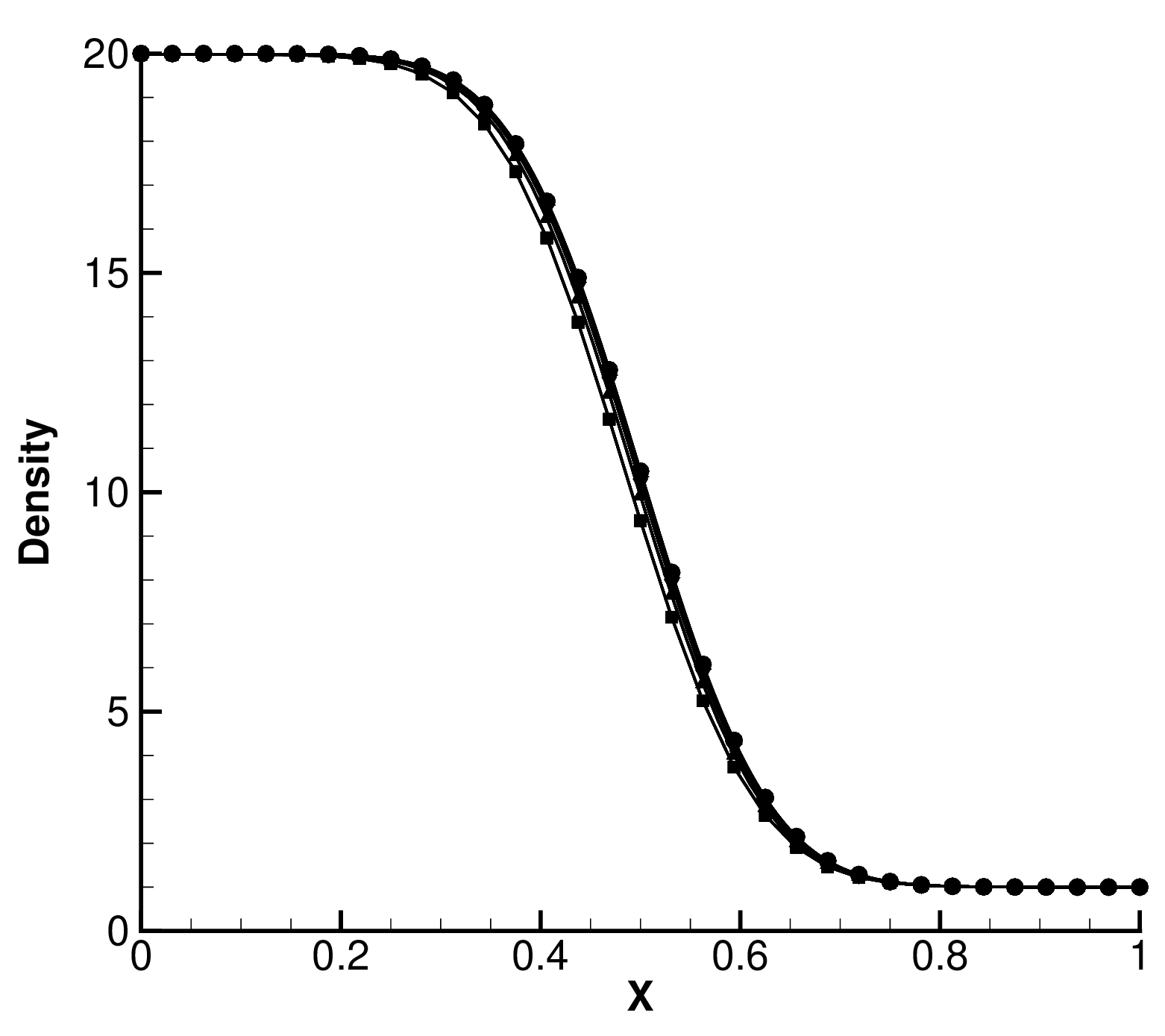}
\includegraphics[width=0.48\textwidth]{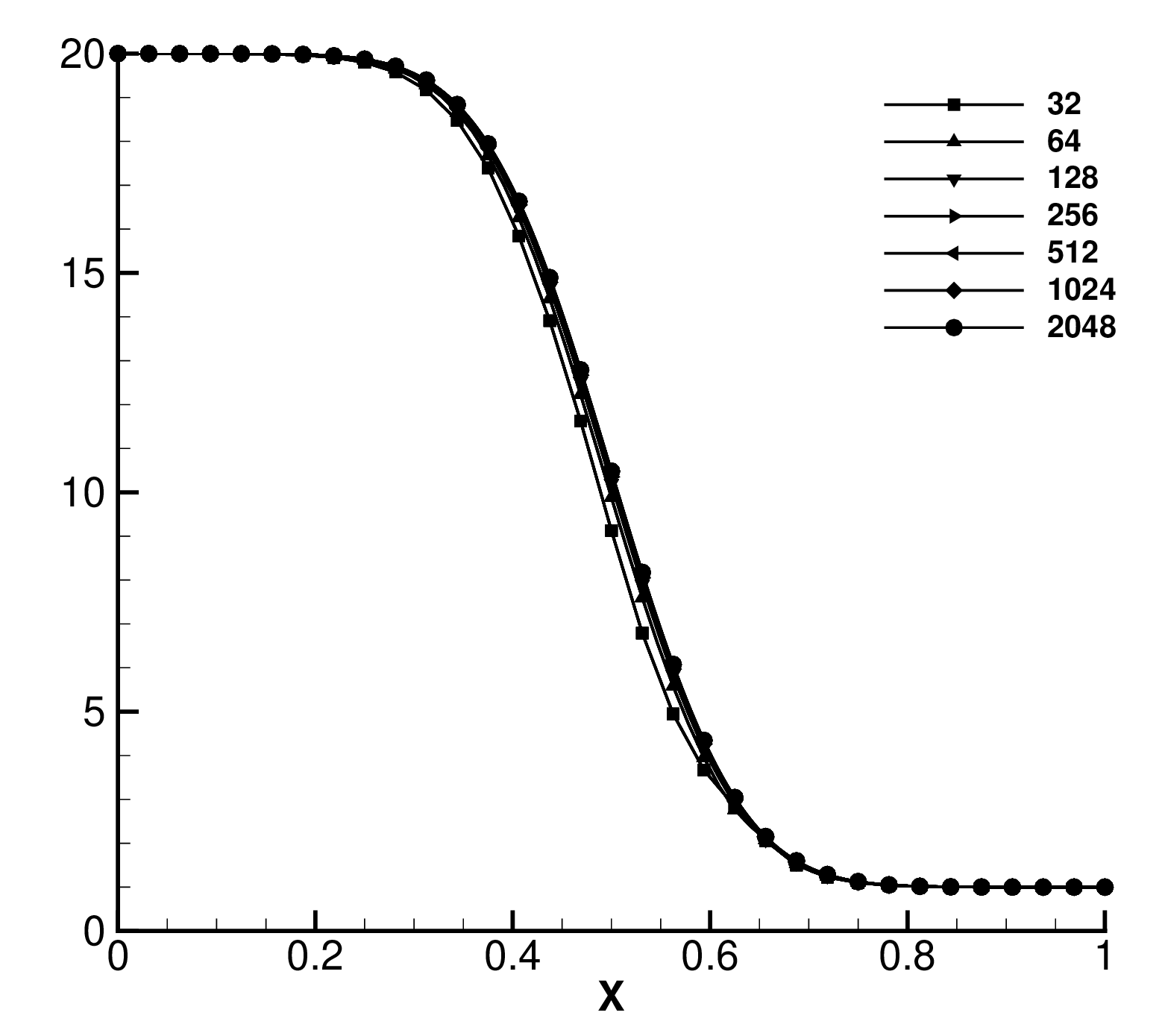}
\includegraphics[width=0.48\textwidth]{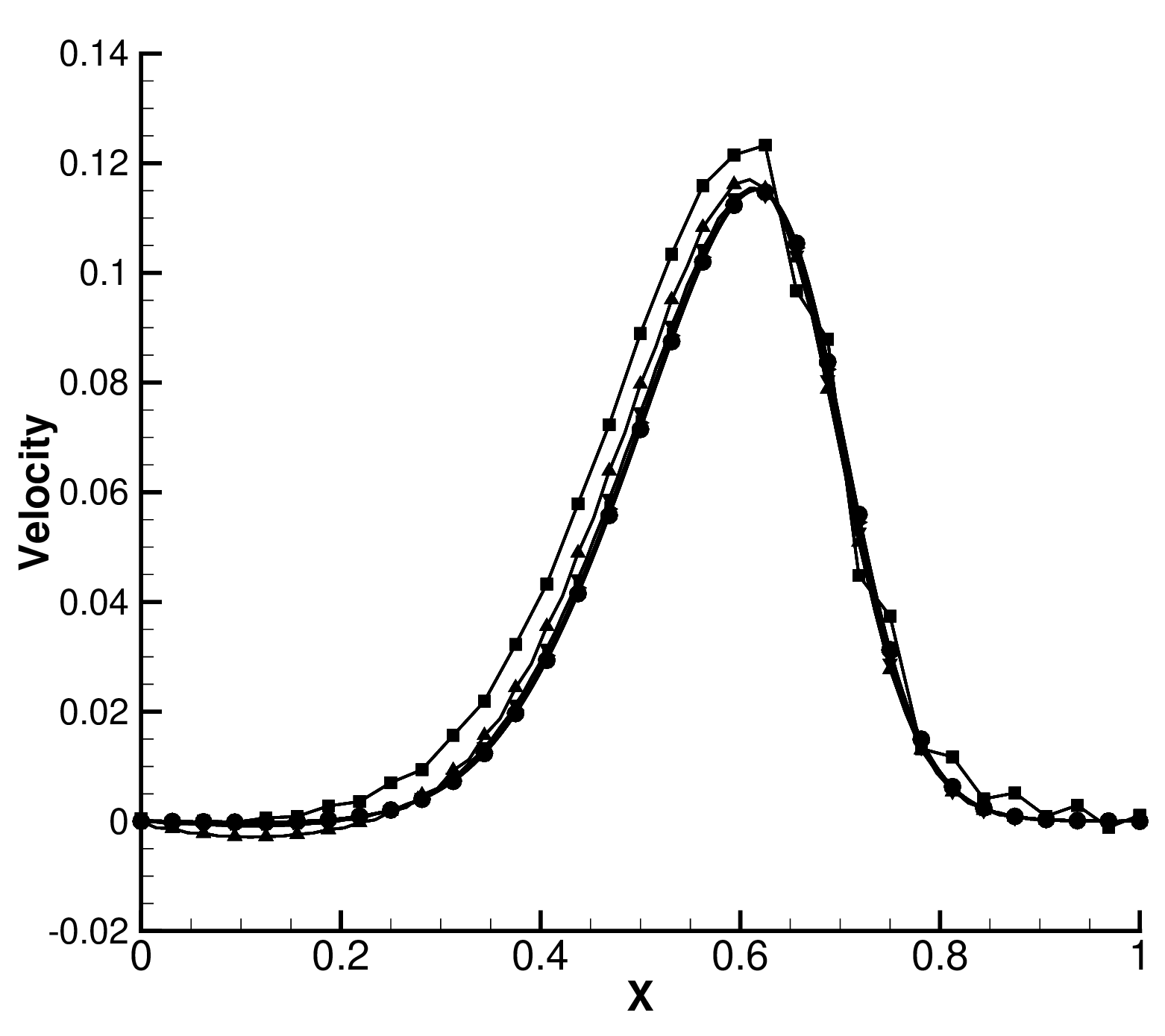}
\includegraphics[width=0.48\textwidth]{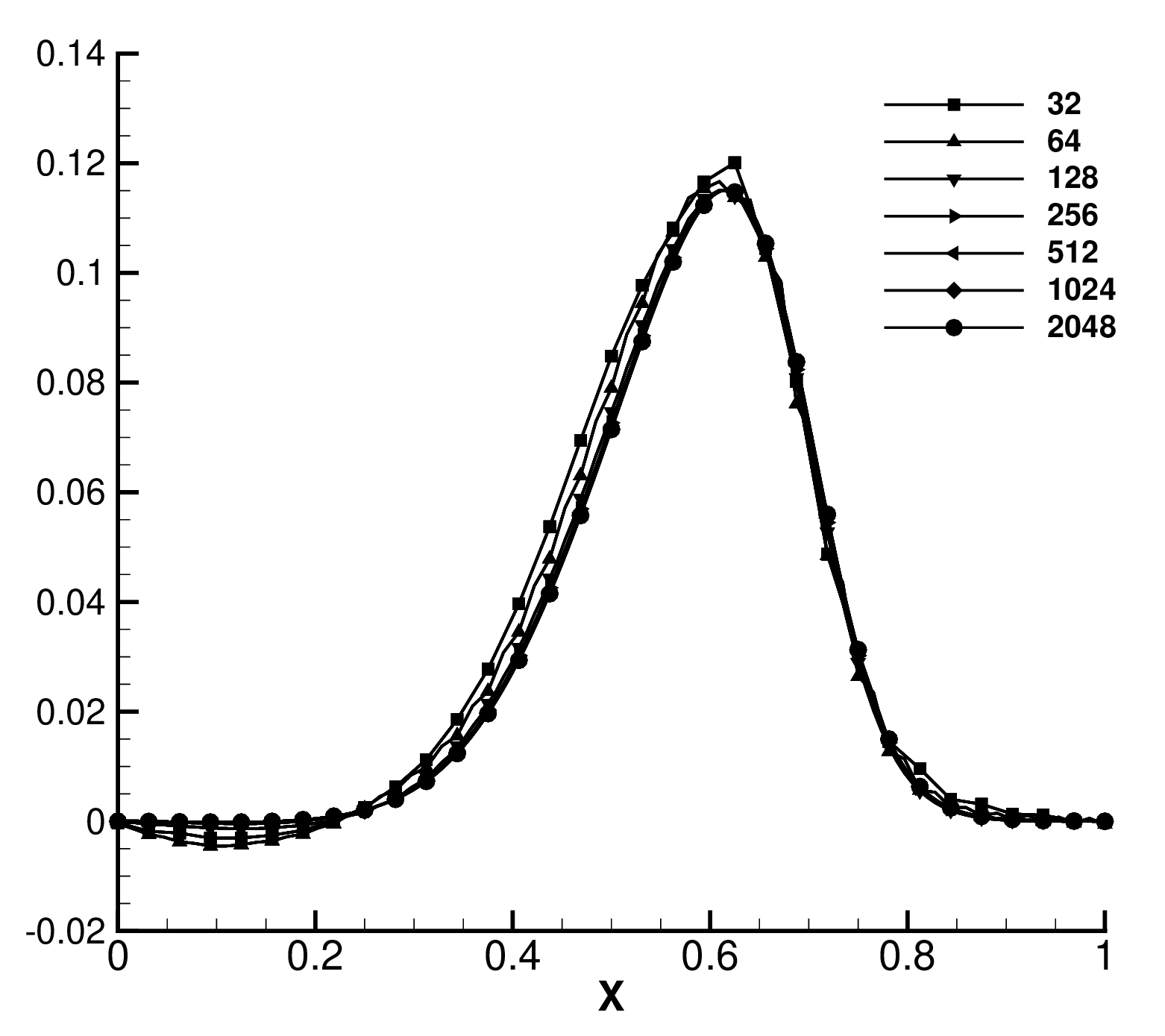}
\includegraphics[width=0.48\textwidth]{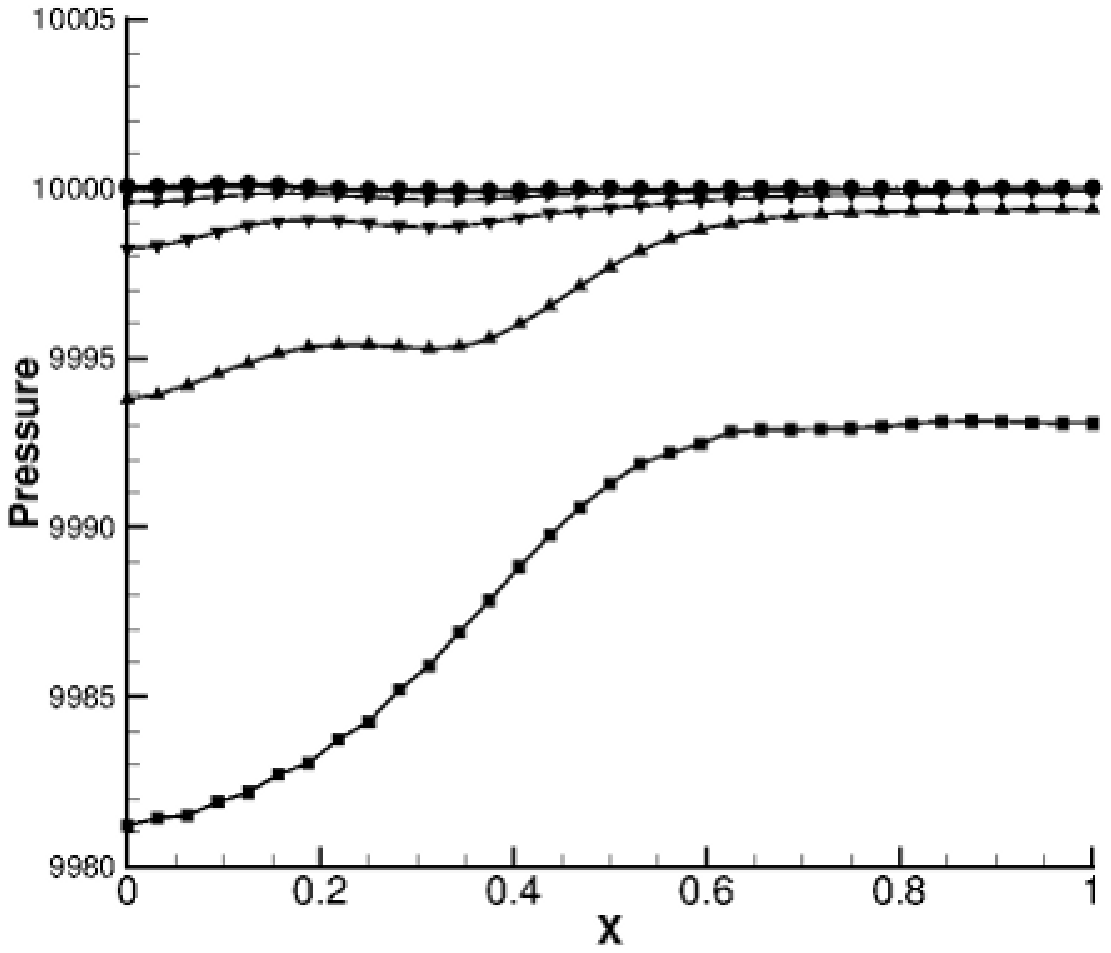}
\includegraphics[width=0.48\textwidth]{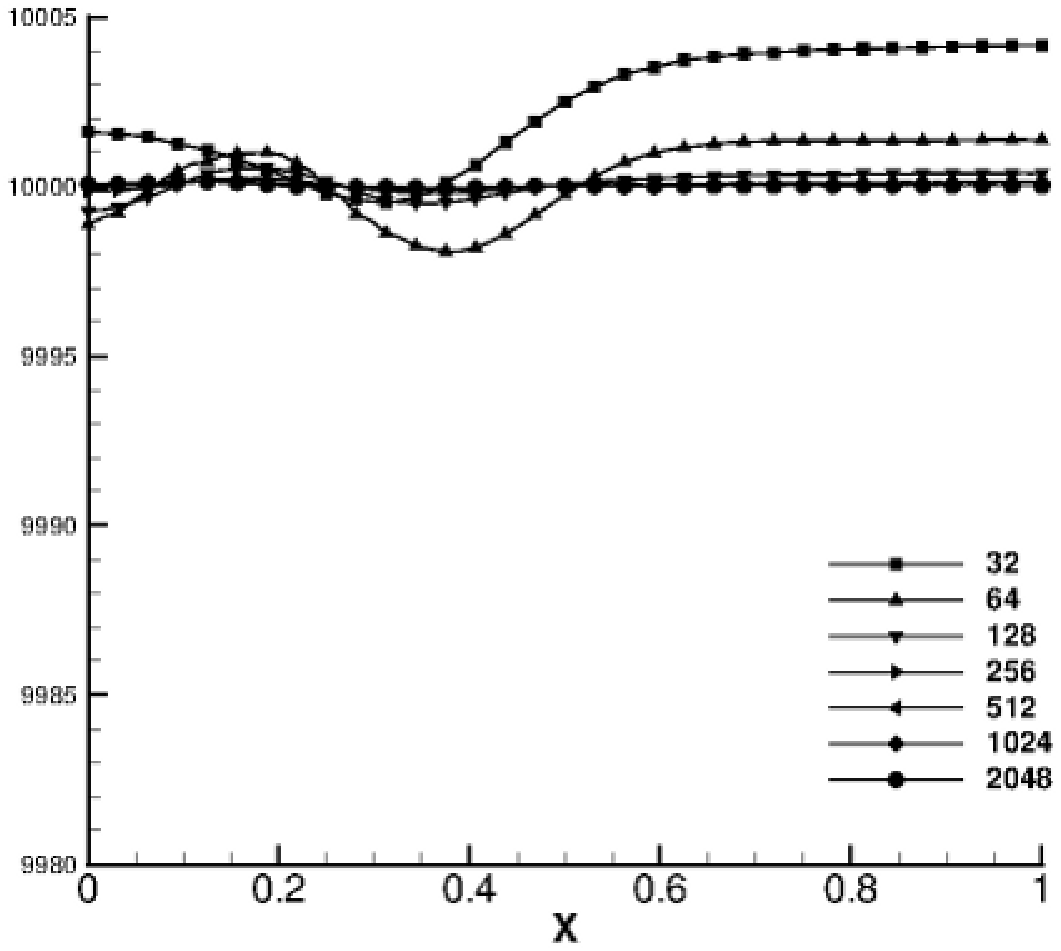}
\includegraphics[width=0.48\textwidth]{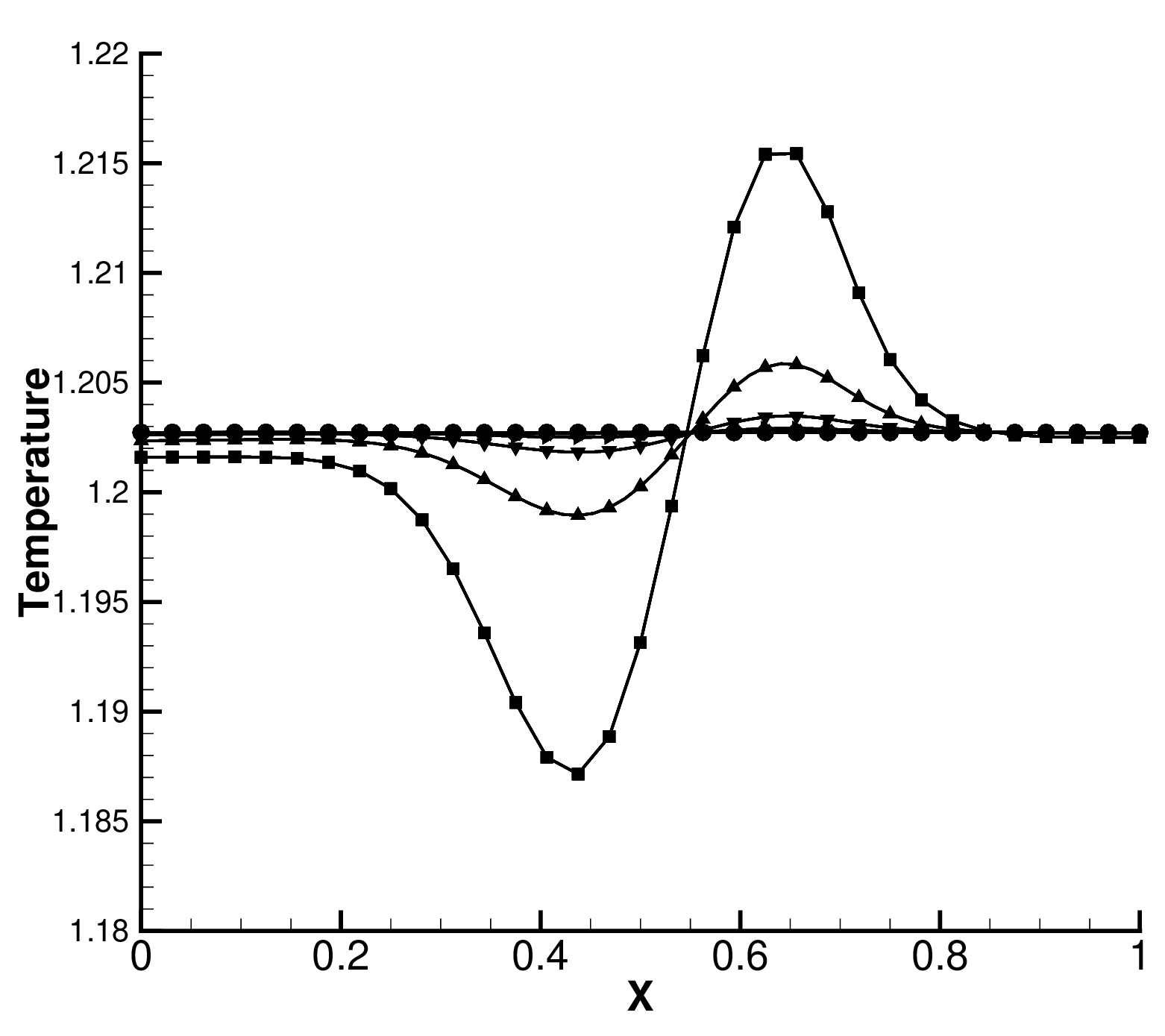}
\includegraphics[width=0.48\textwidth]{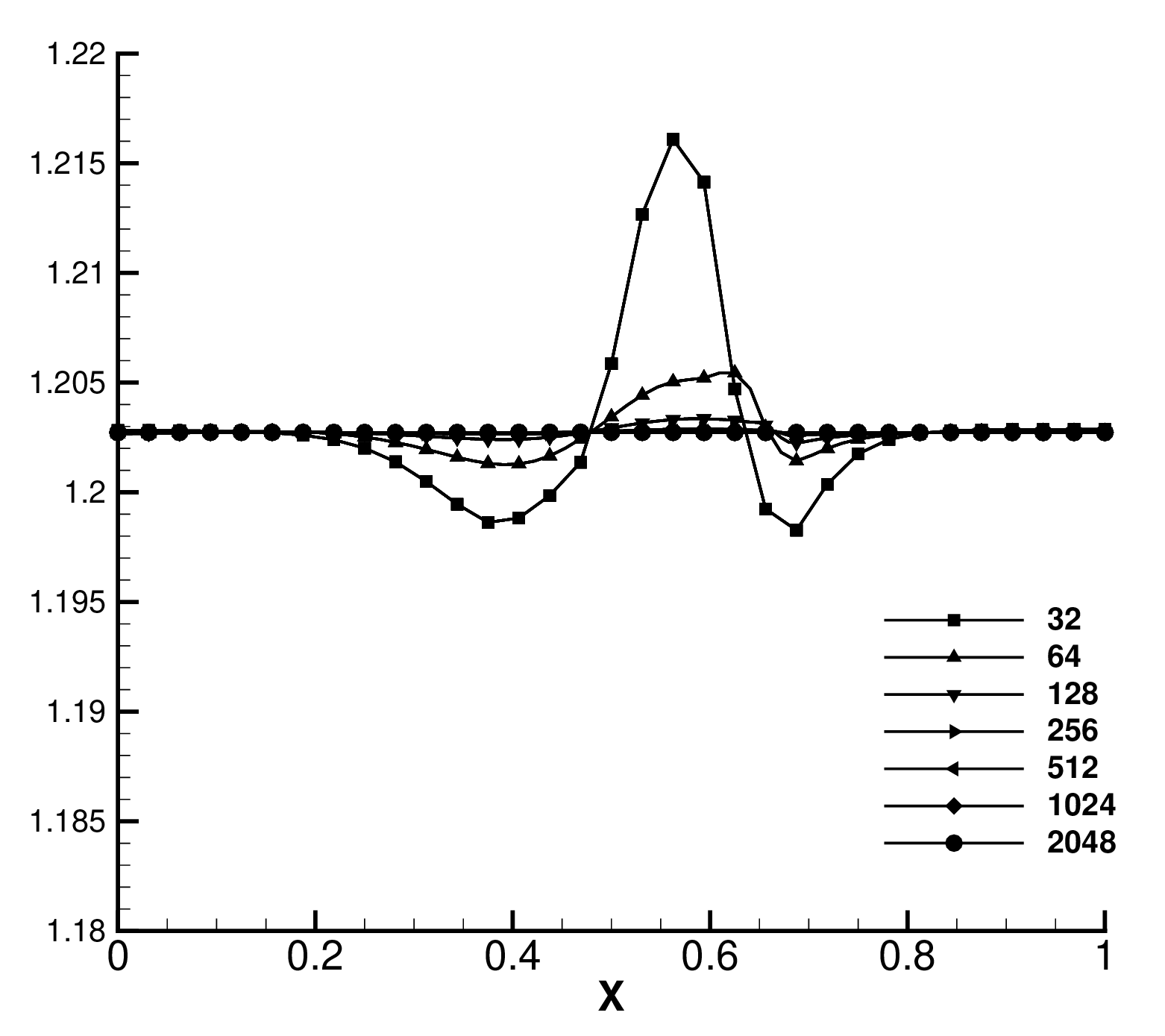}
\caption{Comparison of variables across all grid resolutions for Case 1 with the mass fraction model results on the left and the number fraction model results on the right. \label{fig:Case1Plots}}
\end{centering}
\end{figure}

The results are plotted for all grid resolutions and for both the mass and number fraction formulations in Figure \ref{fig:Case1Plots}. The non-analytical pressure and temperature fields are also shown in these figures, where it can be seen that the number fraction model has substantially lower pressure fluctuations for a given grid resolution, for example, at a grid resolution of 32 points the pressure fluctuation is less than 4Pa in the number fraction model compared to 18Pa in the mass fraction model. This is due to the better treatment of the equation of state in the mixed cells. 

\begin{figure}
\begin{centering}
\includegraphics[width=0.49\textwidth]{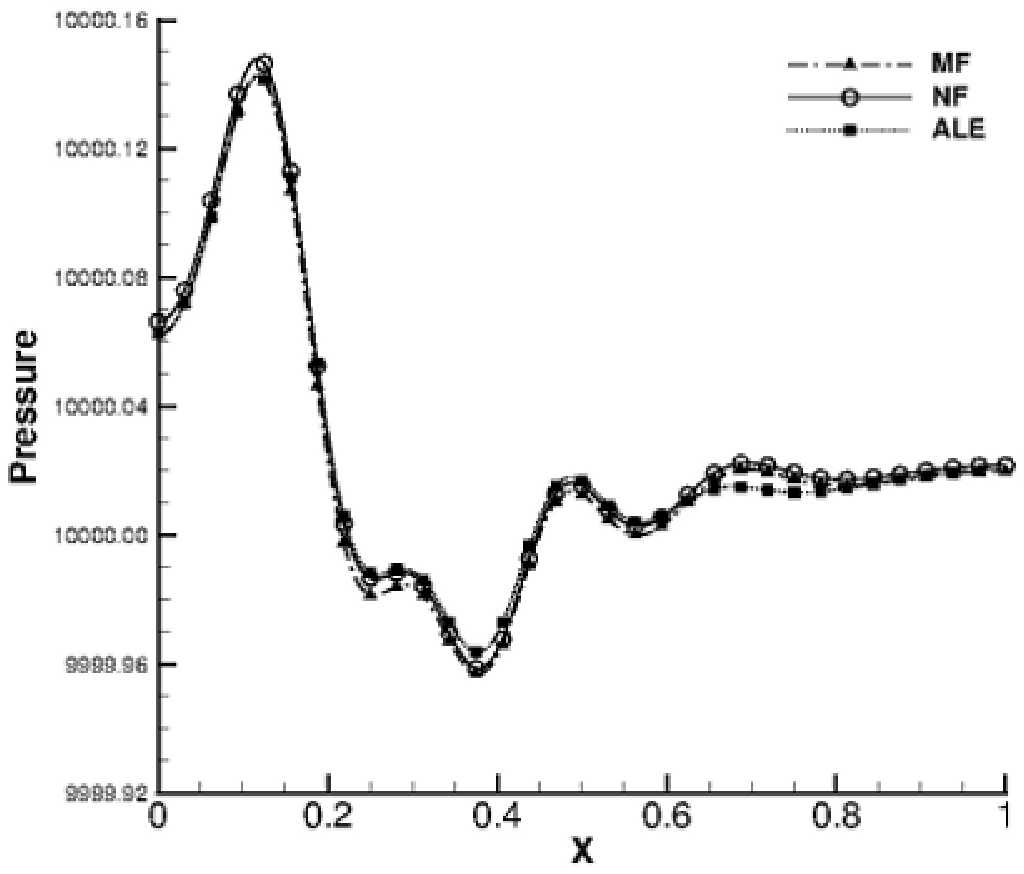}
\includegraphics[width=0.49\textwidth]{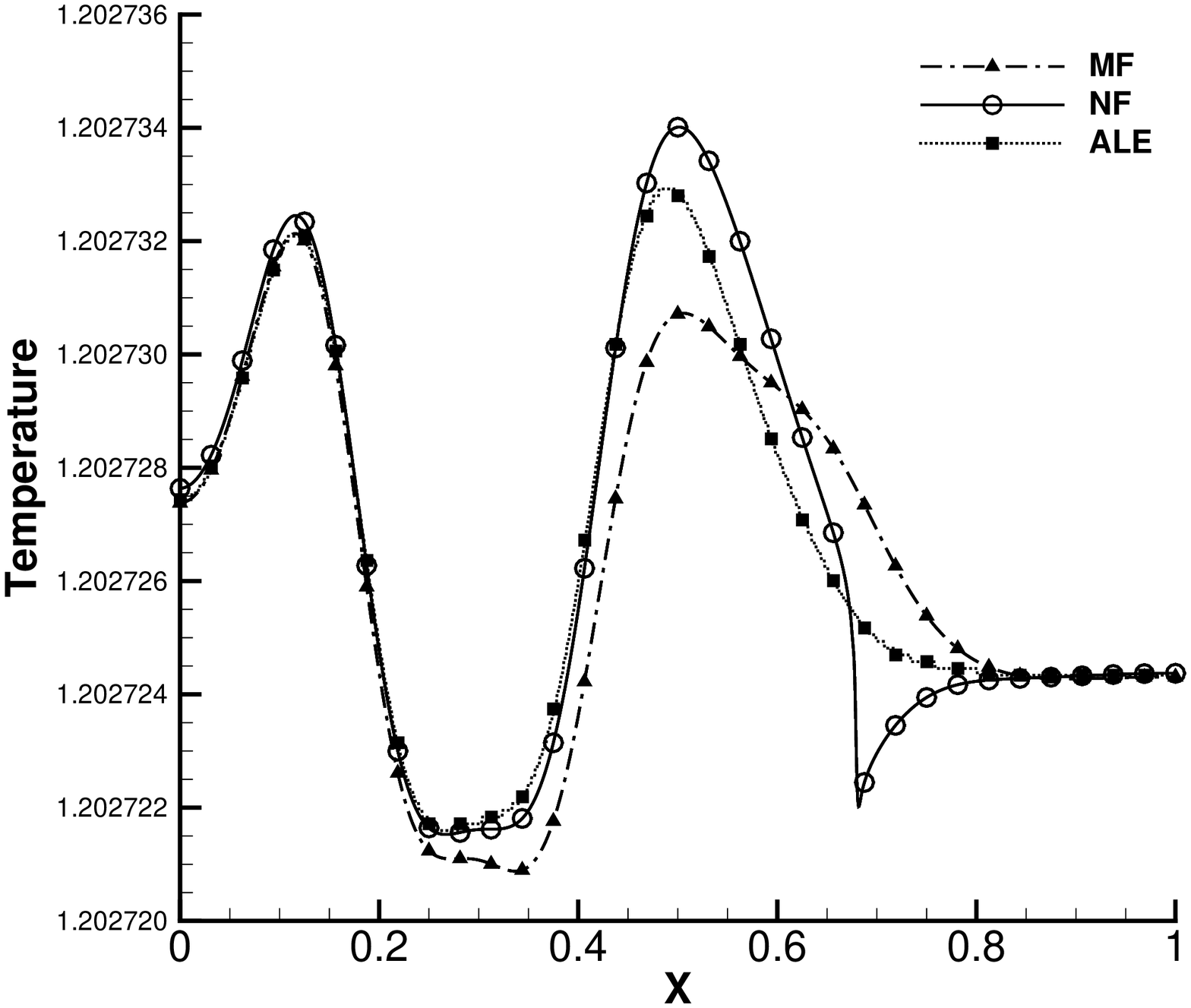}
\caption{Case 1 comparison of non-analytical field variables for the mass fraction (MF) and number fraction (NF) models in Flamenco (2048 cells) as well as the results from the Lagrange-remap code (2048 cells). \label{fig:Case1Comp}}
\end{centering}
\end{figure}

Figure \ref{fig:Case1Comp} shows the converged solutions for the pressure and temperature field. Both are extremely challenging to resolve since t=0.5 represents greater than 30 periodic reflections of extremely small compressible waves in the domain. Here a comparison is plotted at the finest grid resolution with available results from the Lagrange-remap algorithm as a cross-check. The temperature field in the number fraction model is substantially steeper at $x\approx 0.7$, which is a feature of the governing equations. 

\subsubsection{Compressible Diffusion}

\begin{figure}
\begin{centering}
\includegraphics[width=0.48\textwidth]{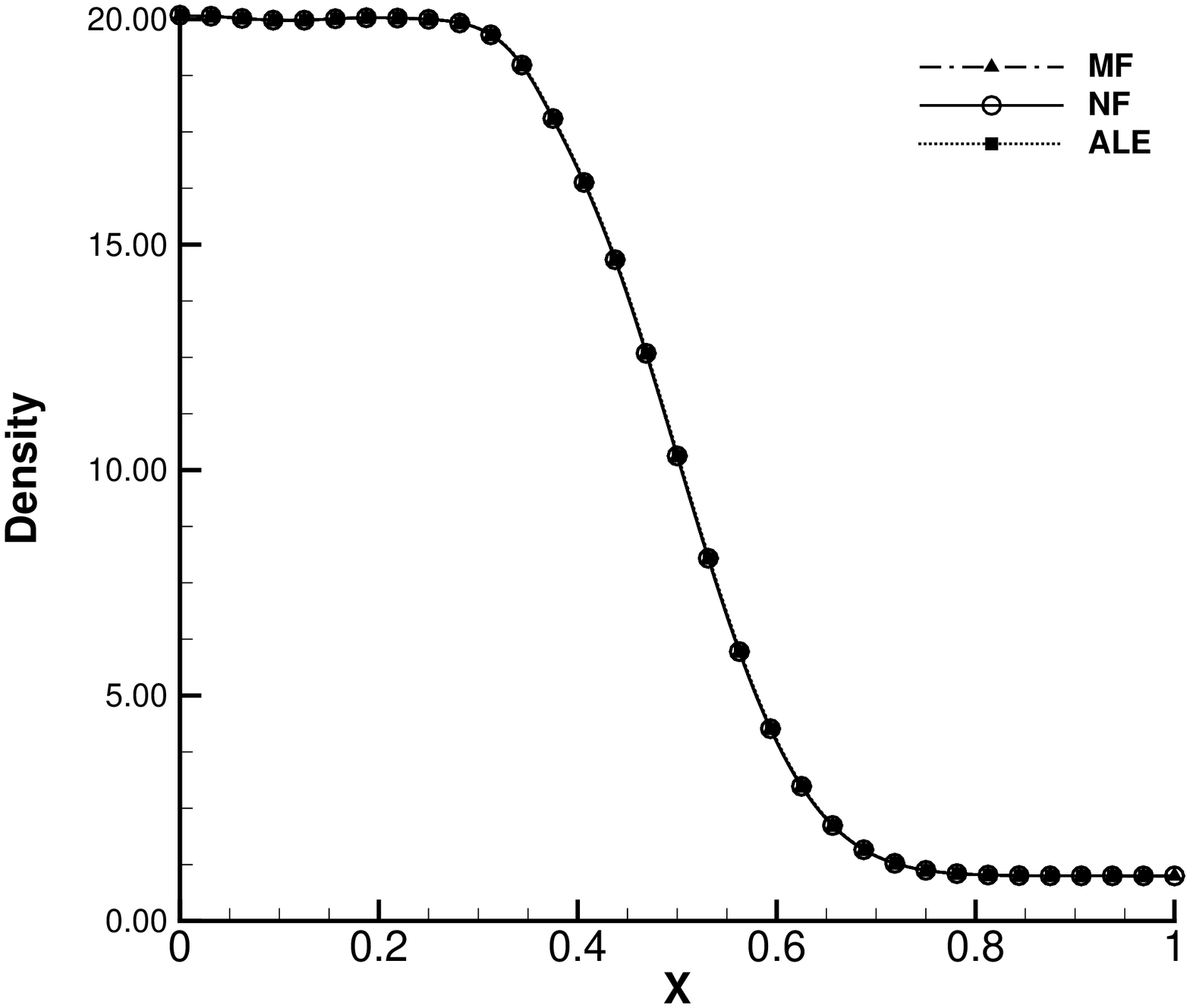}
\includegraphics[width=0.48\textwidth]{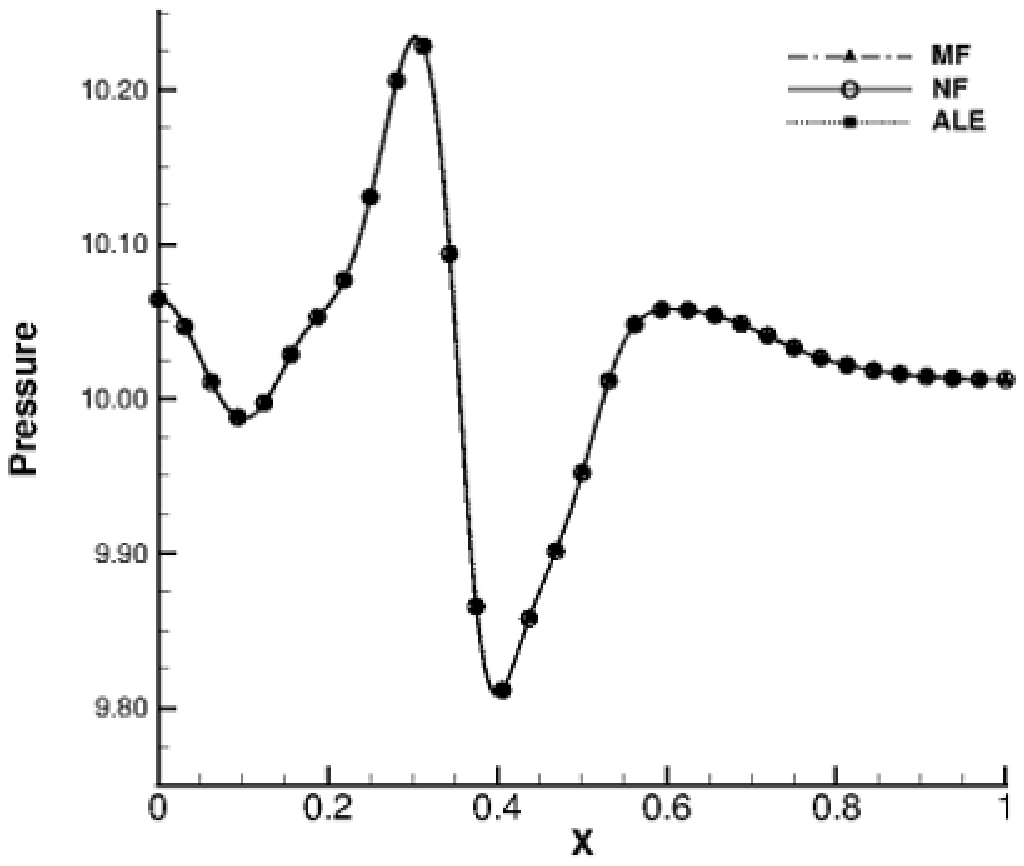}
\includegraphics[width=0.48\textwidth]{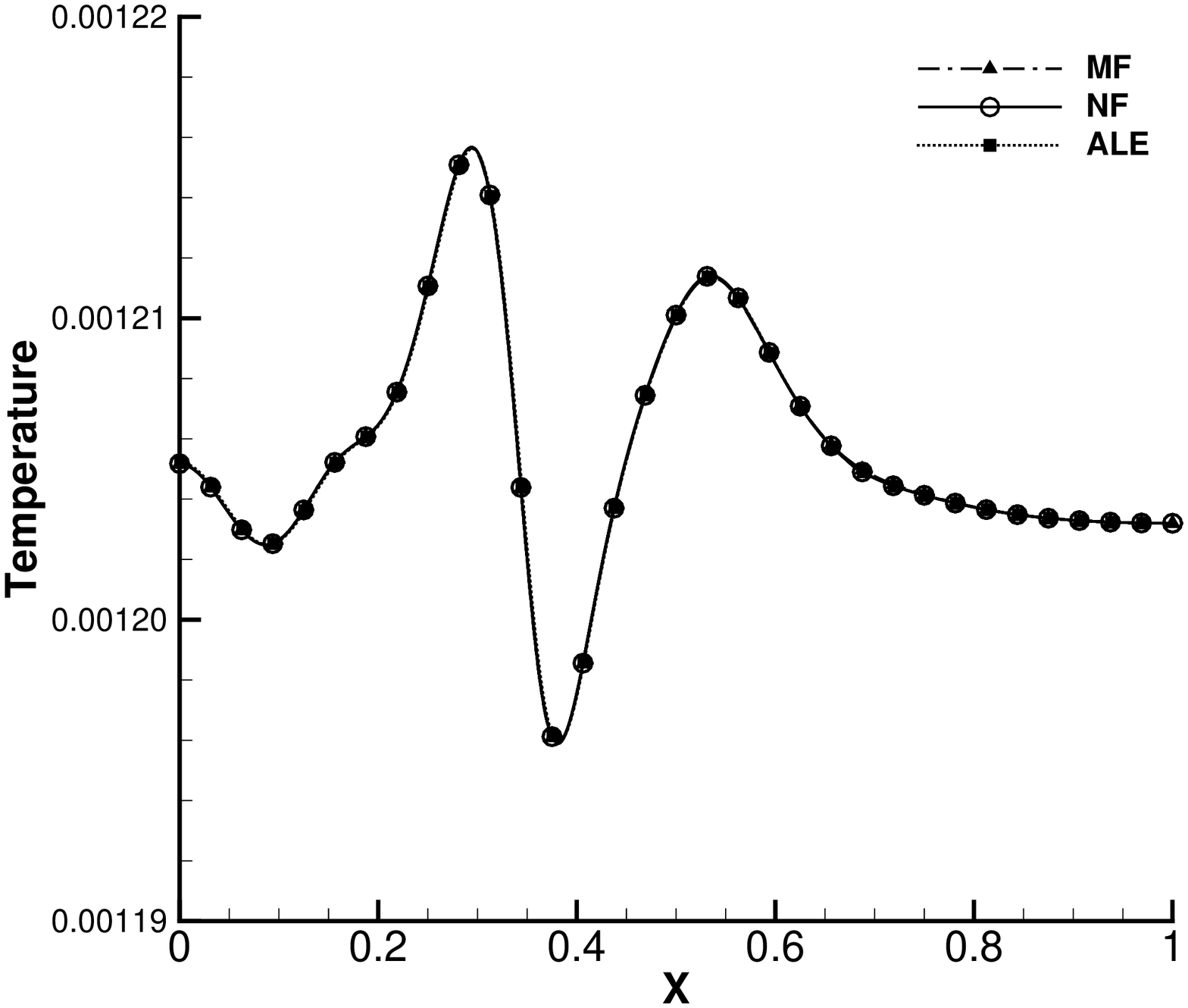}
\includegraphics[width=0.48\textwidth]{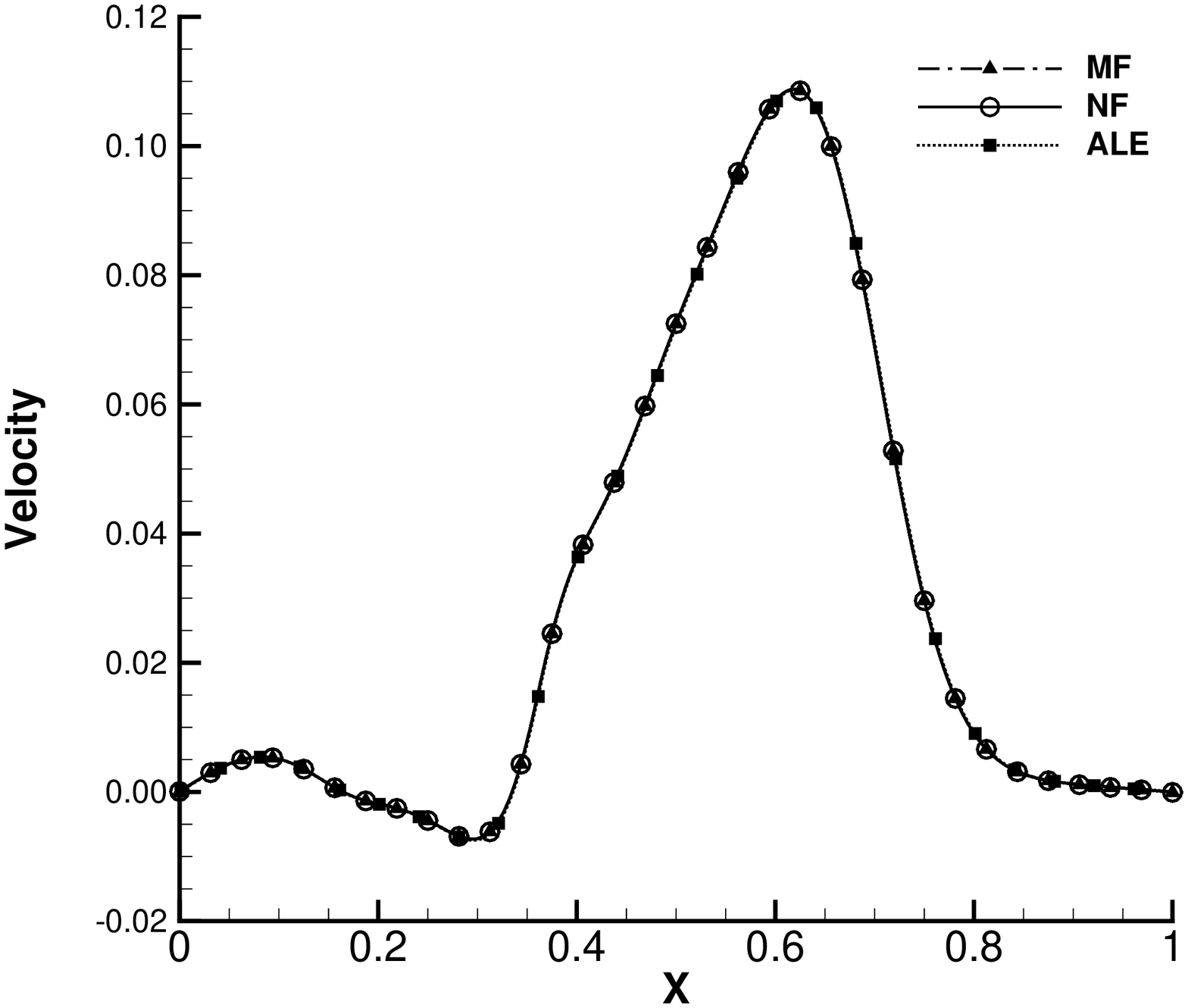}
\caption{Comparison of variables solutions using the mass fraction, number fraction, and Lagrange-remap mass fraction approaches for the compressible diffusion problem Case 1 at 256 cells. \label{fig:Case1P10}}
\end{centering}
\end{figure}

This section focusses on the results for the compressible case where $p_0=10$. In this case there is no analytical solution as the density distribution varies slightly from the analytical, hence Figure \ref{fig:Case1P10} plots the comparison between the mass fraction, number fraction and Lagrange-remap results. The key result here is that all models converge to the same solution in the fully compressible limit using 256 cells. 

\subsection{Case 2: Diffusion of a Contact Surface with a Two Identical Species at two Temperatures}

The computation domain is $0\le x \le 1$, and grid sizes from $32 \rightarrow 2048$ cells have been employed. Reflective boundary conditions are used at the left and right hand boundaries. The two fluids have the properties $\rho_1=20$, $\rho_2=1$, $\gamma_1=\gamma_2=5/3$. The specific heats satisfy the requirement that $c_{v1}=c_{v2}$, and $D=0.01$. A uniform initial pressure is achieved by specifying that $T_2=20T_1$. As previously, $p_0=10000$ $D=0.01$ is used for both the species diffusvity and the heat diffusivity, and the solution is run to $t=0.5$. As the heat diffusivity is constant, the error function solution for the density distribution is again applicable for the quasi-incompressible case,

The initial density and mass fraction distributions are the same Case 1, implying that the initial number fractions are different. Thus the initial cell averages $\overline \rho$, $\overline {\rho Y_1}$ and $\overline {\rho u}$ are the same as Case 1. The initial distribution (and incompressible analytical solution) of $X_1$ is equal now to $Y_1$ and is given by:

\begin{equation}
X_1=\frac{\rho_1 \left[1-\textrm{erf}(\mathcal{Z})\right]}{(\rho_1+\rho_2)+(\rho_1-\rho_2)\textrm{erf}(\mathcal{Z})}
\end{equation}

This function does not have an exact antiderivative, so the initial conditions for $\overline {X_1}$ are computed by a five-point Gaussian quadrature scheme.

In the incompressible limit, the density and mass fraction distributions are the same
as for the uniform temperature case. However, number fractions are different - the
cold gas now expands when it mixes with the hot gas. In the number fraction equation, the term $(\overline{u}-u)\cdot \nabla X_1 \approx 0$, but the $\nabla N/N$ term is essential.

\begin{table*}[]
\centering
\caption{Case 2 convergence rates for the number fraction model.}
\label{tab:Case2VFConvergence}
\begin{tabular}{lllllll}
\hline
N & $L^1$  & $L^2$  & $L^\infty$  & $\mathcal{O} (L^1)$  & $\mathcal{O} (L^2)$  & $\mathcal{O} (L^\infty)$  \\
\hline
32 & 3.8913e-03 & 7.4095e-03 &  2.0769e-02 & - & - & - \\
64 & 1.0073e-03 & 1.8151e-03 & 4.7065e-03  & 1.9497 & 2.0293 & 2.1417 \\
128 & 2.3945e-04 & 4.3487e-04 & 1.2196e-03  & 2.0728 & 2.0614  & 1.9482 \\
256 & 5.8293e-05 & 1.0731e-04 & 3.0899e-04 & 2.0383 & 2.0188 & 1.9808  \\
512 & 1.3589e-05 & 2.4929e-05 & 7.2748e-05 & 2.1009 & 2.1059 & 2.0866 \\
1024 & 2.4534e-06 & 4.5243e-06 & 1.3878e-05 & 2.4696 & 2.4621 & 2.3902 \\
2048 & 9.0045e-07 & 1.7194e-06 & 7.3100e-06 & 1.4460 & 1.3958 & 0.9248 \\
4096 & 9.2548e-07 & 1.5374e-06 & 4.0689e-06 & -0.0396 & 0.1613 & 0.8452 \\
\hline
\end{tabular}
\end{table*}

\begin{table*}[]
\centering
\caption{Case 2 convergence rates for the mass fraction model.}
\label{tab:Case2MFConvergence}
\begin{tabular}{lllllll}
\hline
N & $L^1$  & $L^2$  & $L^\infty$  & $\mathcal{O} (L^1)$  & $\mathcal{O} (L^2)$  & $\mathcal{O} (L^\infty)$  \\
\hline
32 & 4.5145e-03 & 8.9766e-03 &  2.6221e-02 & - & - & - \\
64 & 1.1807e-03	& 2.3650e-03 &	7.0094e-03 & 1.9349 & 1.9243 &	1.9033 \\
128 & 3.0148e-04 & 	6.0560e-04 & 1.7964e-03 & 1.9695 & 1.9654 &	1.9642 \\
256 & 7.5270e-05 &	1.5110e-04 & 4.4891e-04 & 2.0019 & 2.0029 &	2.0006 \\
512 & 1.7961e-05 &	3.5895e-05 & 1.0673e-04 & 2.0672 & 2.0736 &	2.0724 \\
1024 & 3.6458e-06 &	7.1072e-06 & 2.1124e-05 & 2.3006 & 2.3364 &	2.3370 \\
2048 & 8.0986e-07 &	1.3751e-06 & 7.3104e-06 & 2.1705 & 2.3697 &	1.5309 \\
4096 & 1.2701e-06 &	2.4644e-06 & 7.3496e-06 & -0.6492 & -0.8416 &	-0.0077 \\
\hline
\end{tabular}
\end{table*}

\begin{figure}
\begin{centering}
\includegraphics[width=\textwidth]{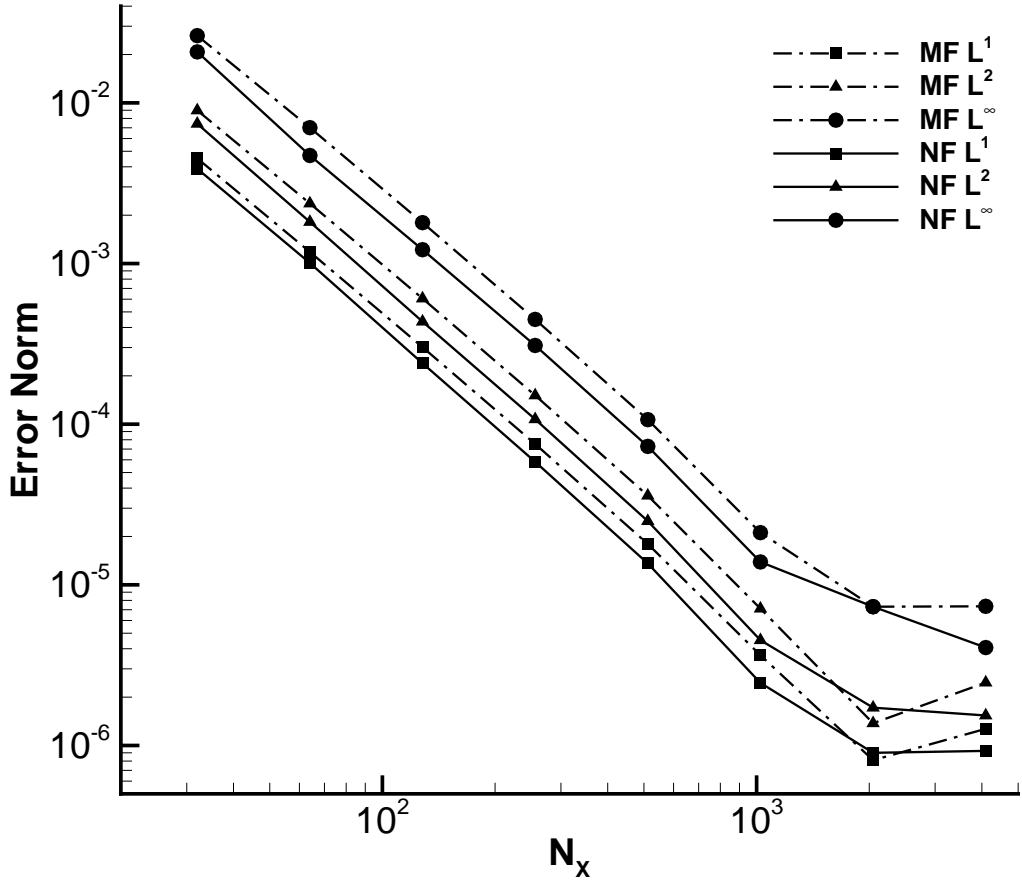}
\caption{Plot of Case 2 convergence rates, showing comparison between the mass fraction and number fraction approaches. \label{fig:Case2Convergence}}
\end{centering}
\end{figure}

Tables \ref{tab:Case2VFConvergence} and \ref{tab:Case2MFConvergence} document the $L^1$, $L^2$ and $L^\infty$ errors in the simulated number fraction profile at $t=0.5$ for the number fraction and mass fraction formulations compared to the incompressible analytical solution, along with the observed convergence rates. Figure \ref{fig:Case2Convergence} plots the error norms as a function of number of points $N_x$ for the mass fraction and number fraction models. 

As this case includes two species with the same ratio of specific heats, then the differences between the mass and number fraction formulations should be lower. This is the case, and both algorithms converge at the expected second order of accuracy up to the point where compressibility impacts the agreement with the incompressible solution. As with Case 1, the errors with the mass fraction model increase at $4096$ points which confirms this hypothesis. The number fraction model again has lower error at a given grid resolution, however the differences are not as substantial as the previous case. This is to be expected as the principle errors in the mass fraction approach are generated when the ratios of specific heat vary. 

\begin{figure}
\begin{centering}
\includegraphics[width=0.48\textwidth]{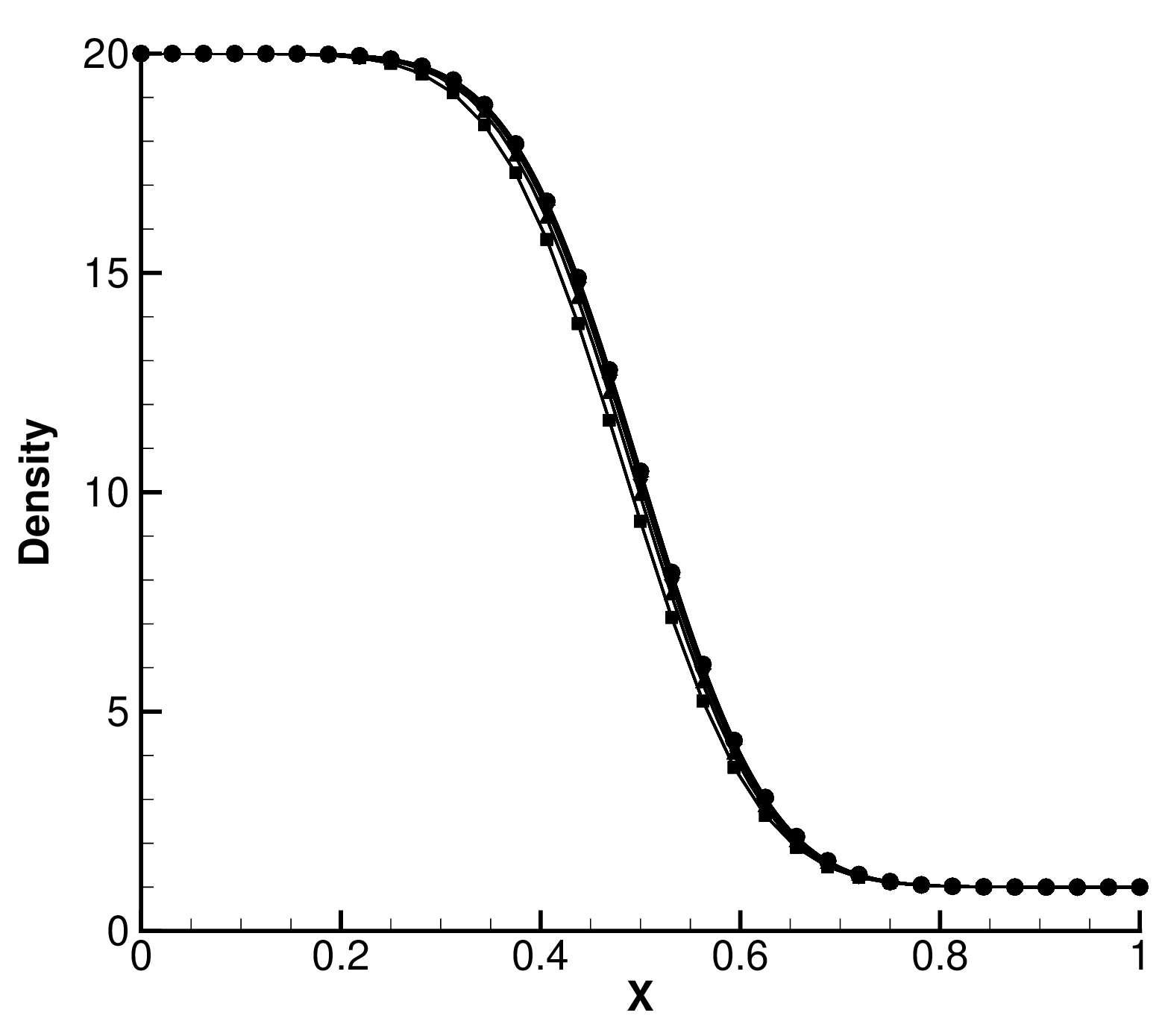}
\includegraphics[width=0.48\textwidth]{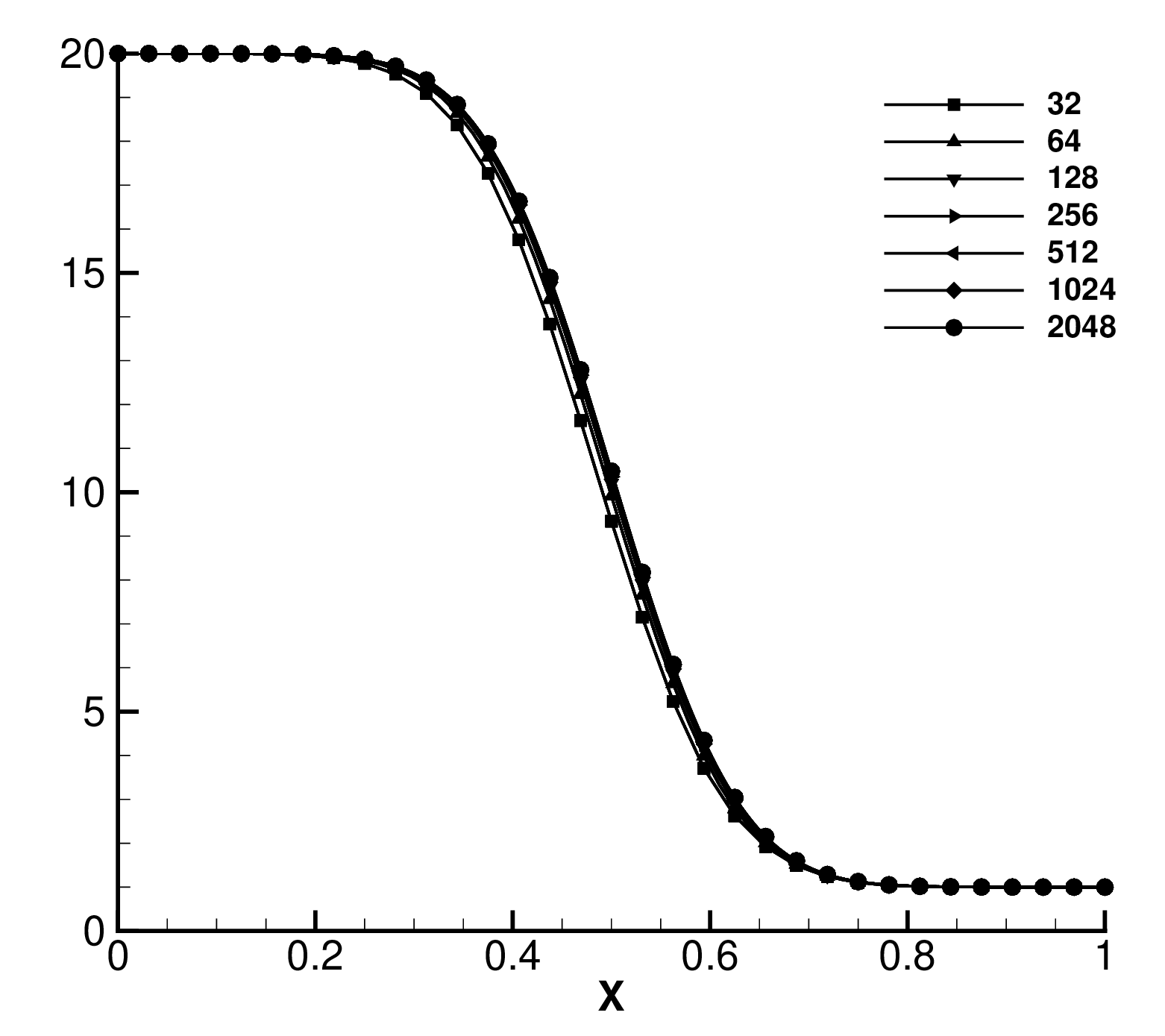}
\includegraphics[width=0.48\textwidth]{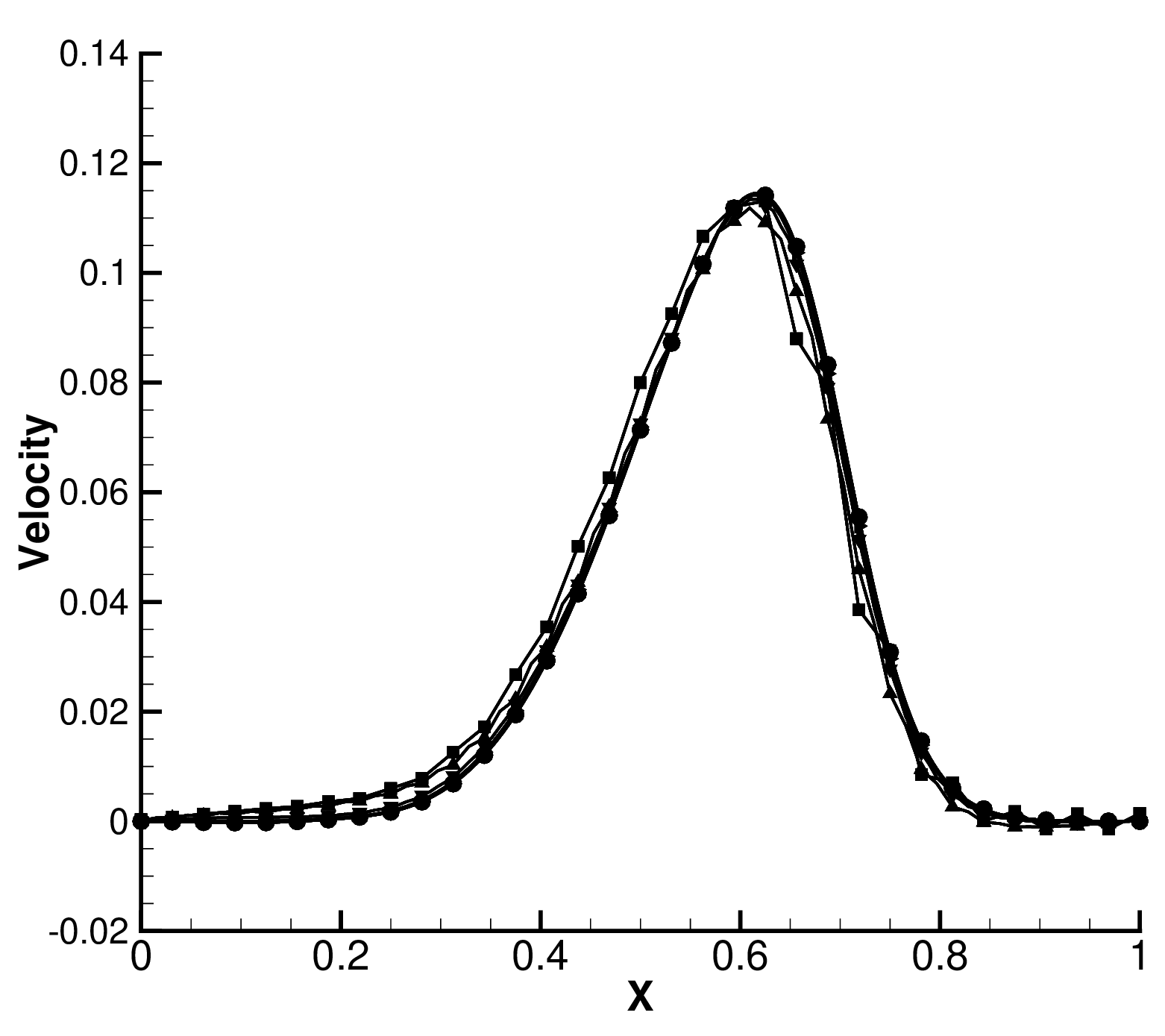}
\includegraphics[width=0.48\textwidth]{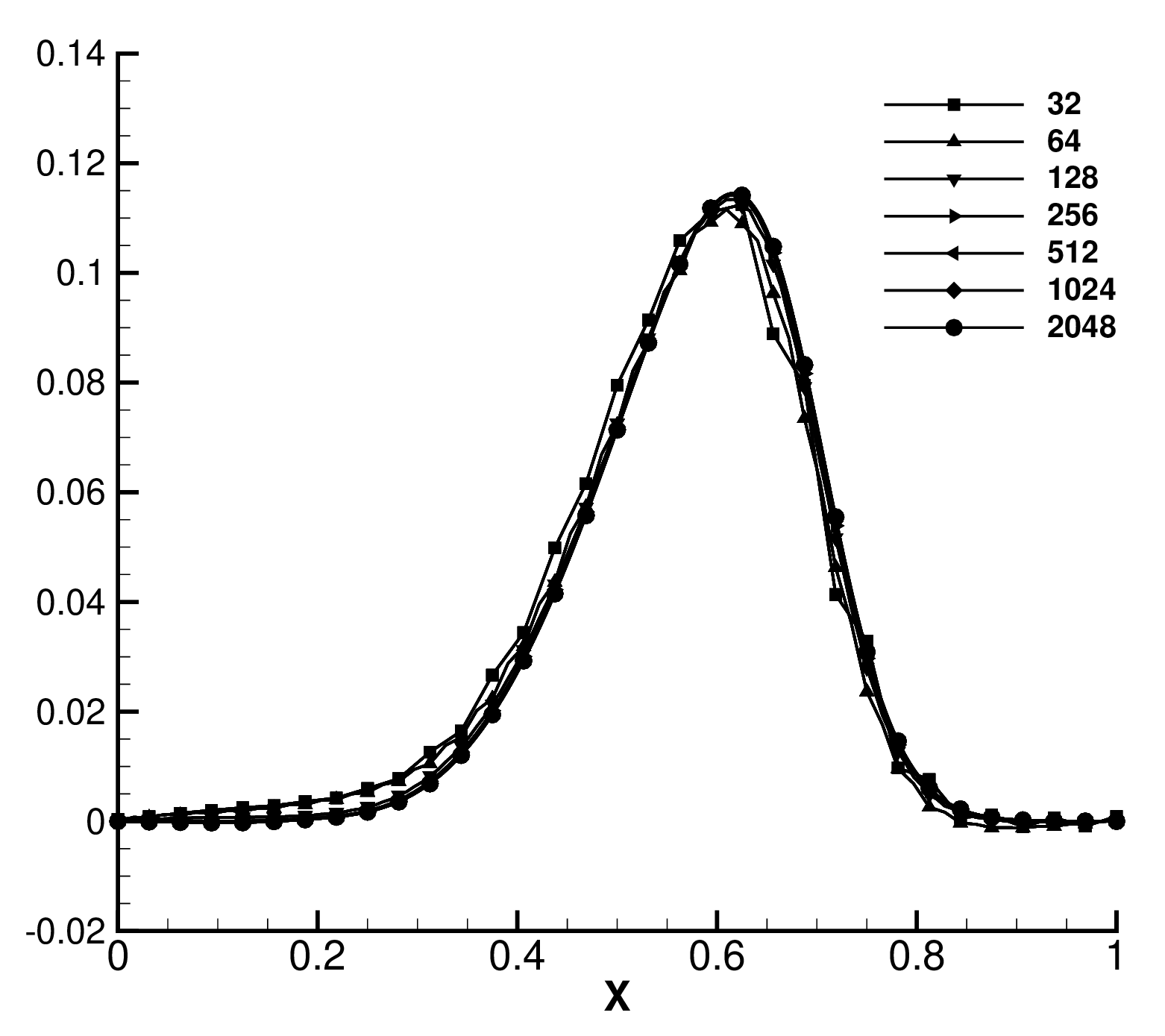}
\includegraphics[width=0.48\textwidth]{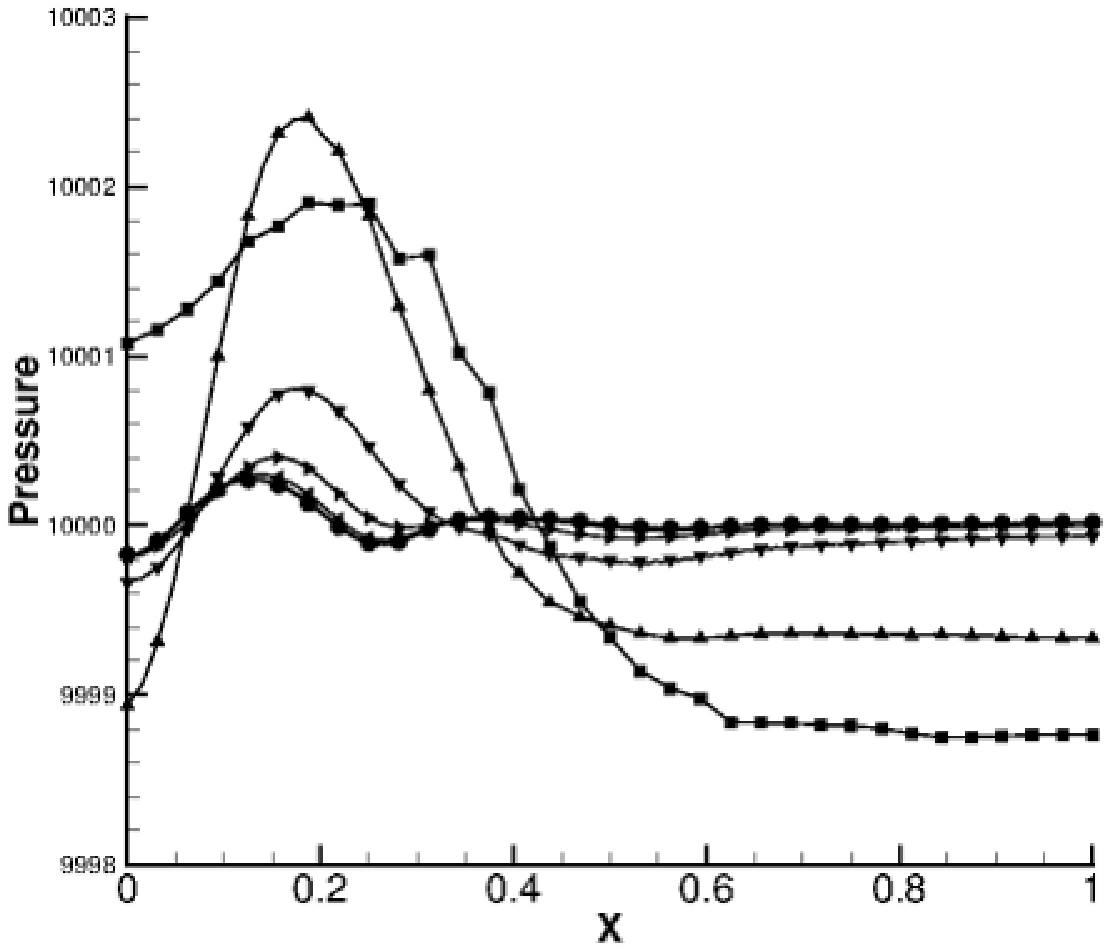}
\includegraphics[width=0.48\textwidth]{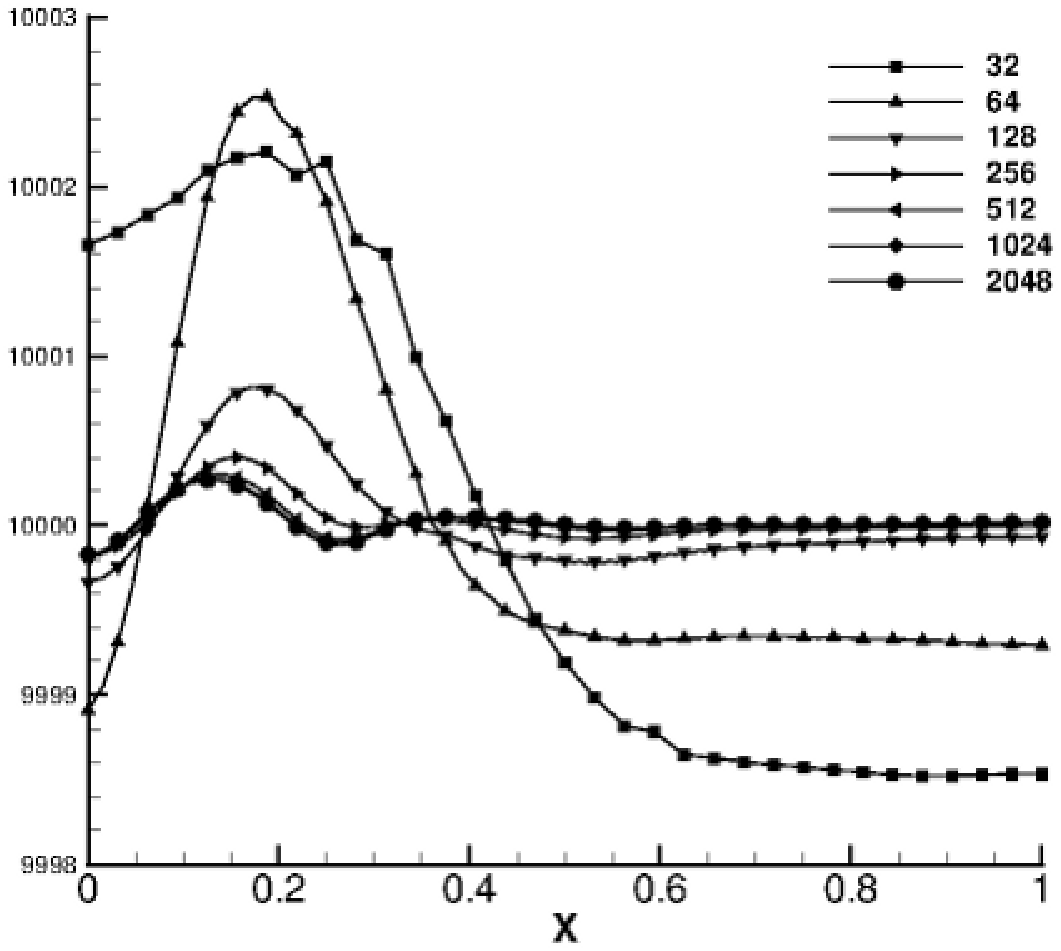}
\includegraphics[width=0.48\textwidth]{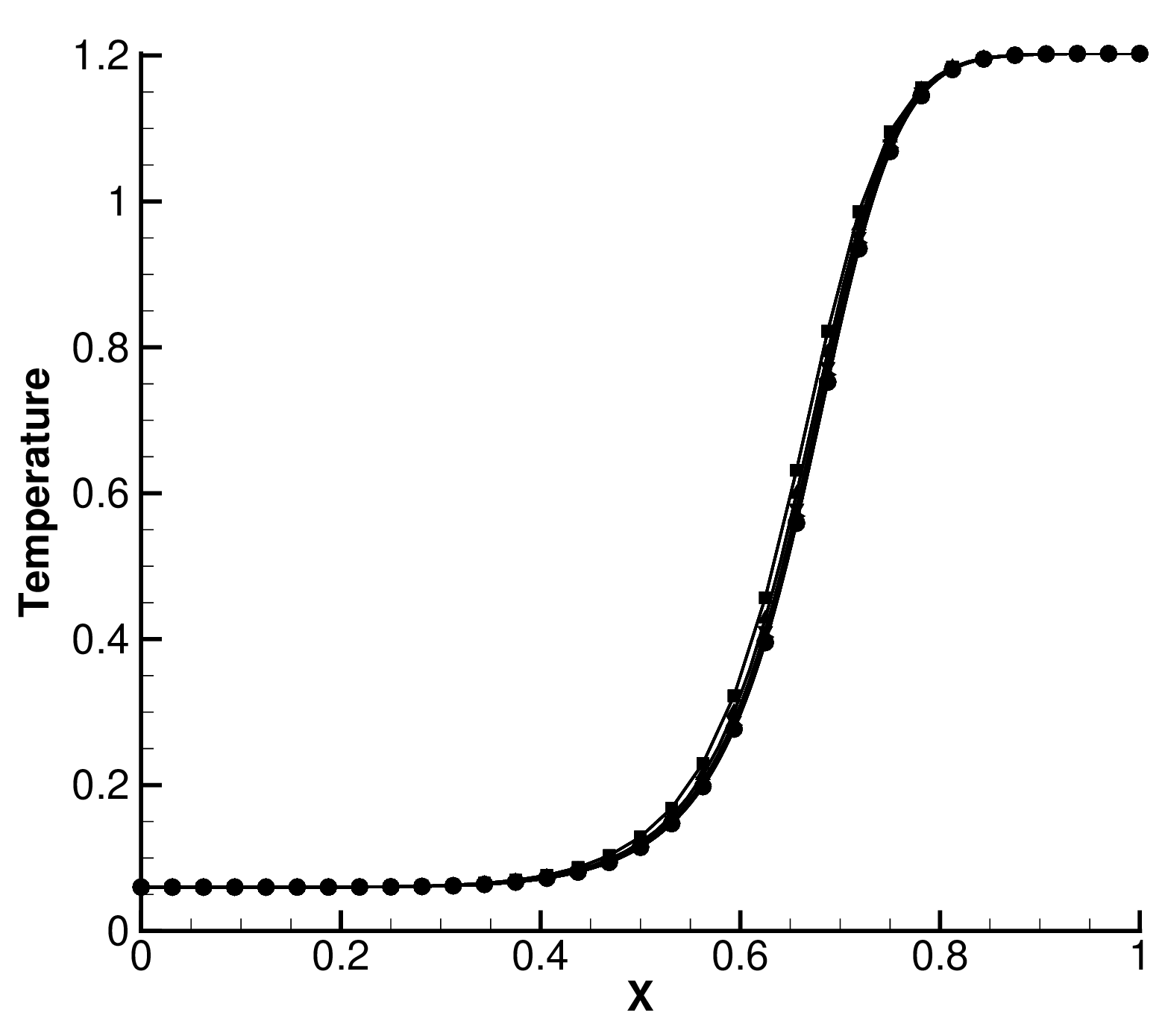}
\includegraphics[width=0.48\textwidth]{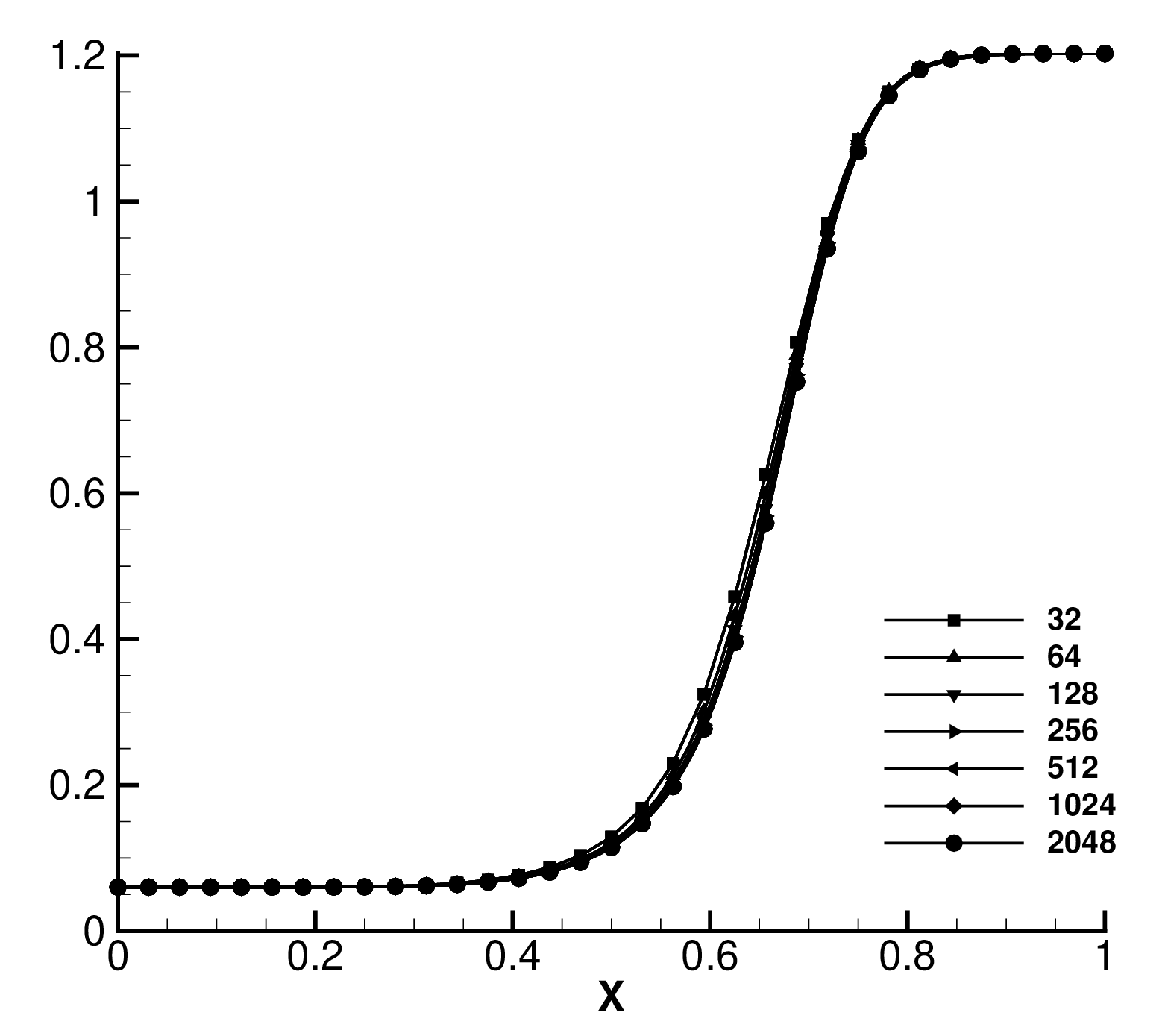}
\caption{Comparison of variables across all grid resolutions for Case 2 with the mass fraction model results on the left and the number fraction model results on the right. \label{fig:Case2Plots}}
\end{centering}
\end{figure}

\begin{figure}
\begin{centering}
\includegraphics[width=0.49\textwidth]{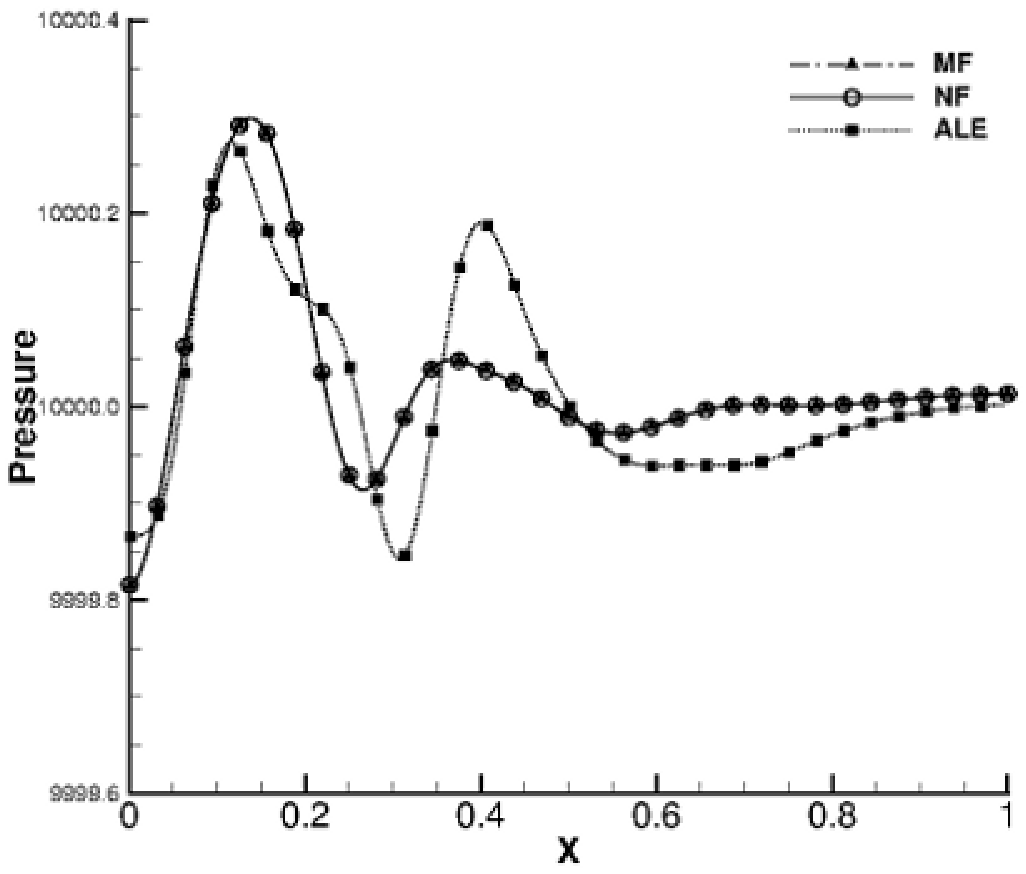}
\includegraphics[width=0.49\textwidth]{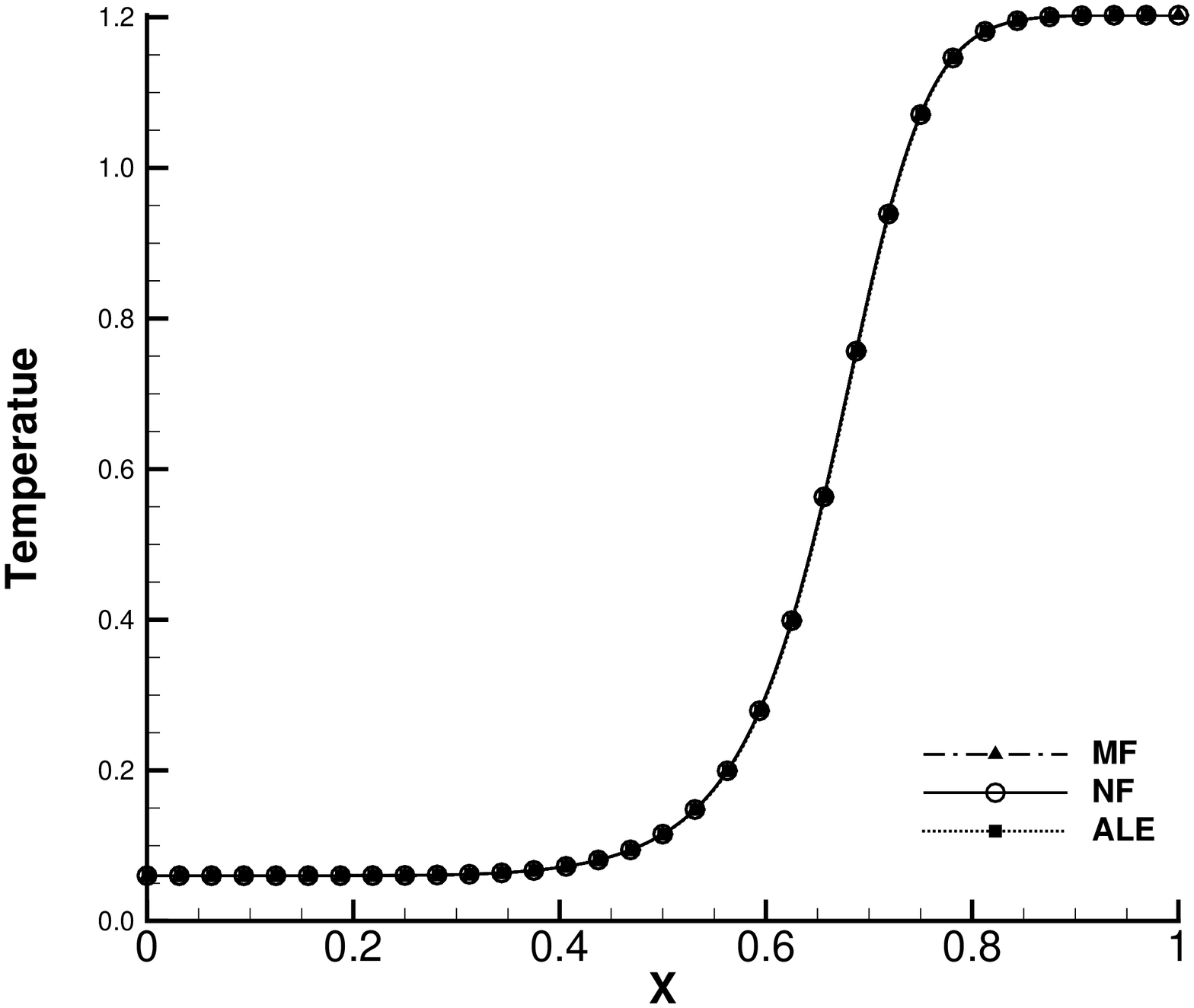}
\caption{Case 2 comparison of non-analytical field variables for the mass fraction (MF) and number fraction (NF) models in Flamenco (512 cells) as well as the results from the Lagrange-remap code (512 cells). \label{fig:Case2Comp}}
\end{centering}
\end{figure}

Figures \ref{fig:Case2Plots} plots the spatial variation of flow properties for both the mass and number fraction formulation, demonstrating that for a specific grid resolution there are only slight variations in the solutions. Figure  \ref{fig:Case2Comp} compares to the Lagrange-remap formulation for $256$ cells showing good agreement for the temperature distribution. The pressure distributions are simulation are similar but are not converged at this mesh resolution. Again this is extreme case where pressure waves travel about $70$ times the domain width at $t=0.5$. 

\subsection{Case 3: Advection and Diffusion of an Isothermal Contact Surface between two Different Species}

As highlighted in the introduction, the key errors in the mass fraction approach appear when a contact surface between two different species at two different temperatures is advected through the computational mesh. Thus Case 3 is a periodic version of Case 1, with a mean velocity specified of 4m/s. This ensures that at $t=0.5$s, the initially diffuse interface returns exactly to the original position.
 
Thus, the computation domain is $0 \le x \le 2$, and grid sizes from $32 \rightarrow 2048$ have been employed. Periodic boundary conditions are used at the left and right hand boundaries. The two fluids have the same properties as Case 1, namely $\rho_1=20$, $\rho_2=1$, $\gamma_1=2$, $\gamma_2=1.4$. The specific heats satisfy the requirement that $(\gamma_1-1)\rho_1 c_{v1}=(\gamma_2-1)\rho_2 c_{v2}$, which gives temperature equilibrium if the two fluids have the same pressure.

For $x\le 1$, the analytical solution for the number fraction distribution is given by 

\begin{equation}
X_1=\frac{1}{2} \left[1-\textrm{erf} \left(\mathcal{Z}\right)\right],\mathcal{Z}=\frac{x-x_0}{\sqrt{4Dt+h_0^2}}.
\end{equation}

\noindent where $x_0=0.5$. The number fraction profile is then mirrored about $x=1$, thus for $1\le x$ the number fraction is given by 

\begin{equation}
X_1=\frac{1}{2} \left[1+\textrm{erf} \left(\mathcal{Y}\right)\right],\mathcal{Y}=\frac{x-x_1}{\sqrt{4Dt+h_0^2}}.
\end{equation}

\noindent where $x_1=1.5$, $h_0=0.02$ and the diffusion coefficient $D=0.01$. The initial cell averaged quantities are gained by simply mirroring the expressions detailed for Case 1 about $x=1$.

\begin{table*}[]
\centering
\caption{Case 3 convergence rates for the number fraction model.}
\label{tab:Case3VFConvergence}
\begin{tabular}{lllllll}
\hline
N & $L^1$  & $L^2$  & $L^\infty$  & $\mathcal{O} (L^1)$  & $\mathcal{O} (L^2)$  & $\mathcal{O} (L^\infty)$  \\
\hline
64 & 3.8866e-03	& 3.9599e-03 &	6.9897e-03 & - & - & - \\
128 & 3.4887e-04 &	3.6400e-04 &	8.0443e-04 &	3.4778 &	3.4434 &	3.1192 \\
256 & 8.8780e-05 &	8.9915e-05 &	1.6597e-04 &	1.9744 & 	2.0173 &	2.2770 \\
512 & 2.4260e-05 &	2.4690e-05 &	4.1986e-05 &	1.8716 &	1.8646 &	1.9829 \\
1024 & 6.8051e-06 &	7.0069e-06 &	1.0922e-05 &	1.8339 &	1.8171 &	1.9427 \\
2048 & 2.6324e-06 &	2.9718e-06 &	5.6445e-06 &	1.3703 &	1.2374 &	0.9523 \\
4096 & 1.8844e-06 &	2.2868e-06 &	4.3834e-06 &	0.4823 &	0.3780 &	0.3648 \\
\hline
\end{tabular}
\end{table*}

\begin{table*}[]
\centering
\caption{Case 3 convergence rates for the mass fraction model.}
\label{tab:Case3MFConvergence}
\begin{tabular}{lllllll}
\hline
N & $L^1$  & $L^2$  & $L^\infty$  & $\mathcal{O} (L^1)$  & $\mathcal{O} (L^2)$  & $\mathcal{O} (L^\infty)$  \\
\hline
64 & 5.2677e-02 &	7.2627e-02 &	1.7744e-01 & - & - & - \\
128 & 1.4011e-02 &	1.8492e-02 &	3.7163e-02 &	1.9106 &	1.9736 &	2.2554 \\
256 & 3.4566e-03 &	4.5593e-03 &	9.2933e-03 &	2.0191 &	2.0200 &	1.9996 \\
512 & 8.7023e-04 &	1.1486e-03 &	2.3790e-03 &	1.9899 &	1.9890 &	1.9659 \\
1024 & 2.1800e-04 &	2.8778e-04 &	5.9729e-04 &	1.9971 &	1.9968 &	1.9938 \\
2048 & 5.4519e-05 &	7.1936e-05 &	1.4825e-04 &	1.9995 &	2.0002 &	2.0104 \\
4096 & 1.3724e-05 &	1.8020e-05 &	3.5851e-05 &	1.9900 &	1.9971 &	2.0480 \\
\hline
\end{tabular}
\end{table*}

\begin{figure}
\begin{centering}
\includegraphics[width=\textwidth]{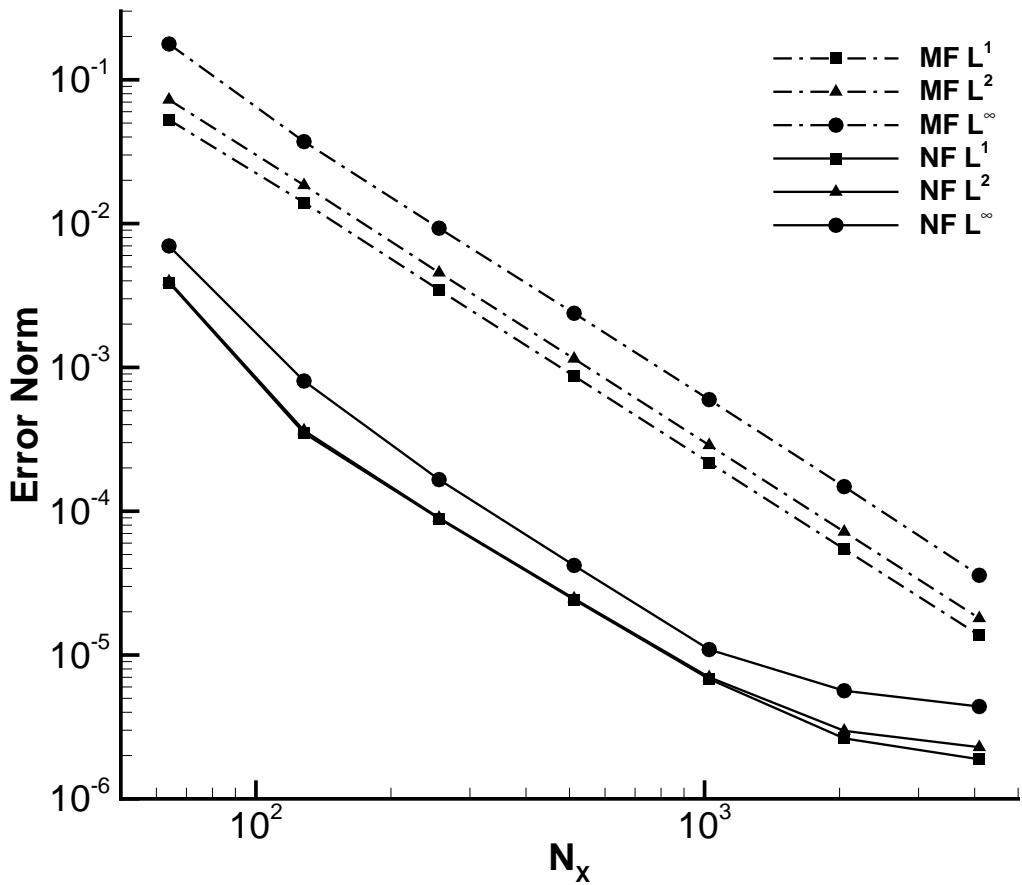}
\caption{Plot of Case 3 convergence rates, showing comparison between the mass fraction and number fraction approaches. \label{fig:Case3Convergence}}
\end{centering}
\end{figure}

Tables \ref{tab:Case3VFConvergence} and \ref{tab:Case3MFConvergence} document the $L^1$, $L^2$ and $L^\infty$ errors in the simulated number fraction profile at $t=0.5$ for the number fraction and mass fraction formulations compared to the incompressible analytical solution, along with the observed convergence rates. Figure \ref{fig:Case3Convergence} plots the error norms as a function of number of points $N_x$ for the mass fraction and number fraction equations. 

Once again, the error norms show approximate second order of accuracy for both choices of governing equation. However, here there is an enormous difference in absolute error, where the number fraction model is at least one order of magnitude more accurate for this problem at a specific grid resolution. 

This highlights perfectly the key motivation for developing this model and discretisation. In practical computations of mixing problems there is usually very little possibility to resolve the diffuse interface between two gases with as many points as is possible in one dimension here. Thus, the 64 or 128 grid resolutions here are most representative of practical computations of mixing flow with turbulent fluctuations. This result implies that for a fixed error in a three dimensional problem the new number fraction model may be run on a mesh almost 4 times smaller in each direction, or 64 times smaller overall. Assuming a CFL restriction on the time step size this translates to a potential computational saving which is on the order of $200$ in three dimensions, compensating for the additional computational expense in the number fraction formulation compared to the mass fraction formulation. 

\begin{figure}
\begin{centering}
\includegraphics[width=0.48\textwidth]{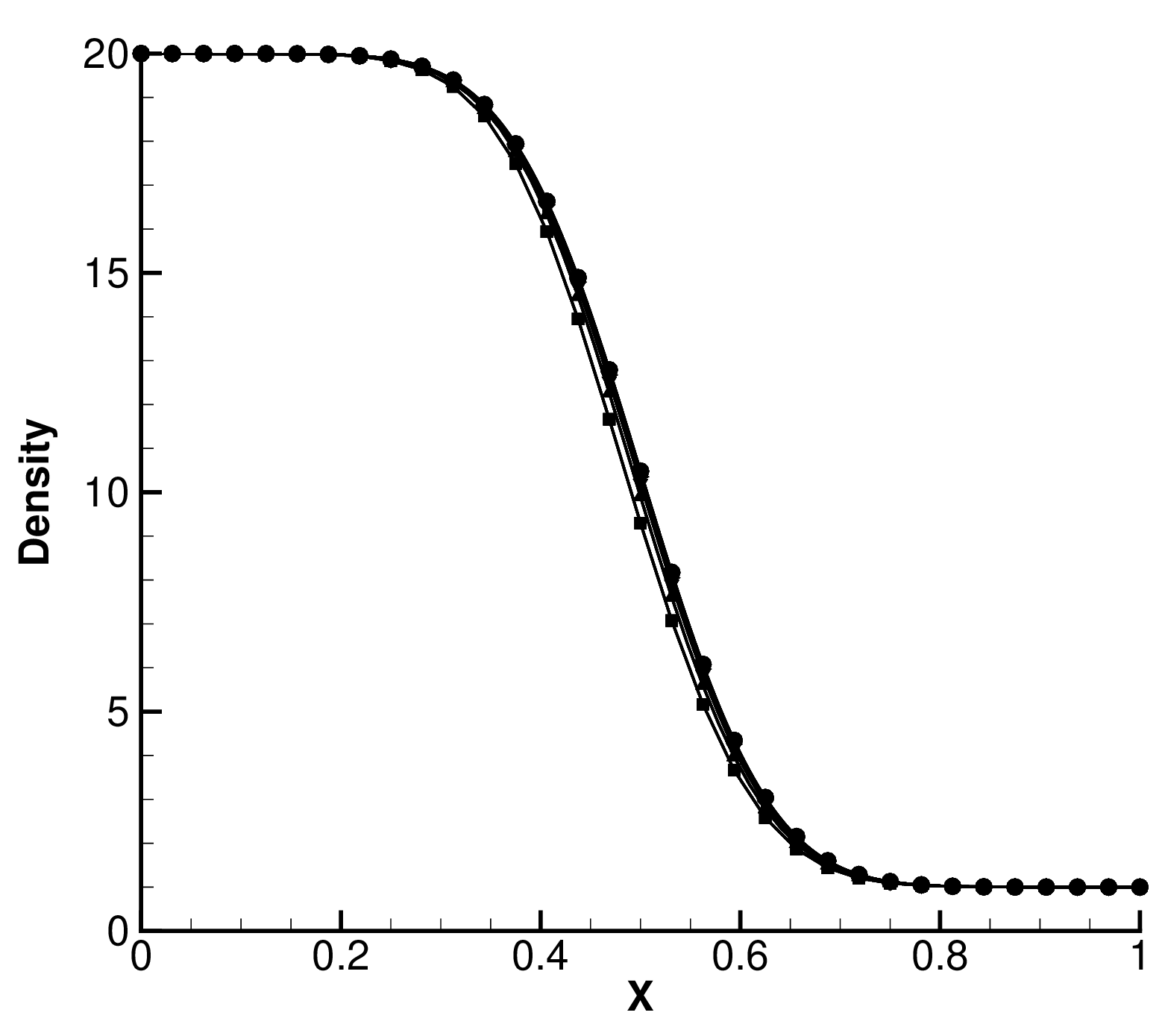}
\includegraphics[width=0.48\textwidth]{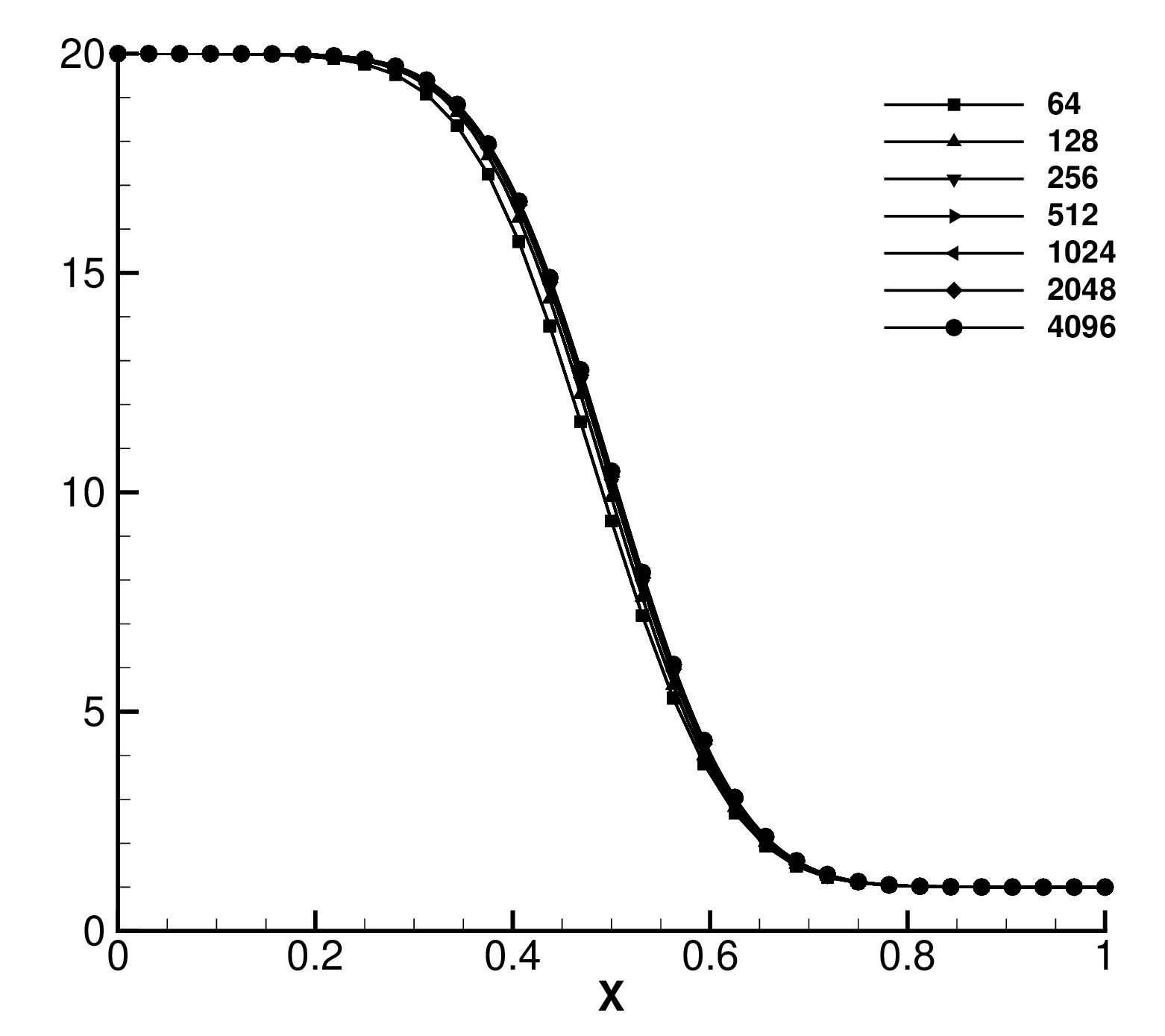}
\includegraphics[width=0.48\textwidth]{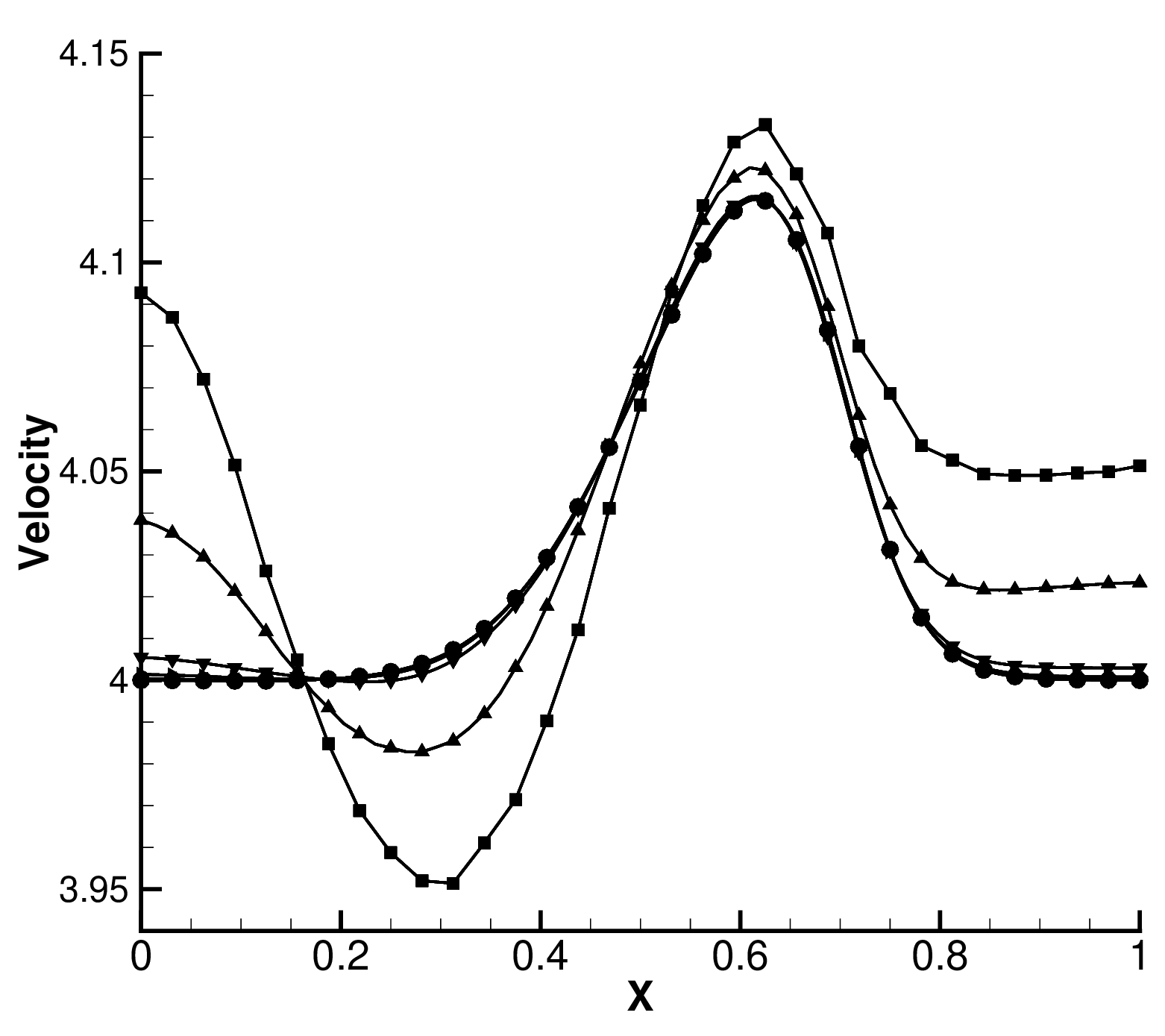}
\includegraphics[width=0.48\textwidth]{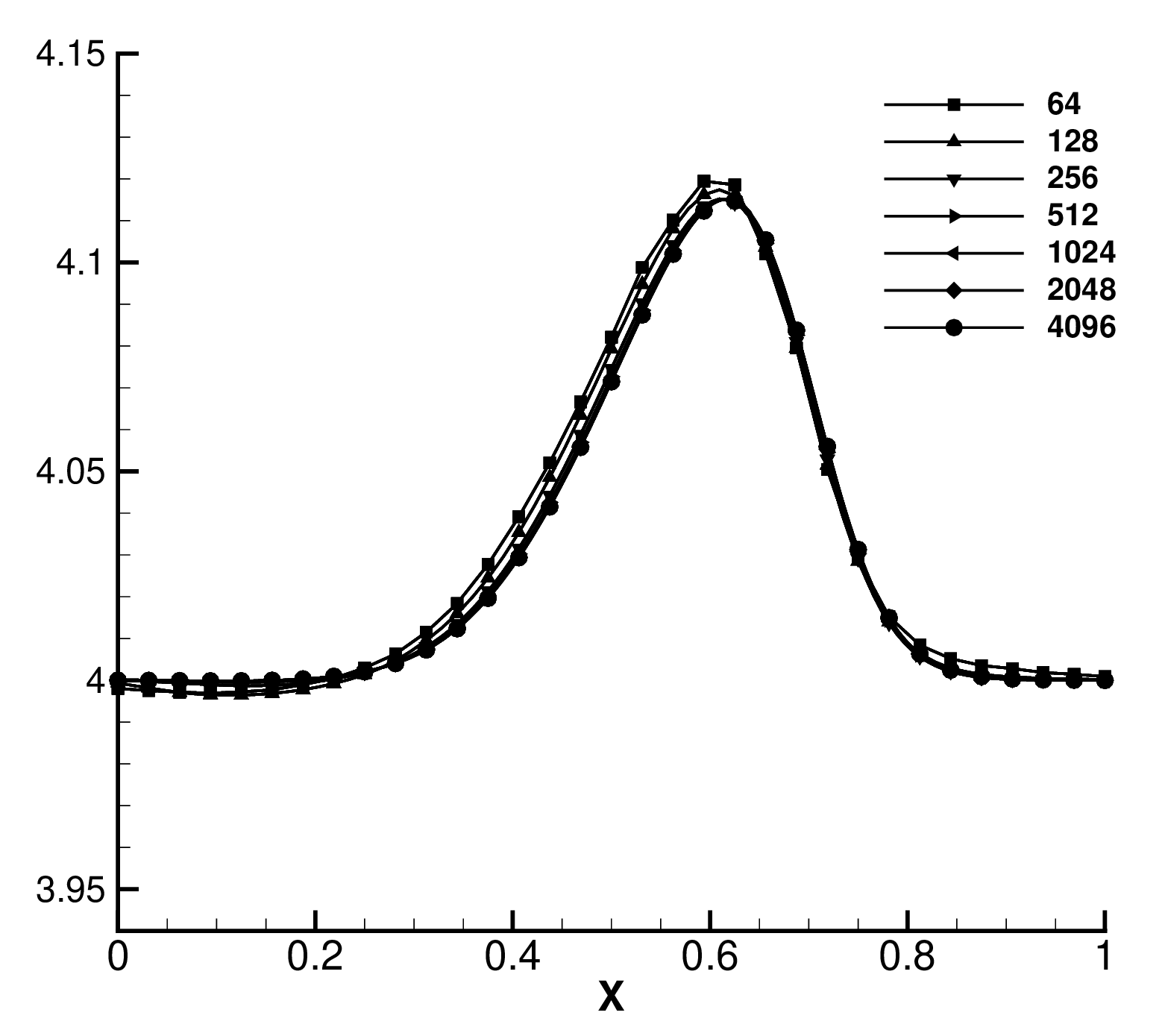}
\includegraphics[width=0.48\textwidth]{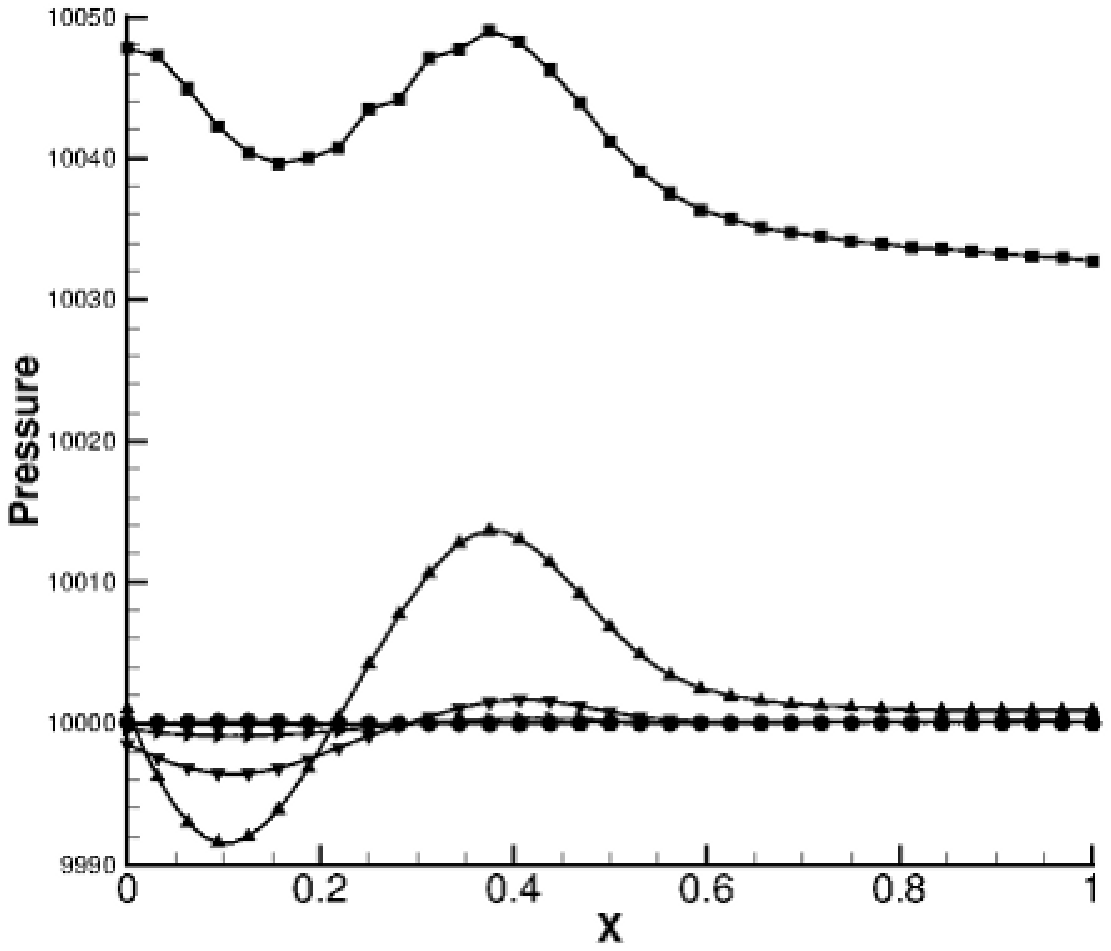}
\includegraphics[width=0.48\textwidth]{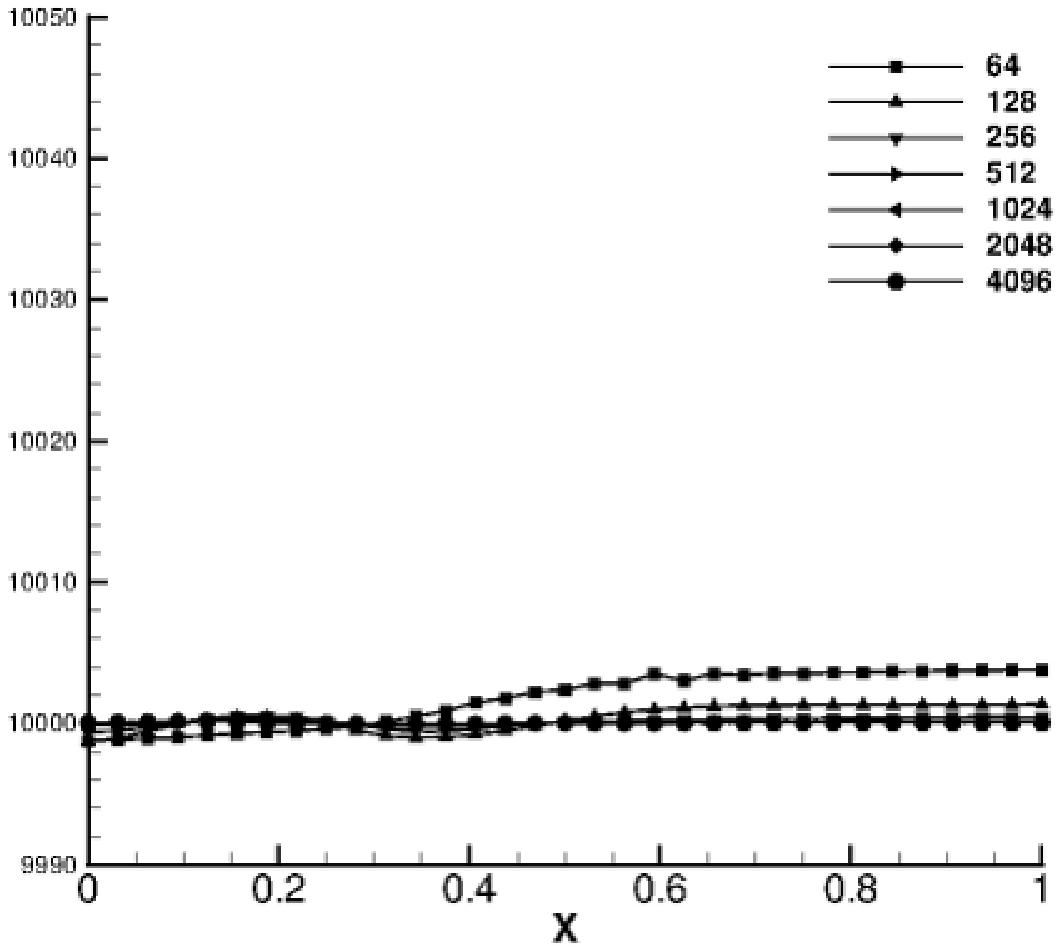}
\includegraphics[width=0.48\textwidth]{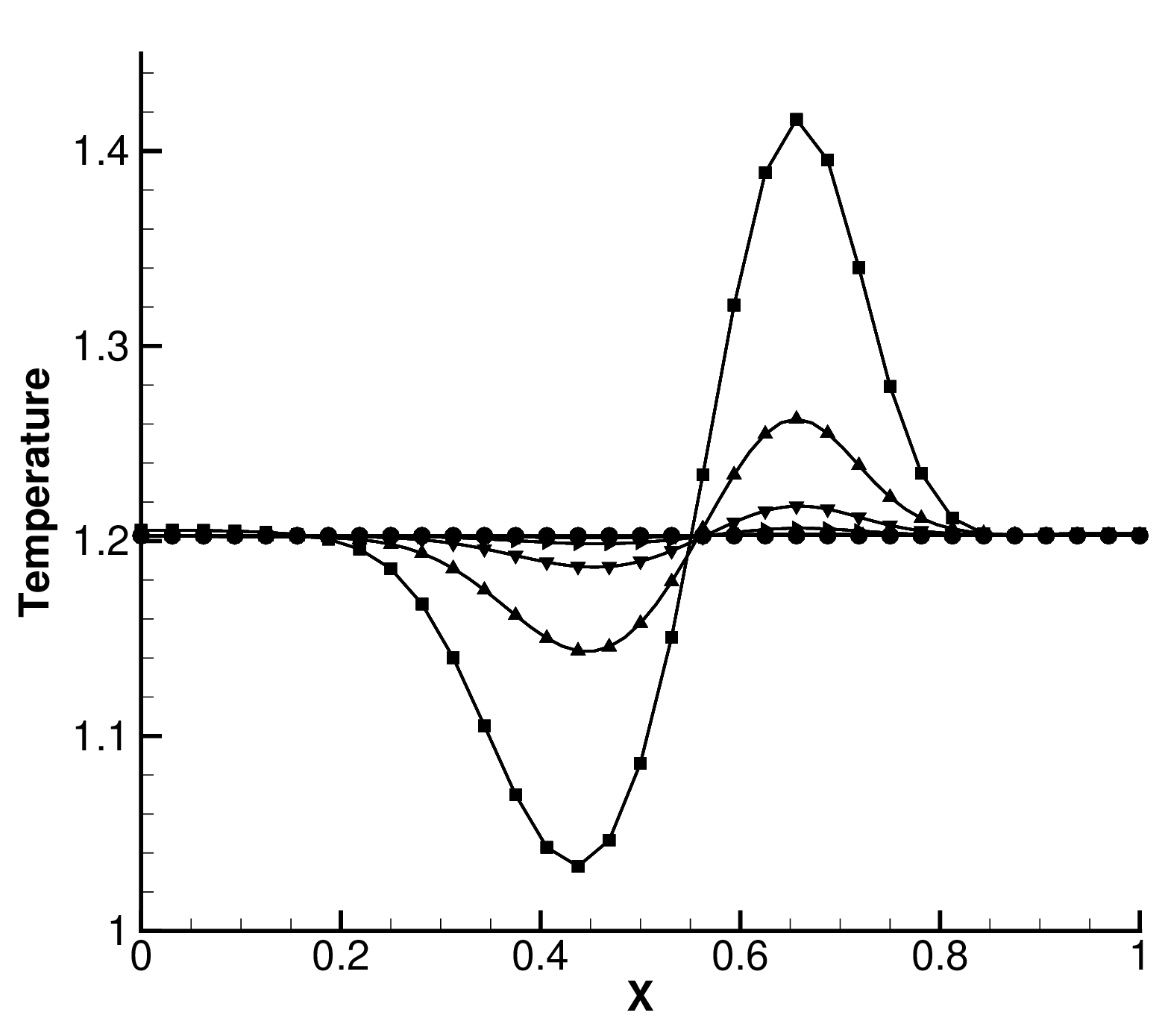}
\includegraphics[width=0.48\textwidth]{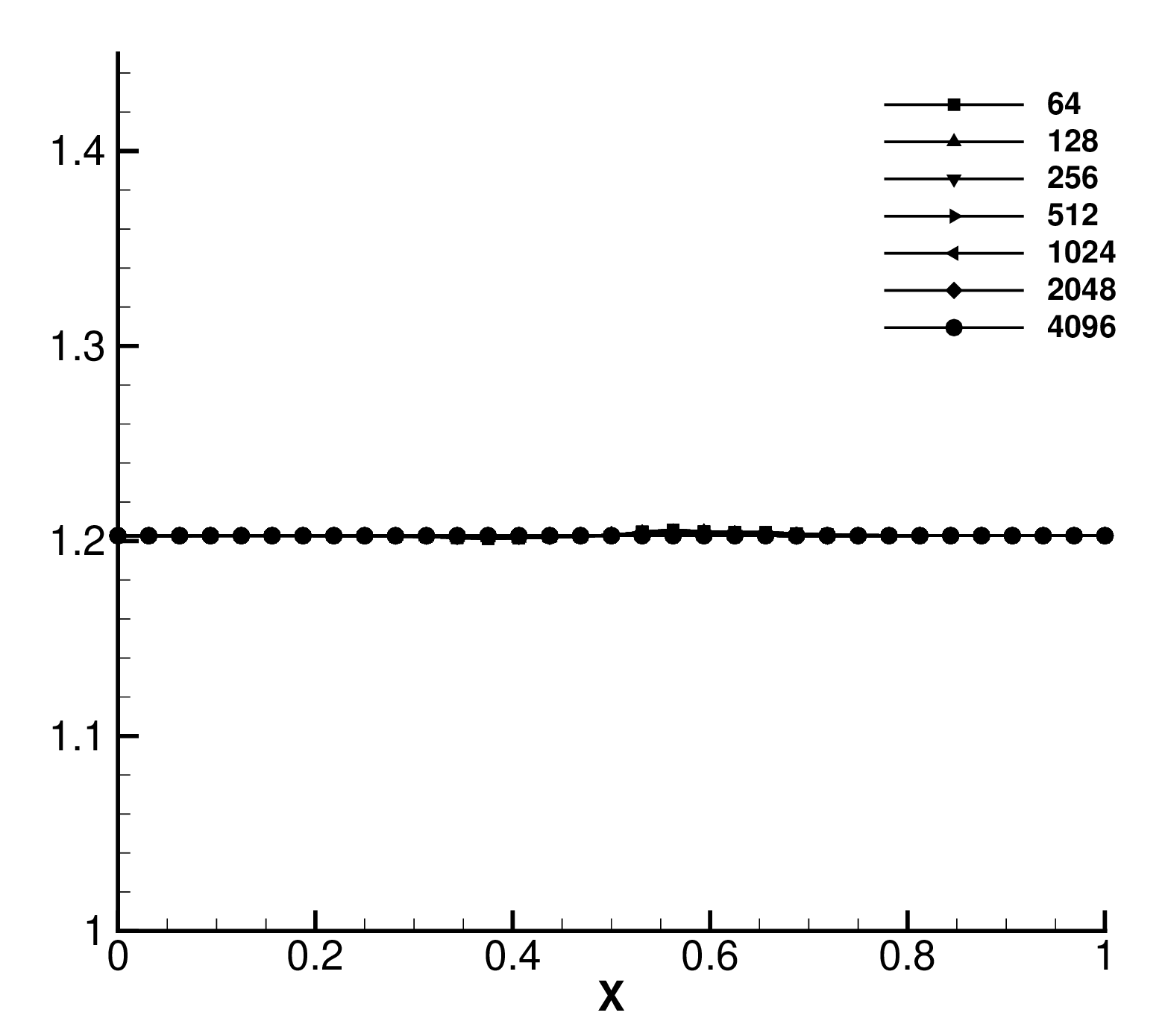}
\caption{Comparison of variables across all grid resolutions for Case 3 with the mass fraction model results on the left and the number fraction model results on the right. \label{fig:Case3Plots}}
\end{centering}
\end{figure}

\begin{figure}
\begin{centering}
\includegraphics[width=0.49\textwidth]{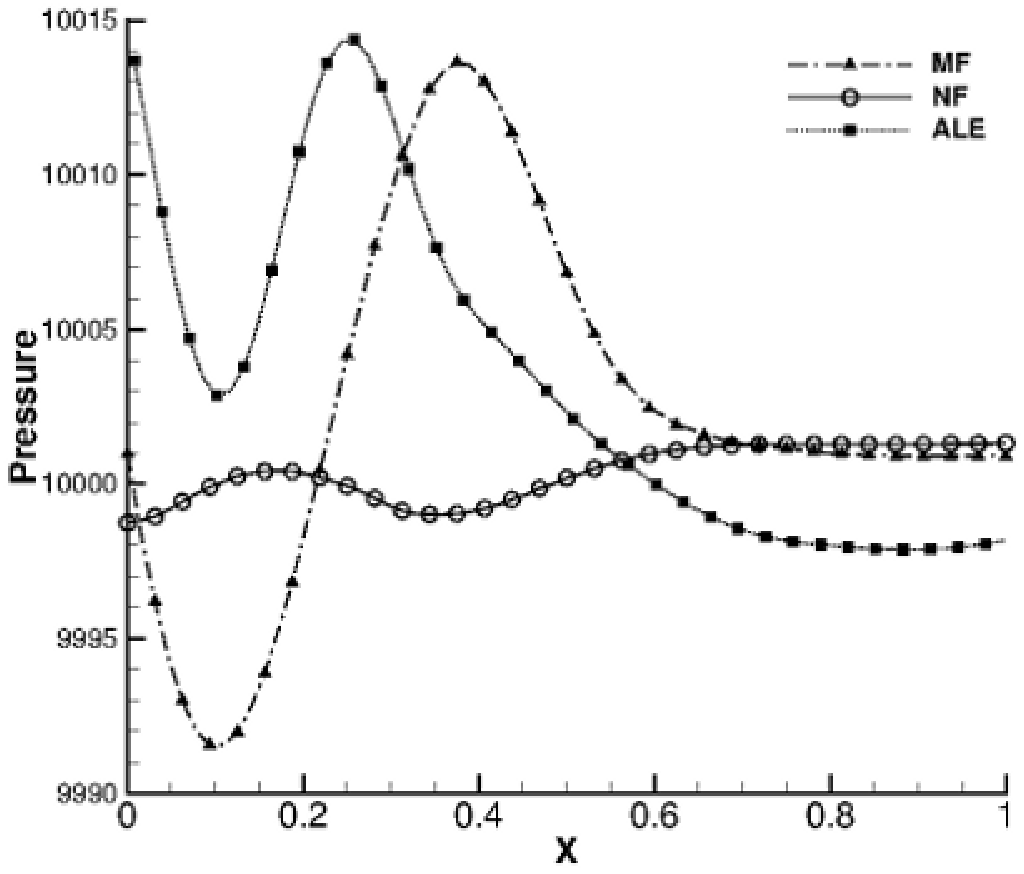}
\includegraphics[width=0.49\textwidth]{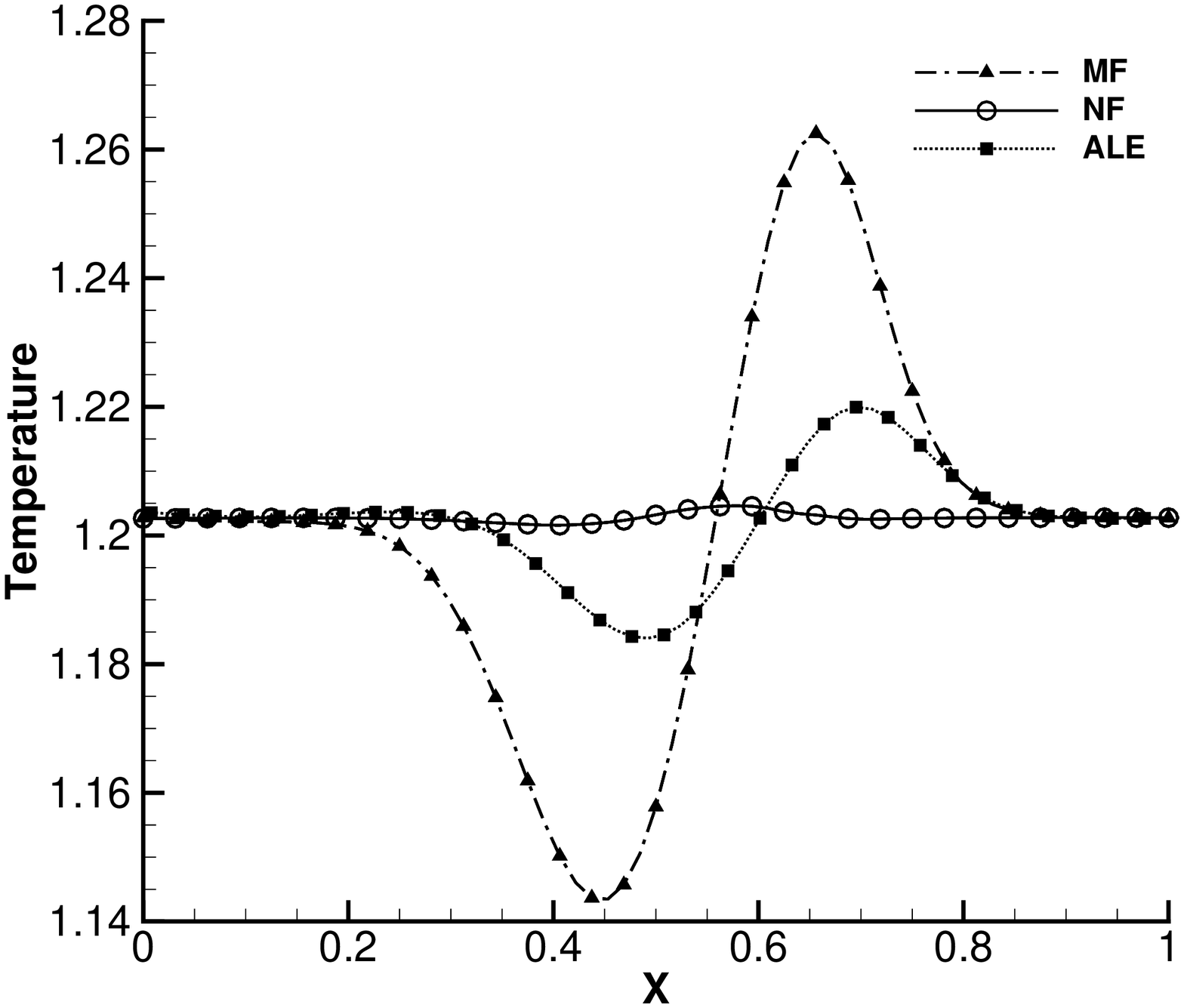}
\caption{Case 3 comparison of non-analytical field variables for the mass fraction (MF) and number fraction (NF) models in Flamenco (128 cells) as well as the results from the Lagrange-remap code Turmoil3D (128 cells). \label{fig:Case3Comp}}
\end{centering}
\end{figure}

Figure \ref{fig:Case3Plots} plots the density, velocity, pressure and temperature distributions at t=0.5s for both models for $0\le x\le 1$ (one half of the symmetric solution). The errors in mass weighted velocity, pressure and temperature when using the mass fraction model are substantial. Figure \ref{fig:Case3Comp} compares the non-analytical fields from the three algorithms at a lower grid resolution (128 cells), highlighting the substantial improvement gained using the number fraction model at realistic resolutions for a three dimensional computation. Note that the large mass fraction errors are present in both the Eulerian and Lagrange remap mass fraction implementations.

This case demonstrates that for misicble mixing of differing gases the single pressure and temperature closure in the mass fraction model greatly impacts the accuracy of the scheme. Allowing volume-averaged species temperatures to vary results in a substantially more accurate numerical solution for the same computational effort.

\section{Two-Dimensional Richtmyer-Meshkov Instability between Air and SF$_6$ \label{2dresults}}

To demonstrate the applicability of the new algorithm to compressible mixing problems, Case 4 undertakes direct numerical simulations of a two dimensional single mode Richtmyer-Meshkov instability triggered by a shock wave passing from air to SF$_6$. The setup for this case is adapted from that proposed by Tritschler {\it et al.} \cite{Tritschler2014pre}. The purpose of this case is to illustrate the advantage of using the proposed number fraction formulation in a more practical setting, since there is widespread experimental use of air-SF$_6$ combinations in shock-induced turbulence (see, for example,\cite{Vetter1995,Holder2003,Jacobs2005,morgan2012late}). 

To conduct an accurate simulation, the initial conditions must be very well resolved. Here, a single mode perturbation of wavelength 0.5mm and initial amplitude 0.025mm is computed. The initial diffuse layer thickness is 0.1mm, sufficiently low to prevent substantial damping of the initially imparted impulse. The initial state of the unshocked gases are $p_0=23000$Pa and $T=298K$. For simplicity a fixed viscosity $\mu=2.243\times 10^{-5}$Pa-s is specified \cite{Tritschler2014}, along with $Sc=1$ and $Pr=1$. Note that inviscid computations have also been run to represent the very high Reynolds number limit. Air is assumed to have $\gamma_{air}=1.4$, and $\gamma_{SF_6}=1.1$, and the molecular weights are $M_{air}=28.964$ and $M_{SF_6}=146.057$ \cite{Tritschler2014pre}. The shock Mach number is 1.5, and the shock has an initial offset of $1$mm from the mean interface position. The interface diffuses slightly prior to shock interaction, increasing the integral width by 2.7\% at the time of shock interaction in the cases incorporating viscosity, diffusion and conduction.  All results are scaled by the modal wavelength $\lambda$ and Richtmyer's velocity.

The details of the initial conditions for this two-dimensional problem are derived as follows. The incompressible limit of mixing by species diffusion gives the following non-zero divergence of velocity \cite{Livescu2013}:

\begin{equation*}
	\nabla\cdot\mathbf{u} = -\nabla\cdot(\frac{D}{\rho}\nabla\rho)
\end{equation*}

\noindent

Where the diffusion velocity is given by:

\begin{equation*}
	D = \frac{\mu}{\rho Sc}
\end{equation*}

\noindent
In 2D this becomes:

\begin{equation*}
\begin{aligned}
\frac{\partial u}{\partial x} + \frac{\partial v}{\partial y} & = -\nabla\cdot(\frac{\mu}{\rho^2Sc}\nabla\rho) \\
& = \frac{2\mu}{Sc\rho^3}\Bigg(\Big(\frac{\partial\rho}{\partial x}\Big)^2+\Big(\frac{\partial\rho}{\partial y}\Big)^2\Bigg) - \frac{\mu}{Sc\rho^2}\Bigg(\frac{\partial^2\rho}{\partial x^2}+\frac{\partial^2\rho}{\partial y^2}\Bigg)\\
\end{aligned}
\end{equation*}

\noindent
Now, let:

\begin{equation*}
\frac{\partial u}{\partial x} = \frac{2\mu}{Sc\rho^3}\Big(\frac{\partial\rho}{\partial x}\Big)^2 - \frac{\mu}{Sc\rho^2}\frac{\partial^2\rho}{\partial x^2}, \,\,\,\frac{\partial v}{\partial y} = \frac{2\mu}{Sc\rho^3}\Big(\frac{\partial\rho}{\partial y}\Big)^2 - \frac{\mu}{Sc\rho^2}\frac{\partial^2\rho}{\partial y^2}.
\end{equation*}

\noindent
Then integrating gives:

\begin{equation*}
u = - \frac{\mu}{Sc\rho^2}\frac{\partial\rho}{\partial x}, \,\,\, v = - \frac{\mu}{Sc\rho^2}\frac{\partial\rho}{\partial y},
\end{equation*}

\noindent which is applied only at the interface (not at the shock). Finally, taking cell averages:

\begin{equation}
\begin{aligned}
\overline{u_i} =  \frac{\mu}{Sc\Delta x \Delta y}\int_{y_{i-\frac{1}{2}}}^{y_{i+\frac{1}{2}}}\Big(\frac{1}{\rho(x_{i+\frac{1}{2}},y)}-\frac{1}{\rho(x_{i-\frac{1}{2}},y)}\Big)dy \\
\end{aligned}
\label{eqn:UCellAvrg}
\end{equation}

\begin{equation}
\begin{aligned}
\overline{v_i} & = \frac{\mu}{Sc\Delta x \Delta y}\int_{x_{i-\frac{1}{2}}}^{x_{i+\frac{1}{2}}}\Big(\frac{1}{\rho(x,y_{i+\frac{1}{2}})}-\frac{1}{\rho(x,y_{i-\frac{1}{2}})}\Big)dx \\
\end{aligned}
\label{eqn:VCellAvrg}
\end{equation}

\noindent
Where:

\begin{equation*}
\rho(x,y) = \rho_1 X_1(x,y)+\rho_2(1-X_1(x,y))
\end{equation*}

\begin{equation*}
X_1(x,y) = \frac{1}{2}\Big(1-\mathrm{erf}\Big(\frac{F(x,y)\sqrt{\pi}}{h_0}\Big)\Big)
\end{equation*}

\noindent
With $F(x,y)$ being the minimum distance from the point $(x,y)$ to the perturbed interface. Equations (\ref{eqn:UCellAvrg}) and (\ref{eqn:VCellAvrg}) are calculated numerically using a 5-point Gaussian quadrature rule, while the minimum distance $F(x,y)$ is also calculated numerically using a bisection method iteration. 

\begin{figure}
\begin{centering}
\subfigure[Density for the viscous models]{\begin{overpic}[width=0.49\textwidth]{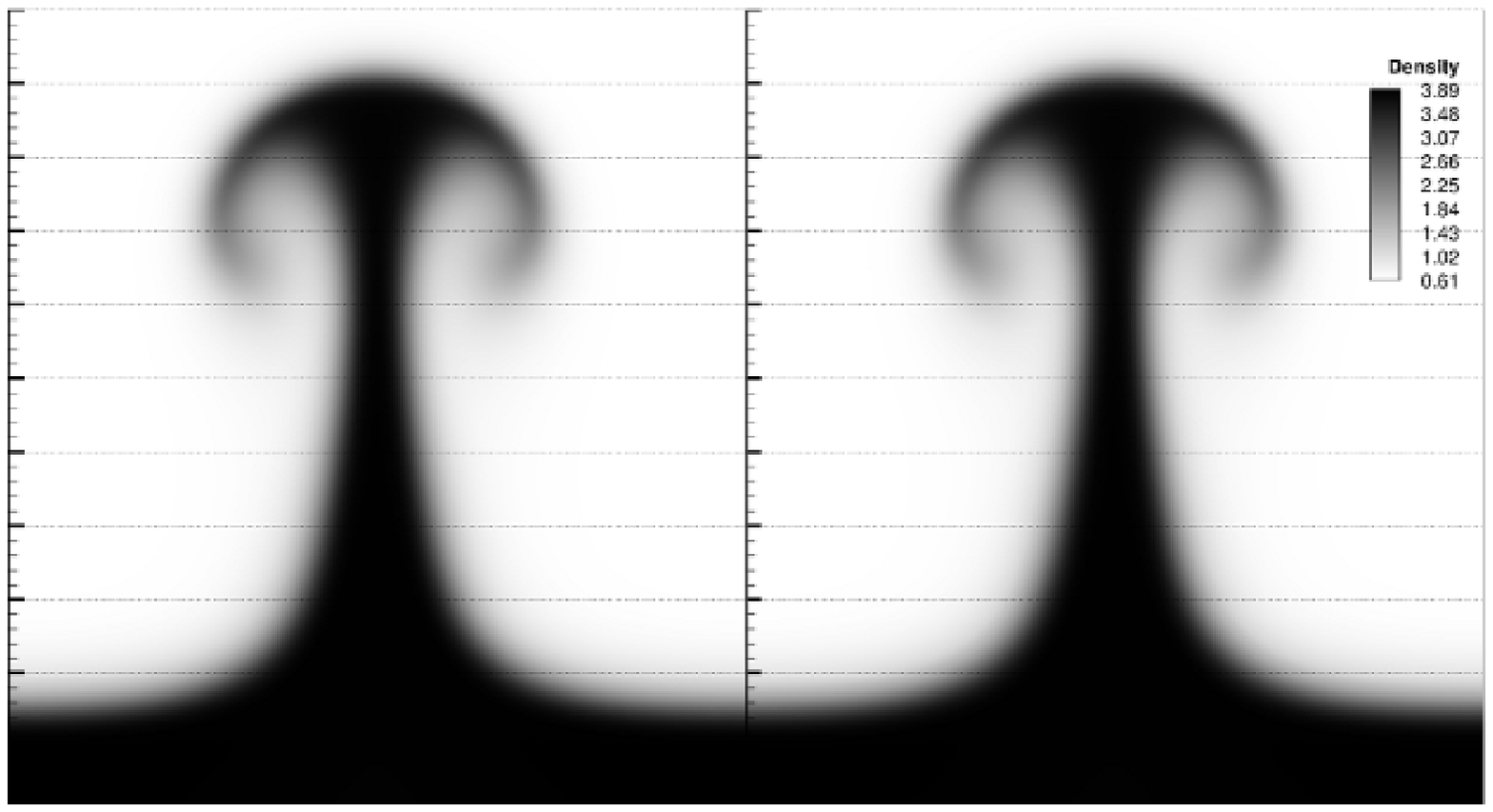}
    \small
    \put(105,96){NF}
    \put(5,96){MF}
\end{overpic}
}
\subfigure[Density for the inviscid models]{\begin{overpic}[width=0.49\textwidth]{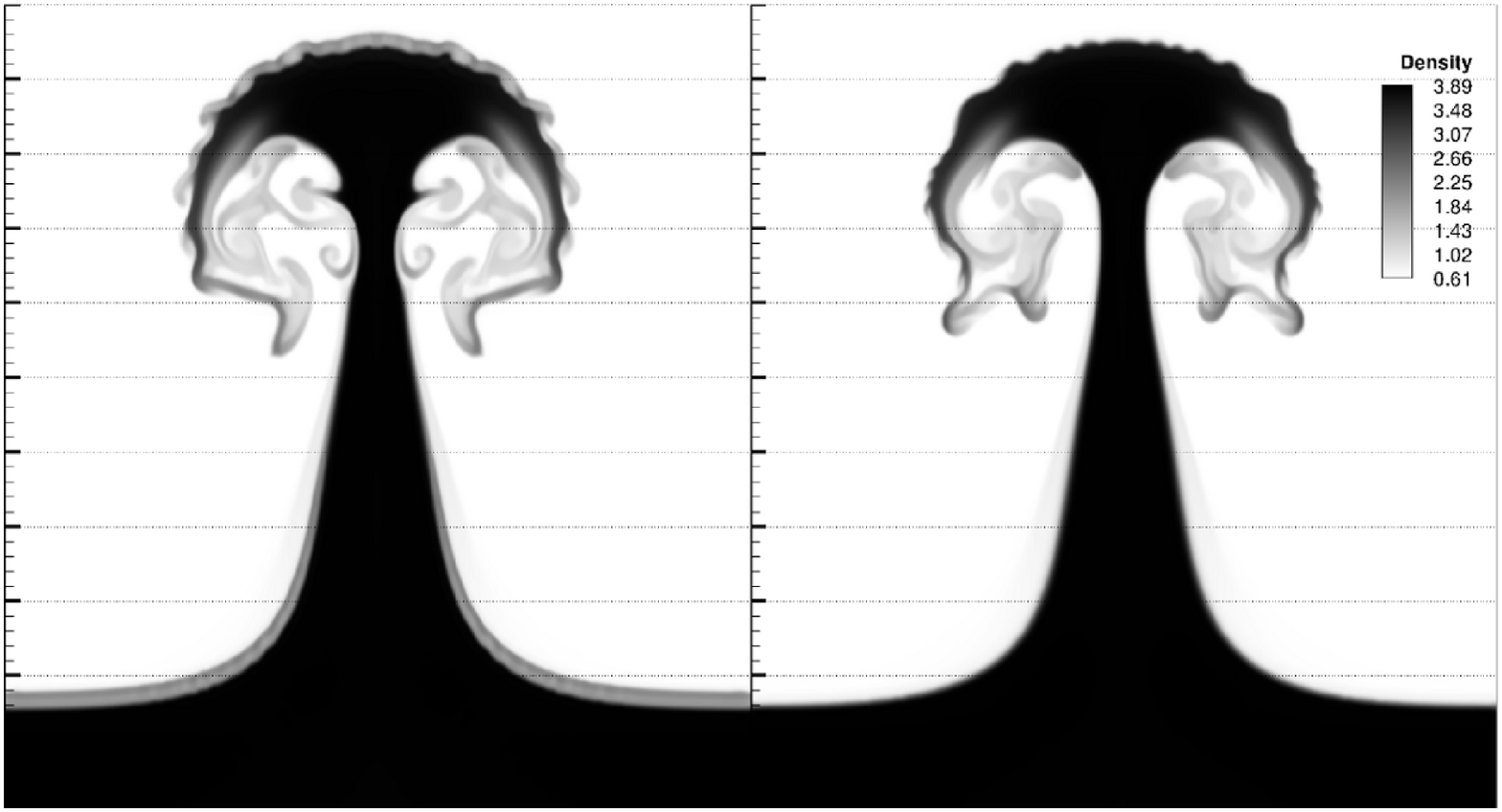}
    \small
    \put(105,96){NF}
    \put(5,96){MF}
\end{overpic}}
\subfigure[Temperature for the viscous models]{\begin{overpic}[width=0.49\textwidth]{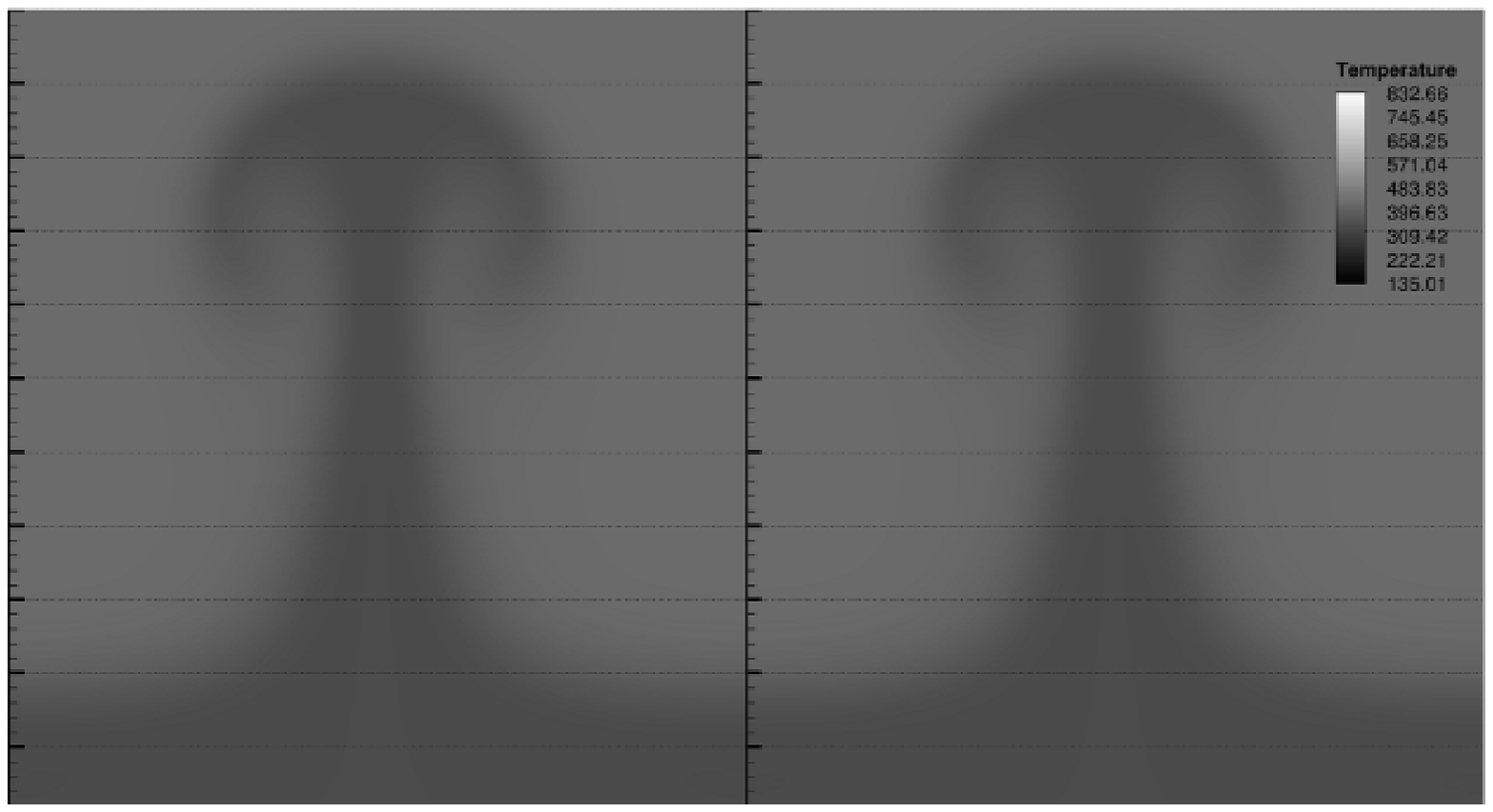}
    \small
    \put(105,96){NF}
    \put(5,96){MF}
\end{overpic}}
\subfigure[Temperature for the inviscid models]{\begin{overpic}[width=0.49\textwidth]{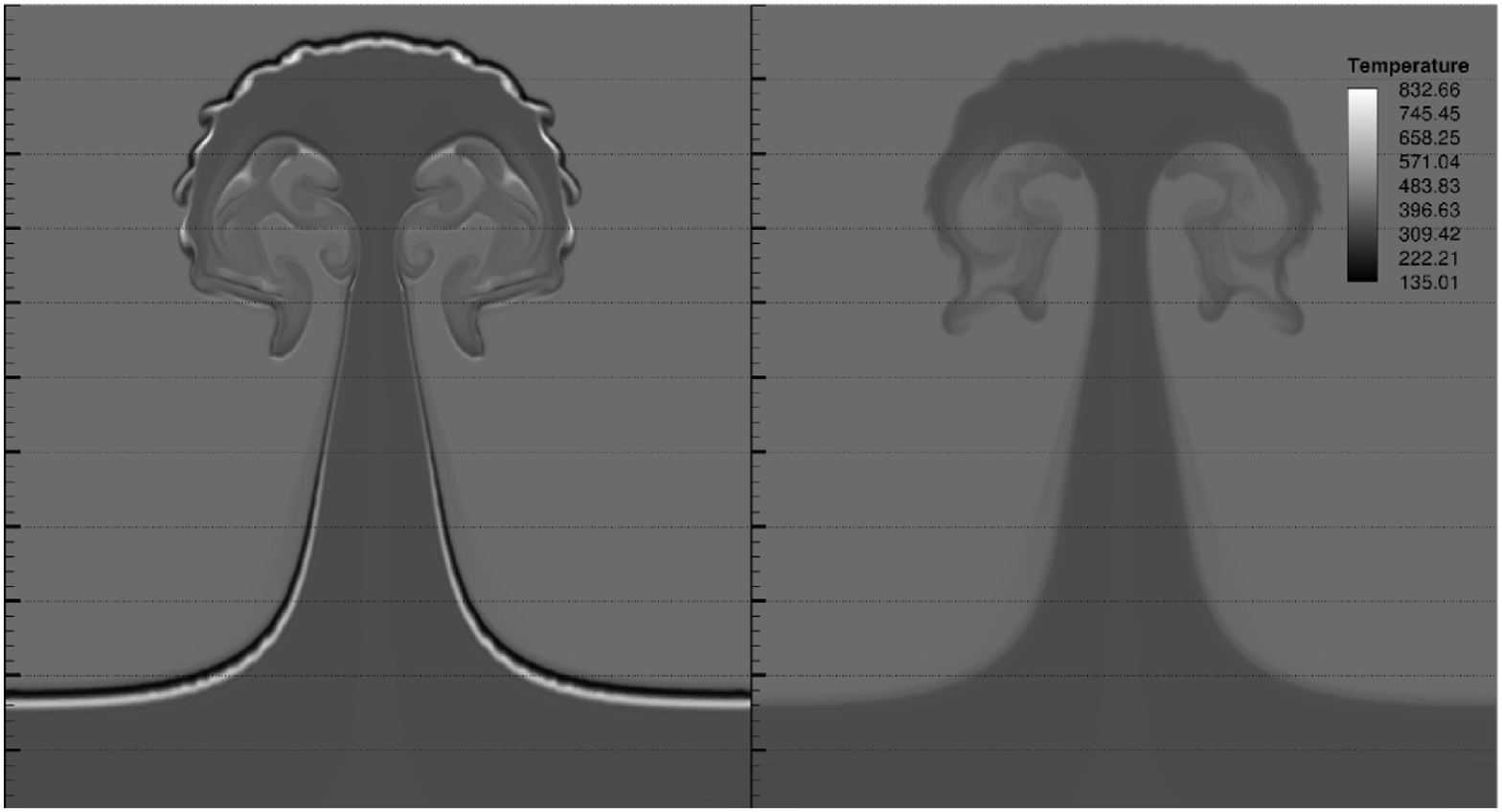}
    \small
    \put(105,96){NF}
    \put(5,96){MF}
\end{overpic}}
\caption{Density (top) and temperature (bottom) contours for a single mode RM instability from SF$_6$ to air. For each image, the left hand panel is a computation for the mass fraction (MF) model, the right for the number fraction (NF) model \label{Case4Vis}}
\end{centering}
\end{figure}

Figure \ref{Case4Vis} shows visualisations of inviscid and viscous computations with the number fraction and mass fraction models. In this problem, the inviscid computations highlight the worst of the numerical errors in the mass fraction model, where temperature errors of several hundred Kelvin appear across the layer. The inviscid computation clearly shows the benefit of the number fraction approach for inviscid advection. 

The addition of viscosity, diffusion and thermal conductivity reduces the error in the mass fraction model results, such that both models converge to the same solution as expected. The converged solutions are visualised in Figure \ref{Case4Vis} (a) and (c) and are identical for both models.

However, the spurious temperatures produced in the inviscid component of the mass fraction model are expected to impact on the convergence of the solution, as highlighted in the previous periodic advection case. Here, three quantitative measures are employed to explore this expectation. Three useful common measures of the time evolution of the layer are the integral mixing width $W$, molecular mixing fraction $\Theta$ and the mixing parameter $\Xi$ (see, for example \cite{Youngs1991,Youngs1994,Cook2002,Latini2007,Thornber2010}) are defined as:

\begin{equation}
W=\int^{L_x}_0 \langle X_1\rangle \langle X_2\rangle dx,\hspace{0.25cm} \Theta=\frac{\int \langle X_1 X_2 \rangle dx}{\int
\langle X_1 \rangle \langle X_2 \rangle dx}
\label{intwidtheq}
\end{equation}

\begin{equation}
\Xi=\frac{\int \langle \min(X_1,X_2) \rangle dx}{\int
\min(\langle X_1 \rangle,\langle X_2 \rangle) dx},
\end{equation}

\begin{figure}
\begin{centering}
\includegraphics[width=0.48\textwidth]{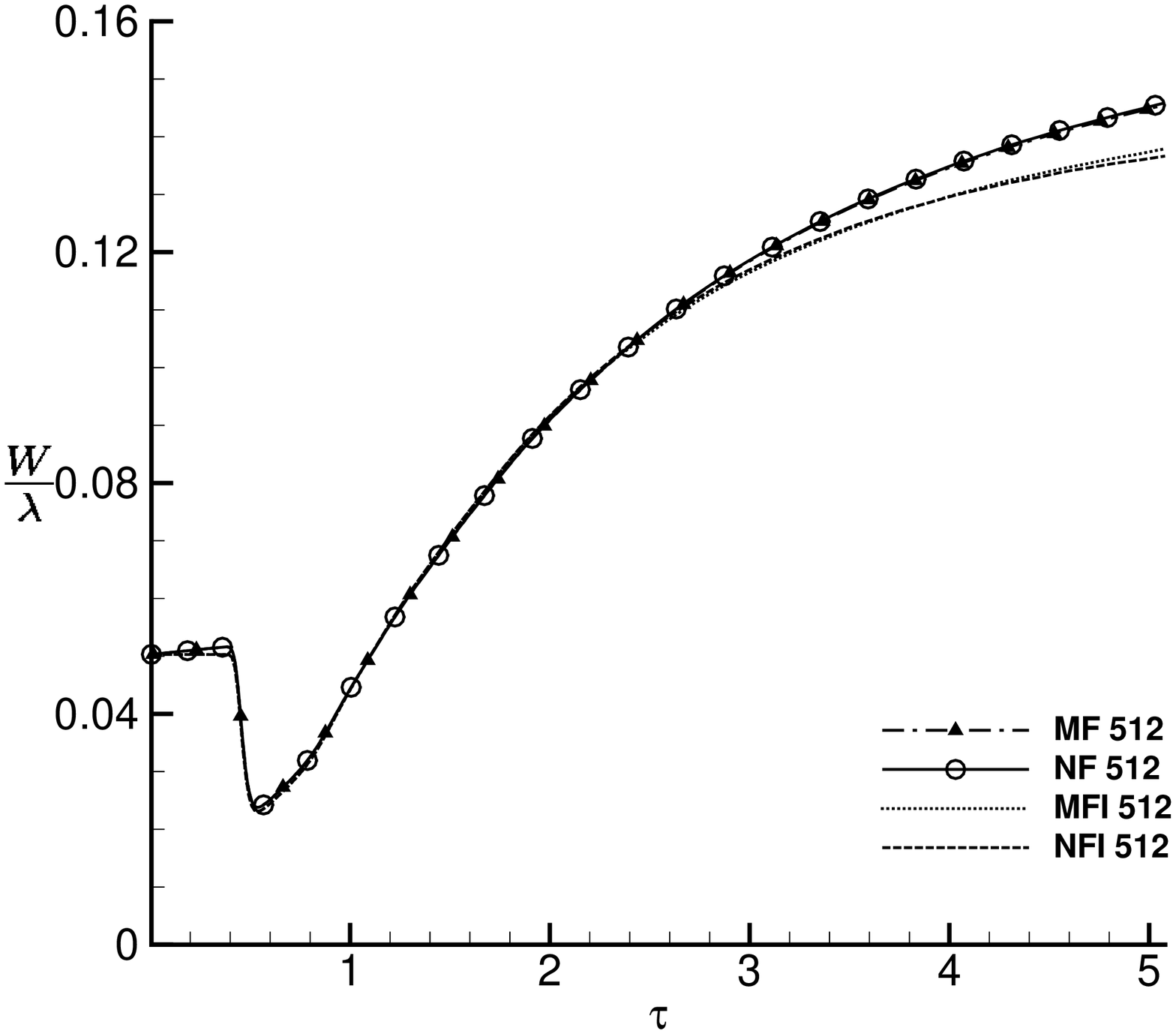}
\includegraphics[width=0.48\textwidth]{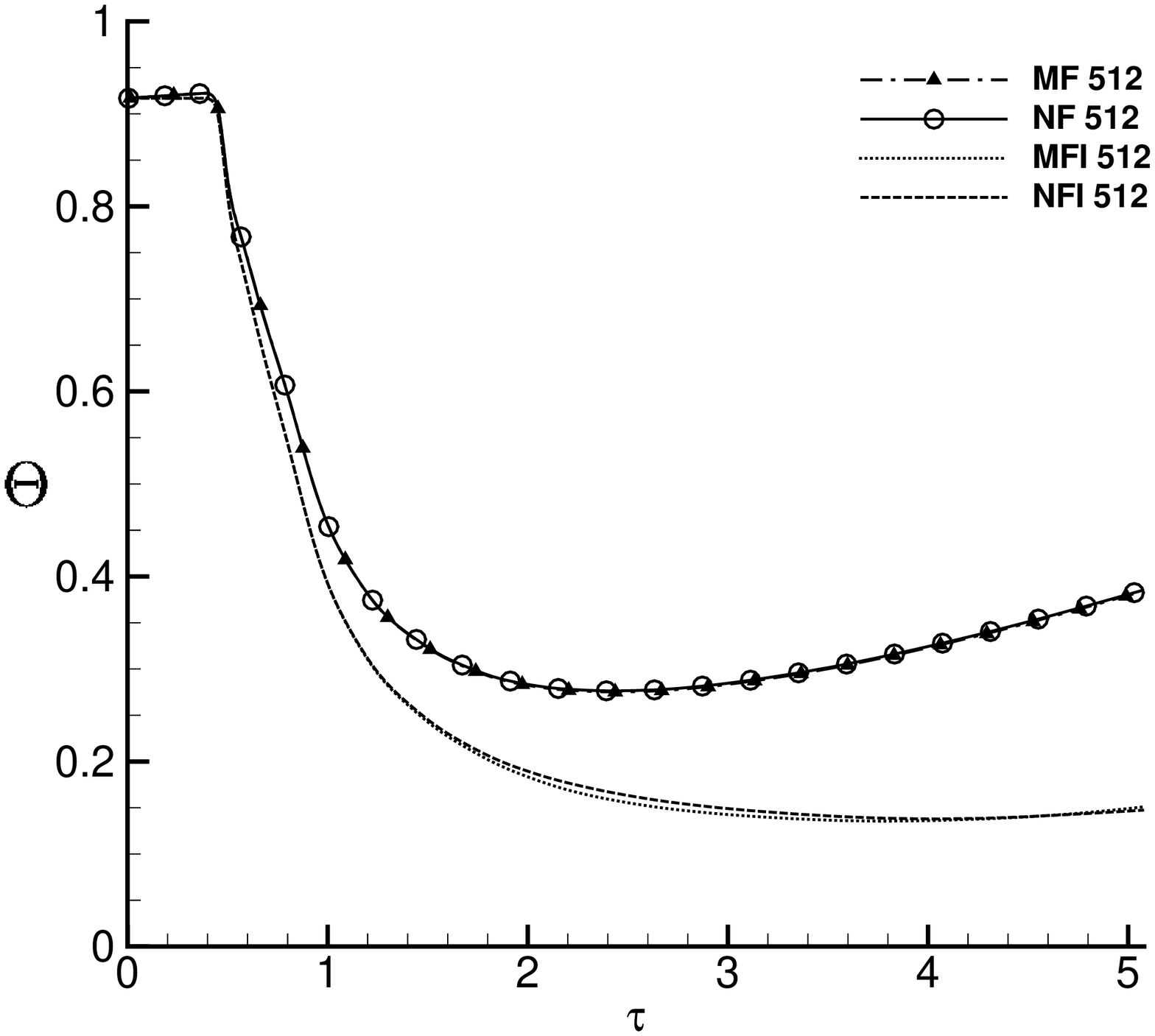}
\includegraphics[width=0.48\textwidth]{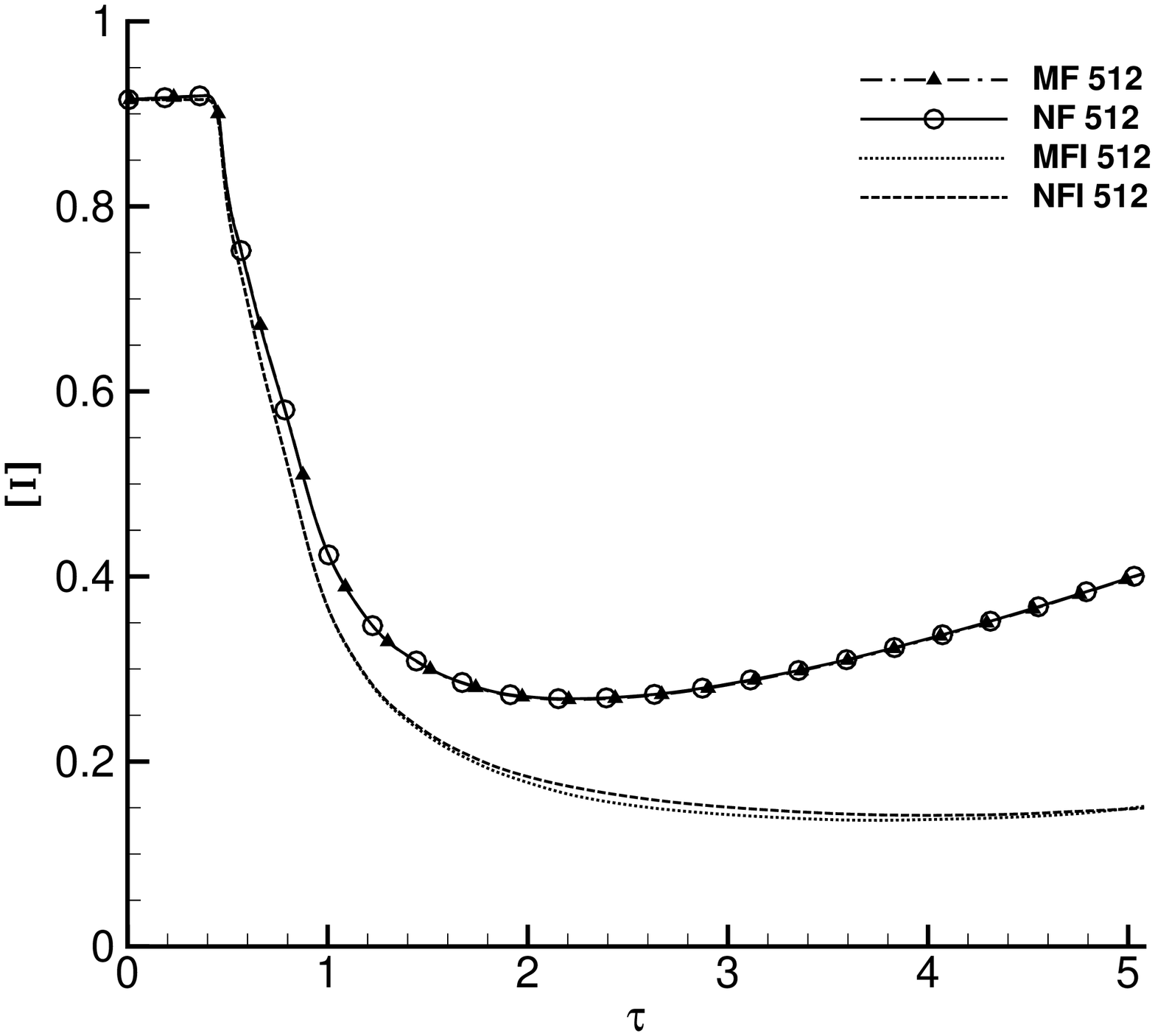}
\caption{Comparison of integral properties at the converged grid resolution for the mass fraction (MF) and number fraction (NF) models. Also plotted are the mass fraction inviscid (MFI) and number fraction inviscid (NFI) results. \label{fig:Case4Plots2}}
\end{centering}
\end{figure}

These quantities are plotted in Figure \ref{fig:Case4Plots2} for the mass fraction and number fraction models at the finest grid resolutions, showing that the respective quantities have converged to the same solution regardless of model. There are slight differences in the inviscid solution, however the general trend is that the inviscid case has lower width at late times, and lower mixing parameters. 

\begin{figure}
\begin{centering}
\includegraphics[width=0.48\textwidth]{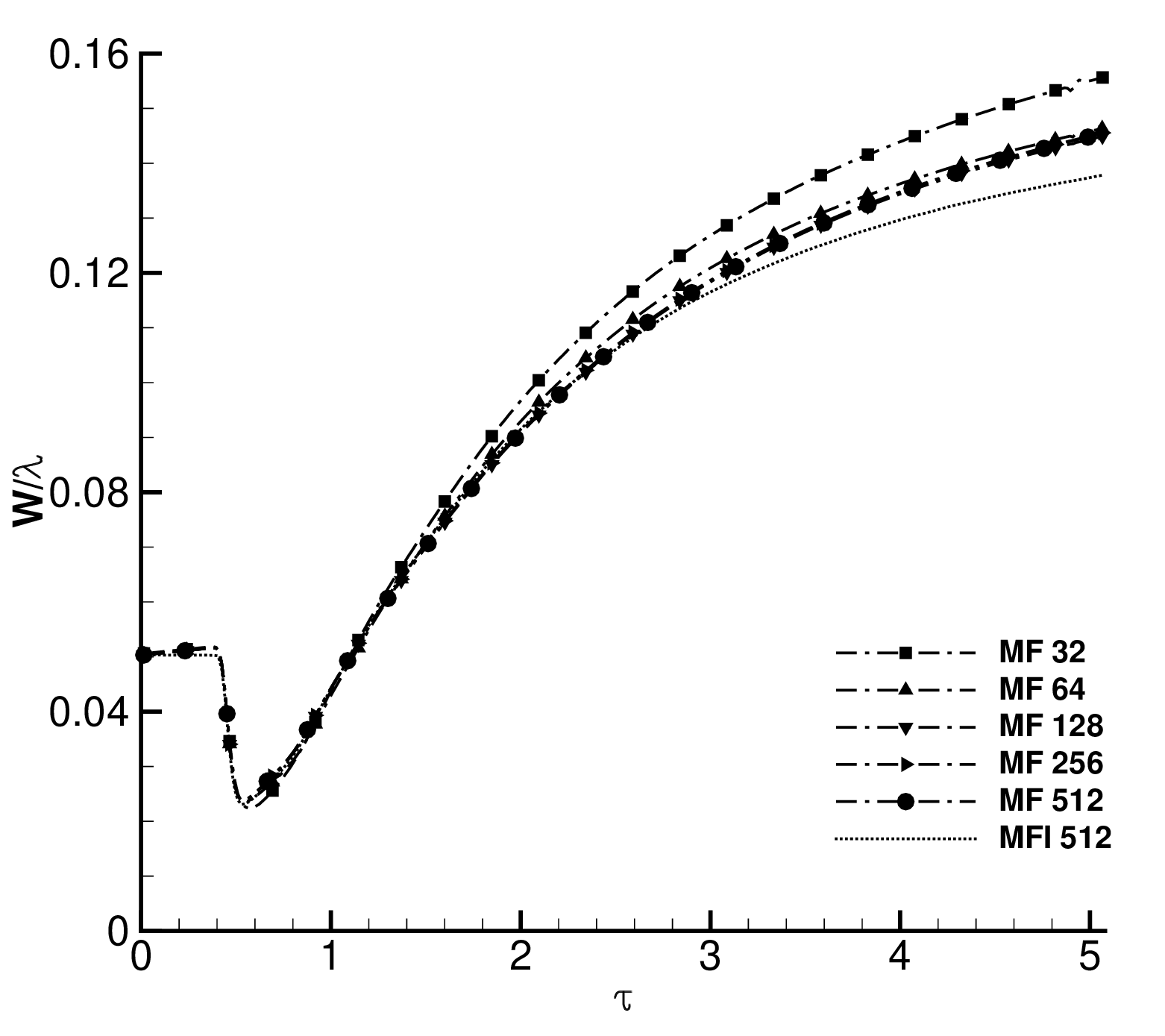}
\includegraphics[width=0.48\textwidth]{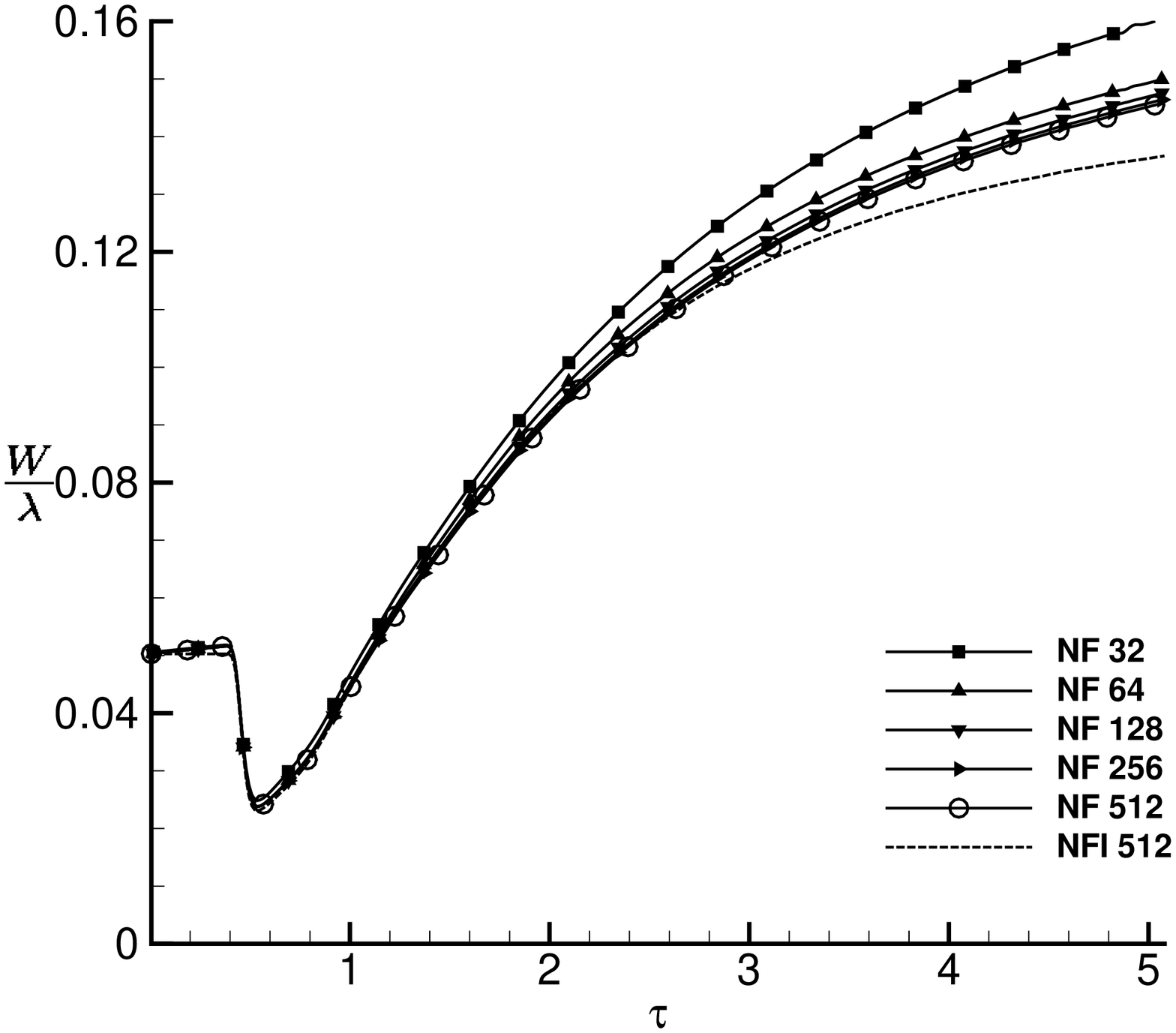}
\includegraphics[width=0.48\textwidth]{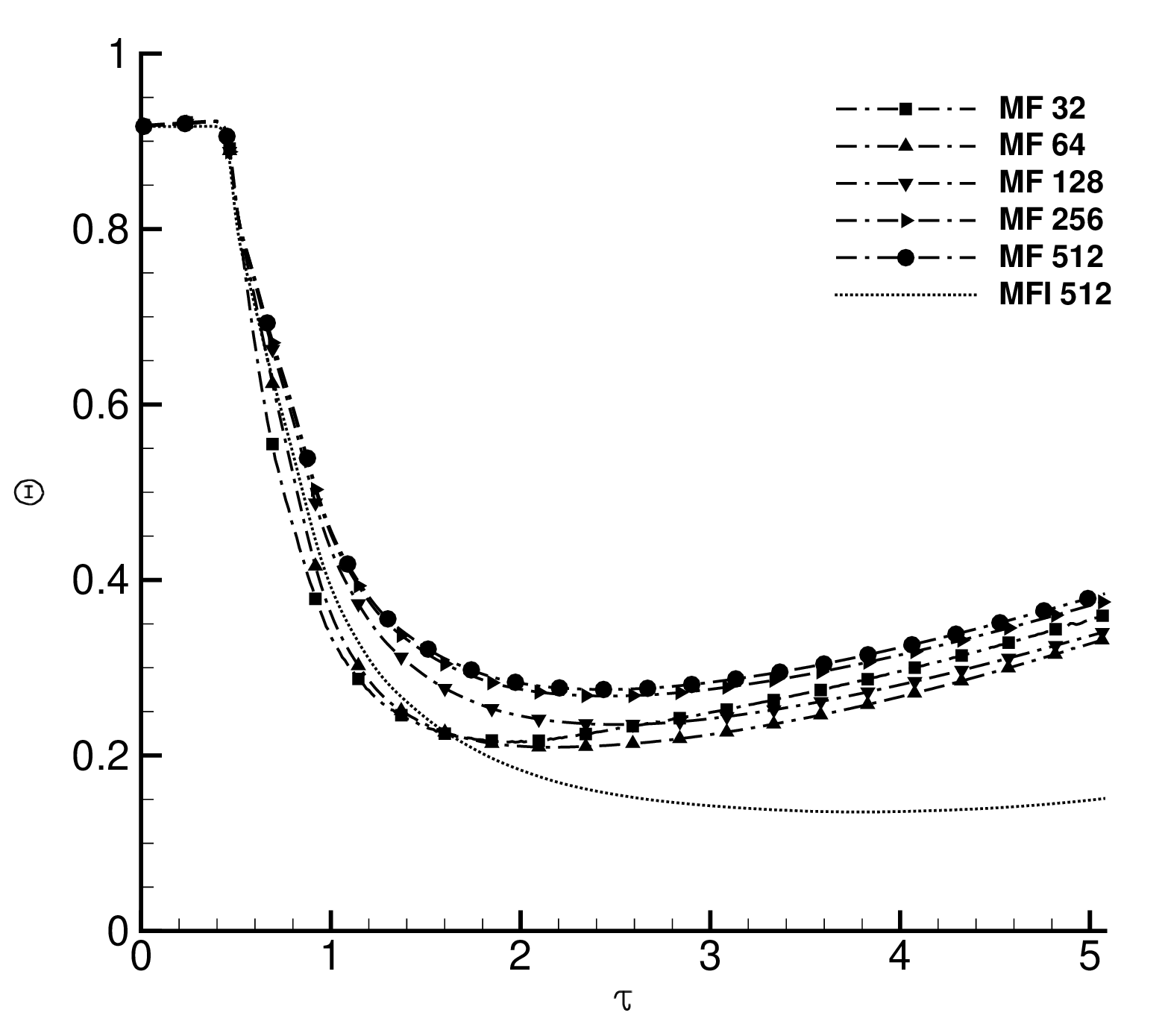}
\includegraphics[width=0.48\textwidth]{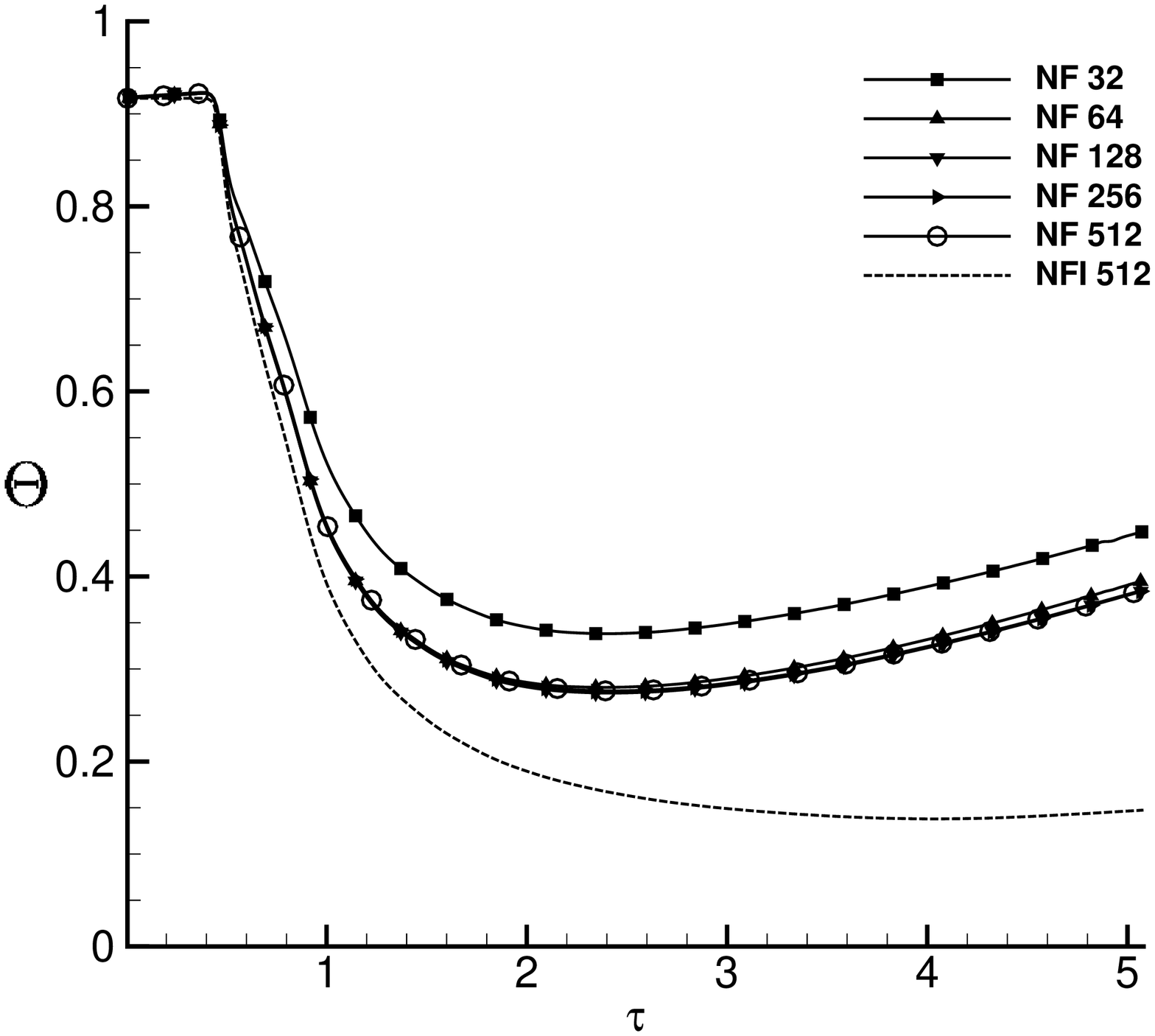}
\includegraphics[width=0.48\textwidth]{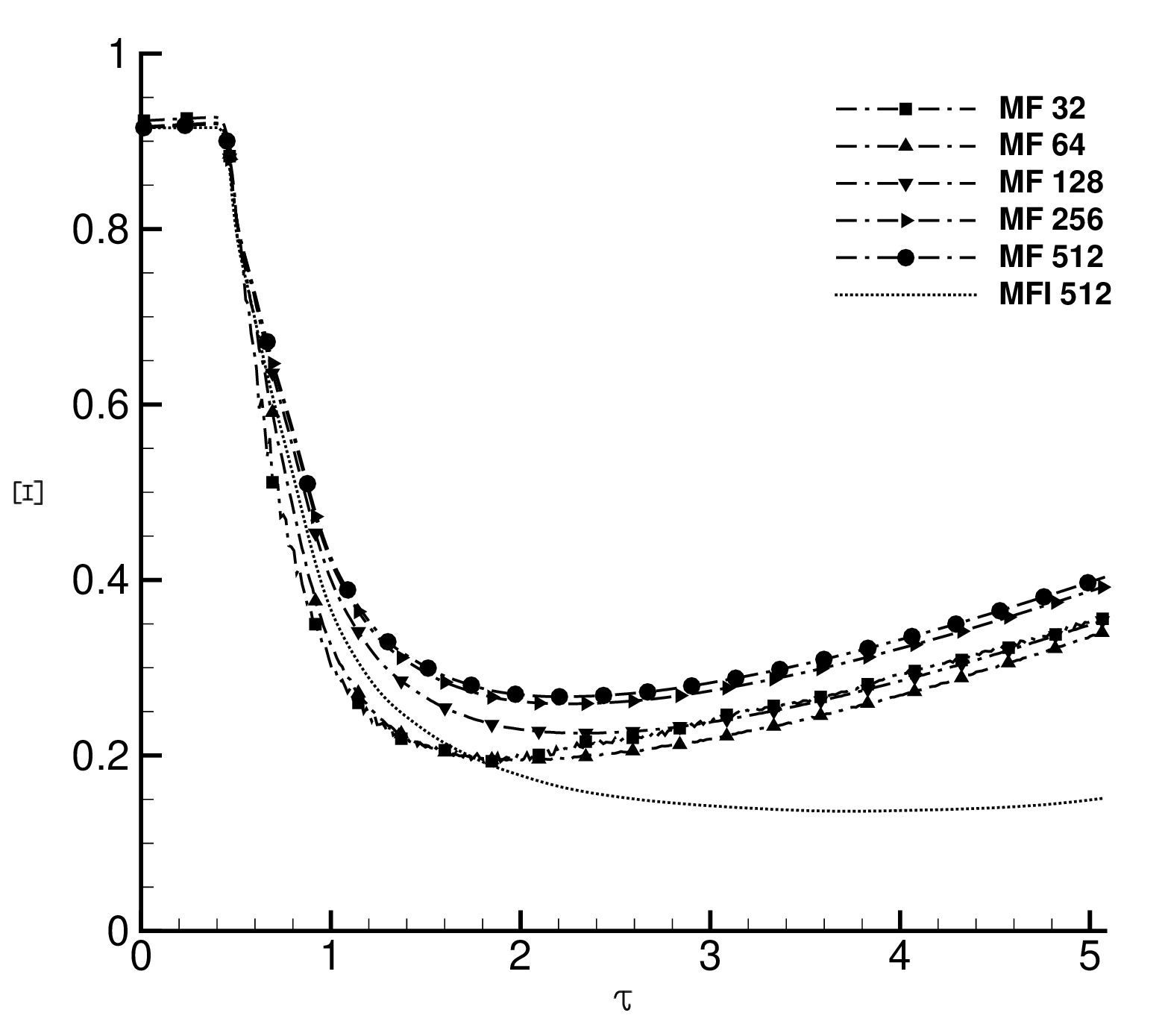}
\includegraphics[width=0.48\textwidth]{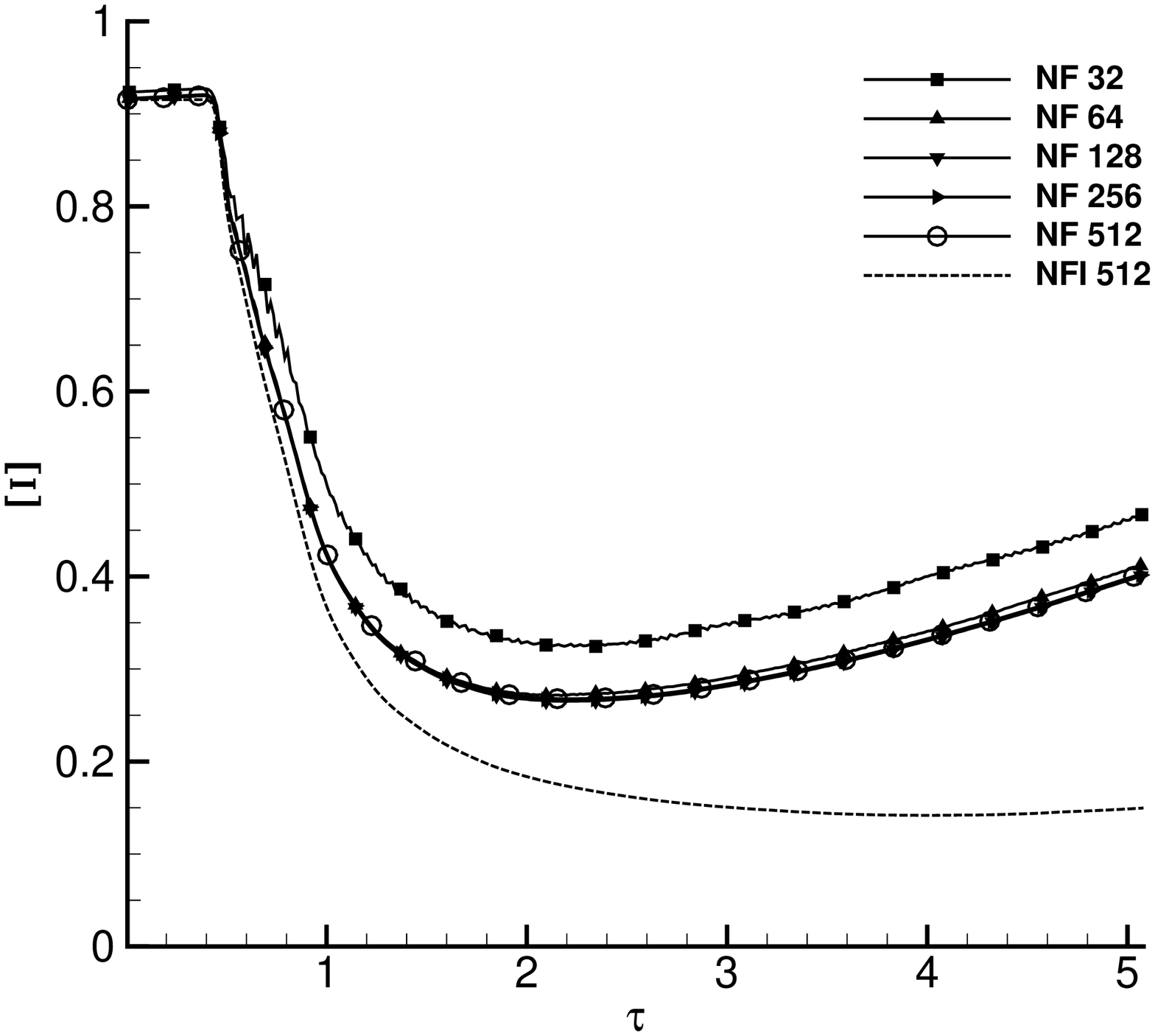}
\caption{Comparison of variables across all grid resolutions (points per wavelength) for Case 4 with the mass fraction (MF) model results on the left and the number fraction (NF) model results on the right, including mass fraction inviscid (MFI) and number fraction inviscid (NFI) results. \label{fig:Case4Plots1}}
\end{centering}
\end{figure}

Figure \ref{fig:Case4Plots1} plots $W$, $\Theta$ and $\Xi$ for cross-sectional resolutions from $32$ to $512$ points per wavelength (ppw) for the mass and number fraction models. Firstly, the integral width $W$ converges at $128$ ppw for both schemes. Rather surprisingly, the mass fraction model appears to converge faster. The mixing parameters $\Theta$ and $\Xi$ follow the expected trends, the number fraction model converging at $64$ ppw  while the mass fraction model varies substantially until $256$ ppw. 

\begin{figure}
\begin{centering}
\includegraphics[width=0.48\textwidth]{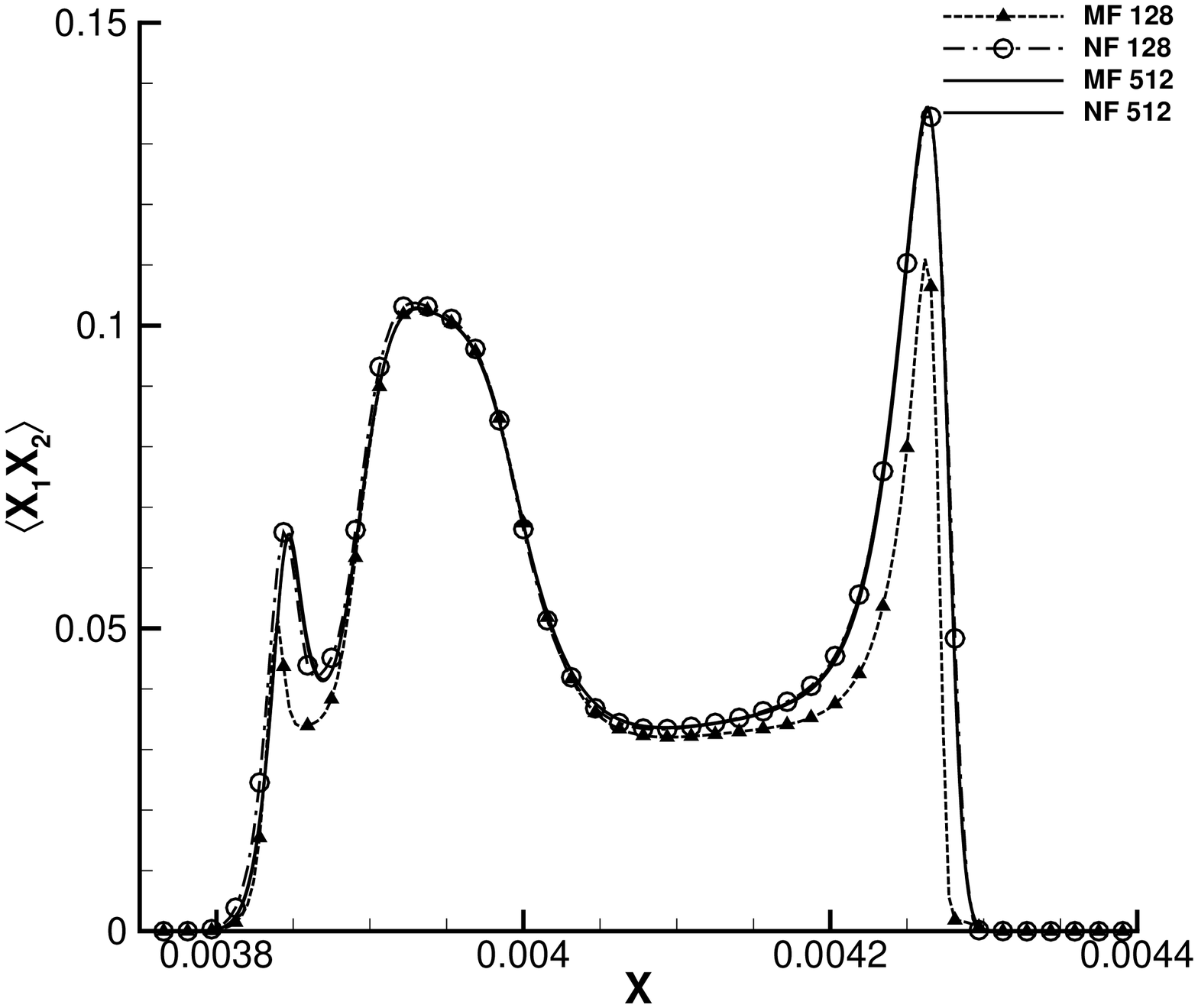}
\includegraphics[width=0.48\textwidth]{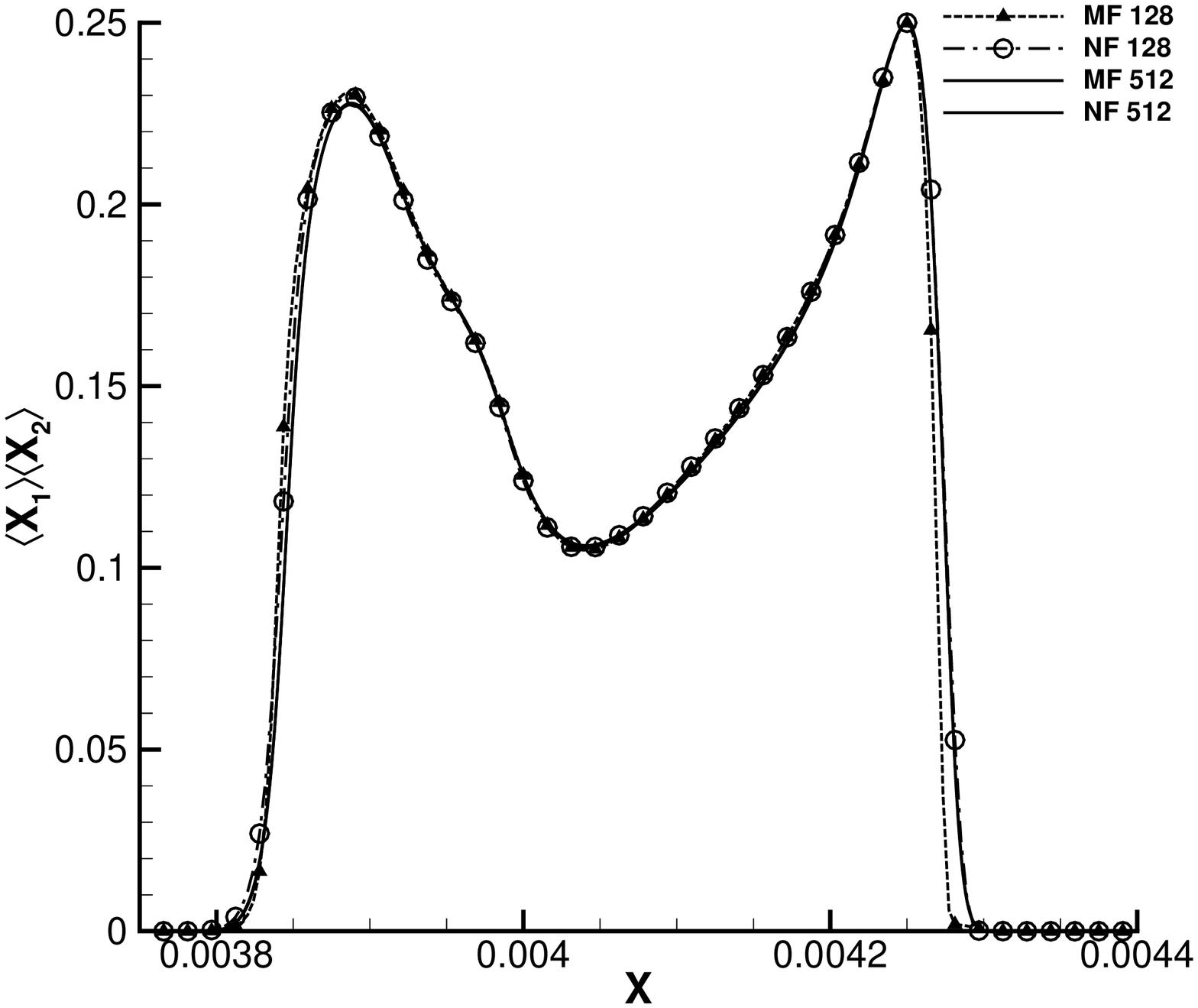}
\caption{Plot showing the specific terms employed to compute the integral properties for the mass fraction (MF) and number fraction (NF) models at converged (512 points) and non-converged (128 cells) grid resolutions. \label{fig:Case4PlaneAverages}}
\end{centering}
\end{figure}

Thus the observations of errors in the prediction of $\Theta$ and $\Xi$ are as expected from the previous cases, however $W$ is not. Figure \ref{fig:Case4PlaneAverages} plots the $x$ variation of the planar averages $\langle X_1\rangle \langle X_2 \rangle$ and $\langle X_1  X_2 \rangle$ for the mass fraction model and number fraction model both at the converged grid resolution ($512$ ppw) and at an intermediate resolution ($128$ ppw). 

The integral width is the area under the curve $\langle X_1\rangle \langle X_2 \rangle$. From this Figure it can be seen that the mass fraction model has larger errors in $\langle X_1 X_2 \rangle$ than the number fraction model, however those errors are approximately equal on both sides of the mixing layer. Thus although the integral width appears to converge faster for the mass fraction model, it is simply that the errors are cancelling each other. 

Turning to $\langle X_1\rangle \langle X_2 \rangle$, the integral of which is the numerator in the $\Theta$ equation, there is a very large error when using the mass fraction model compared to the number fraction model at $128$ ppw.  

Overall, the solution of the number fraction model at $64$ ppw has an equivalent error to the mass fraction model $256$ ppw computed using identical variable reconstruction, Riemann solver and discretisation of the viscous terms. This represents an enormous computational saving for equivalent error, a factor of $16$ fewer points in 2D along with a factor of $4$ fewer time steps. Even with an observed (unoptimised) increase in computational cost of $\approx 50$\%, the computations are on the order of $40$ times faster for equivalent accuracy. For three-dimensional flows the gain would be expected to be greater than $100$. This clearly demonstrates the superiority of the newly proposed number fraction based governing equations. 

\section{Conclusions \label{concl}}

This paper has presented a new number fraction based five-equation model for miscible fluids incorporating viscosity, species diffusivity and thermal conduction. An equation is solved for the variable $\overline{X}$ which reduces to the usual definition of volume fraction when there is a sharp interface in a computational cell and to the species number fraction if there is homogeneous mixing within the cell. A numerical scheme has been derived to solve these equations which is second order accurate in space and time, and respects the underlying physics. 

Four test cases have been proposed, and computations have been run using the standard mass fraction model and the newly proposed number fraction model.These test cases were designed such that they validate and verify all key terms in the governing equations. Three of the cases have analytical solutions in the incompressible limit which were used to compute the observed order of accuracy.  

The discretisation of the number fraction and mass fraction models were both demonstrated to be second order accurate in space. This was shown to be valid up to the point at which compressibility effects became important. The first two cases are relatively straightforward for both models, having either constant temperature or constant ratio of specific heats. However, the errors of the  number fraction model were consistently two or three times lower than that of the mass fraction model for the same grid.

The third test case was designed to be more challenging. The third test case consists of a diffusing contact surface between fluids of different thermodynamic properties moving with a mean velocity. Although both models converge at second order accuracy, the number fraction model is one order of magnitude more accurate for a given grid resolution. 

The final test case is of a commonly employed gas combination in experiments of shock-induced turbulent mixing. A shock wave impinges on to a two-dimensional perturbed interface between  air and SF$_6$, triggering the growth of a Richtmyer-Meshkov instability. In this test case, the number fraction model is approximately converged on all measures at $128$ cross-sectional grid resolution, whereas the mass fraction model is not converged until $512$ points. This represents a computational saving of approximately $40$ times for the equivalent accuracy.

Based on these results, the number fraction model is clearly superior to the standard mass fraction approach for the computation of compressible turbulent mixing problems of miscible fluids with distinct thermodynamics properties. This formulation is directly applicable to Direct Numerical Simulations (as undertaken here) and Large-Eddy-Simulations.

\section{Acknowledgements}

This research was supported under Australian Research Council's Discovery Projects funding scheme (project number DP150101059). The authors would like to acknowledge the computational resources at the National Computational Infrastructure through the National Computational Merit Allocation Scheme which were employed for all cases presented here. Ben Thornber would also like to acknowledge the kind hospitality of Prof. Toro at the University of Trento where portions of this research were undertaken.

\bibliography{bibliography}

\end{document}